\documentclass[9pt,twoside]{rmaa-rho}
\RMxAAtemplatetype{\RMxAA} 

\usepackage{graphicx} 
\usepackage{amsmath} 
\usepackage{caption}
\usepackage{rotating}
\usepackage{siunitx}
\usepackage{longtable}
\usepackage{xtab}

\vol{100}
\pages{1000-1006}
\thisyear{2026}
\doi{\href{https://doi.org/10.22201/ia.01851101p.20XX.XX.XX.XX}{https://doi.org/10.22201/ia.01851101p.20XX.XX.XX.XX}}

\title{High-Angular-Resolution Survey of Stars within 18 pc of the Sun: Part I}

\author[1]{F.~C.~Rojas-Galindo \orcidlink{0009-0002-9628-4230}}
\author[1]{V.~G.~Orlov \orcidlink{0000-0002-4650-2984}}
\author[2]{R.~A.~Mendez \orcidlink{0000-0003-1454-0596}}
\author[3]{W.~Brandner \orcidlink{0000-0003-1939-6351}}

\affil[1]{Universidad Nacional Autónoma de México, Instituto de Astronomía, CDMX 04510, México}
\affil[2]{Departamento de Astronomía, Universidad de Chile, Casilla 36-D Santiago, Chile}
\affil[3]{Max-Planck-Institut für Astronomie, Königstuhl 17, 69117 Heidelberg, Germany}

\leadauthor{Rojas-Galindo et al.}
\smalltitle{LaTex Macro RMxAA}

\corres{F. Carolina Rojas-Galindo}
\email{fcrojas@astro.unam.mx}

\received{May 13, 2026}
\accepted{September 21, 2026}

\license{Texto de la licencia aquí}

\setbool{rho-abstract}{true} 
\setbool{rho-resumen}{true} 

\begin{abstract}
    We present the rationale, design, methodology, and observational facilities of a high-angular-resolution imaging survey of stars within 18~pc of the Sun. Our comprehensive sample consists of 2047 stars compiled from \textit{Gaia} eDR3 and complementary catalogs. The main goal is to observe as many stars from this sample as possible using high-angular-resolution techniques, considering their inherent limitations, to perform astrometry of known binary systems and determine detection limits for secondary components for each observed star. This approach may also allows us to identify previously unresolved stellar companions. Observations were carried out using the 2.1-m telescope at OAN-SPM in the North and the 3.58-m NTT in the South, employing high-speed cameras for speckle imaging. We obtained observations of 703 stars with $V < 14$~mag at OAN-SPM, and 645 stars with $I < 16$~mag at the NTT, of which 140 stars were observed in both hemispheres, for a total of 1208 unique stars observed. In this paper: (1) we present the construction of the catalog of 2047 stars within 18 pc; (2) using a control sample of 70 known close binaries within 18 pc with separations < \SI{1}{\arcsecond}, we show that selecting targets based on \textit{Gaia} quality parameters commonly used as binarity indicators--RUWE, {\tt ipd\_frac\_odd\_win}, and {\tt ipd\_frac\_multi\_peak}--can miss a significant fraction of binaries, justifying our unbiased, volume-limited survey; and (3) we present the results of pilot observations conducted with the 2.1 m telescope at OAN-SPM in July 2022. A full analysis of the complete sample will be presented in of forthcoming papers.
\end{abstract}


\begin{resumen}
    Presentamos el razonamiento, diseño, metodología e instalaciones observacionales de un estudio de imágenes de alta resolución de estrellas dentro de 18 pc del Sol. Nuestra muestra integral consta de 2047 estrellas recopiladas de \textit{Gaia} eDR3 y catálogos complementarios. El objetivo principal es observar tantas estrellas de esta muestra como sea posible usando técnicas de alta resolución angular, considerando sus limitaciones inherentes, para realizar astrometría de sistemas binarios conocidos y determinar los límites de detección de componentes secundarios para cada estrella observada. Este enfoque podría también permitirnos identificar componentes estelares no resueltos previamente. Las observaciones se realizaron usando el telescopio de 2.1 m del OAN-SPM en el norte y el NTT de 3.58 m en el sur, empleando cámaras de alta velocidad para imágenes speckle. Obtuvimos observaciones de 703 estrellas con $V < 14$~mag en OAN-SPM, y 645 estrellas con $I < 16$~mag en el NTT, de las cuales 140 estrellas fueron observadas en ambos hemisferios, para un total de 1208 estrellas únicas observadas. En este artículo: (1) presentamos la construcción del catálogo de 2047 estrellas a 18 pc; (2) usando una muestra de control de 70 binarias cercanas (dentro de 18 pc) con separaciones $<$ \SI{1}{\arcsecond} mostramos que seleccionar objetos con base en parámetros de calidad de \textit{Gaia} comúnmente usados como indicadores de binariedad (RUWE, {\tt ipd\_frac\_odd\_win}, y {\tt ipd\_frac\_multi\_peak}) puede excluir una fracción significativa de binarias, lo que justifica nuestro survey no sesgado y limitado por volumen; y (3) presentamos los resultados de las observaciones piloto, realizadas con el telescopio de 2.1 m de OAN-SPM en julio de 2022. El análisis de la muestra completa será presentado en una serie de próximos artículos.
\end{resumen}

\begin{document}

\maketitle
\pagestyle{fancy}\thispagestyle{firststyle}

\nolinenumbers
 
\section{INTRODUCTION}

The study of nearby stars is of significant importance, as accurately estimating their diverse astrophysical parameters can improve our understanding of the Galactic disk. For example, their mass distribution provides an initial estimate of the mass density near the Galactic plane. The distribution of their ages serve as a record of the star formation history within the disk, while their metallicities trace its chemical enrichment history \citep{Nordstrom2004}. Among these nearby stars, a significant fraction are members of binary or multiple systems, making the study of the stellar multiplicity an important component of understanding the solar neighborhood.

It is well known that stars tend to form in pairs and in higher-order hierarchies. Almost all stars have been, or are, gravitationally bound to one or more companions since their formation \citep{Lee2019}, and a large fraction of these systems have survived throughout their dynamical evolution \citep{Moe2017}. The study of binary and multiple star systems as well as binarity fraction and multiplicity fraction is essential for determining the frequency, general characteristics, and dependence of stellar systems on the initial mass. Additionally, such studies provide a direct link between theoretical stellar models and observational parameters--for example, the mass-luminosity relation, which is derived from the measured masses of binary components \citep{Andersen1991,Henry1999}. However, the scope of this work is not to compute and study multiplicity statistics in general, but rather to provide the high-angular-resolution observational database that may serve as a useful resource that can contribute to achieving that scope in future multiplicity studies.

It is well established that the presence of one or more stellar components can significantly impact stellar evolution. Therefore, it is crucial to confirm whether the observed stars are truly single or part of multiple systems. High-resolution imaging techniques have become increasingly vital for confirming or ruling out the multiplicity of the stars.  

In addition, high-resolution techniques are critical for detecting unresolved objects within space missions which have carried out extensive surveys of the solar neighborhood; as their wide fields of view and large pixel sizes often result in low spatial resolution. A representative example of this is the TESS (Transiting Exoplanet Survey Satellite) space telescope--dedicated to the search for exoplanets, especially Earth-sized planets within or near the habitable zone, where planetary temperatures are suitable for the existence of liquid water on the surface--which although covers a wide field of view, it has very large pixels: 21 arcsec/pixel \citep{Ricker2015}. This means that light is collected from any star within the photometric aperture, but due to the large pixel size, the target star can be contaminated by companions or field stars that remain unresolved. When transit-like events are detected, it is often impossible to determine with certainty which star produced the signal. Consequently, such an event is categorized as a candidate exoplanet transit, and the star is designated as a TOI (TESS Object of Interest), requiring follow-up and validation to determine its true nature. High-resolution techniques improve the validation of exoplanets and are vital for characterizing host stars; this is essential for accurately deriving exoplanet parameters, as the existence of unresolved stellar companions can significantly bias the resulting properties \citep{Ciardi2015,Furlan2017}. The TESS mission includes millions of targets identified in the TESS Input Catalog (TIC), with the TOI catalog representing the subset of TIC stars that exhibit potential exoplanet signals. TESS thus exemplifies the need for high-angular-resolution follow-up of wide-field survey missions; in its particular case, such observations are essential to rule out companions and validate exoplanet candidates. Among nearby stars, a few have been identified as TOIs, making high-angular-resolution follow-up of these targets especially relevant.

Similarly, although the \textit{Gaia} mission has significantly improved angular resolution, it still faces limitations when resolving tight stellar companions. Consequently, stars that appear to be single may, in fact, be binaries or even multiple systems \citep{Howell2021}. The \textit{Gaia} mission provides parameters that can serve as binarity indicators for unresolved binary stars. However, these parameters are statistical in nature and do not have a one-to-one correspondence with binarity. Therefore, selecting targets based solely on \textit{Gaia} binarity indicators would cause an important fraction of unresolved binaries to be missed as they appear as single sources according to these indicators (see Section \ref{sec:param}). This motivates the need for an unbiased\footnote{The term "unbiased" refers specifically to our target selection strategy: we observe every star in the volume-limited sample that satisfies only the physical constraints inherent to our observations--the magnitude limit and declination range imposed by the observational techniques and geographic locations of the telescopes, respectively--regardless of any astrophysical property or statistical indicator (see Sec \ref{subsec:SB}).} high-angular-resolution survey of all stars within a volume-limited sample, regardless of any pre-selection based on \textit{Gaia} parameters.  

Volume-limited samples of nearby stars have been compiled from multiple catalogs regardless of physical or statistical parameters \citep{Kirkpatrick2024,Golovin2023}. However, high-angular-resolution observational surveys have typically been restricted to specific spectral types or mass ranges \citep{Raghavan2010,Delorme2012,Janson2012,Winters2021,Clark2024,Vrijmoet2026}, or to stars selected by specific \textit{Gaia} parameters or known binary status \citep[e.g., the speckle interferometry program at SOAR\footnote{Southern Astrophysical Research (SOAR) telescope.} described in][]{Tokovinin2025}. In contrast, our survey observes all accessible stars across both hemispheres independently of such restrictions--an approach that, to our knowledge, has not been previously adopted in any high-angular-resolution observational survey. The most comparable approach to ours is \citet{Salama2022}; however, that survey is limited to the northern hemisphere (Dec $> -13^{\circ}$) and the diffraction limit of their telescope (\SI{0.15}{\arcsecond}) is approximately twice that of our 2.1-m telescope (70 mas) at the Observatorio Astronómico Nacional San Pedro Mártir (OAN-SPM), meaning that our survey can resolve closer companions than the survey of \citet{Salama2022}. 

The goal of conducting an unbiased, volume-limited, high-angular-resolution survey is to observe as many accessible stars as possible within 18 pc of the Sun, taking into account the inherent magnitude limits of the observational techniques and the geographic locations of the telescopes. Our scientific objectives are: 1) to perform astrometry on all known close binary systems that we can observe, providing new data points to improve existing orbital solutions and extending the temporal baseline of the measurements; 2) to determine detection limits for secondary components for each observed star, establishing the faintest companion detectable given the observational constraints; and 3) as a further result, to identify previously undetected close binary stars, knowing that, given the proximity of our sample to the Sun, it is possible that no new close binary stars are detected. Given the scope of the survey, our strategy prioritizes single-epoch observations of each individual star, in order to cover as many targets of the sample as possible across the entire celestial vault. However, several stars were observed in more than one observing run, mainly due to the weather or data-quality issues. Moreover, although multiple epochs are generally desirable for astrometric orbit determination, a single new epoch is sufficient to extend the temporal baseline of known binary stars, which is one of the primary contributions of this survey. Additionally, in the case of finding a new, close double star, it is more likely to be a bona fide binary due to its small angular separation ($\rho <$ \SI{1}{\arcsecond}), thus a single observation could be sufficient for a preliminary identification.

The spirit of this survey originated in the northern hemisphere, where the observations were initially planned to be conducted at OAN-SPM. The distance cutoff of 18 pc was chosen based on the diffraction limit of the 2.1-m telescope at OAN-SPM--70 mas at a wavelength of 550 nm (corresponding to the $V$-band)--which allows us to resolve double stars at 18 pc with angular separations $\rho > 70$ mas (corresponding to physical separations greater than $\sim$1.2 AU), within the detection limit of each observed star. Subsequently, the survey was extended to include observations acquired with the 3.58-m New Technology Telescope (NTT) at ESO's La Silla Observatory in the southern hemisphere. This extension improved the angular resolution--the NTT achieves a diffraction limit of 56 mas at 800 nm (corresponding to the $I$-band), i.e., physical separations of $\sim$1.0 AU at 18 pc--and increased the number of accessible targets in the observational sample. The instruments installed on both the northern and southern telescopes allow us to reach the diffraction limits of each telescope, enabling us to resolve companions at separations that \textit{Gaia} cannot detect\footnote{The minimum separation achievable for \textit{Gaia} EDR3 depends of the magnitude difference and the orientation of the components with respect to the dominating scan direction. Completeness for small separations is high above \SI{0.7}{\arcsecond}, and only a few pairs were resolved between the hard limit of \SI{0.18}{\arcsecond} and \SI{0.4}{\arcsecond} \citep{Fabricius2021}.}, as well as to complement the observations of nearby stars that are limited by the large pixel size of TESS, thus addressing the limitations of both missions.

The \textit{Gaia} eDR3 database serves as the starting point for building our sample of nearby stars, as it contains the most complete list of stars in the solar neighborhood with parallax uncertainties smaller than 1\% at 18 pc \citep{Gaia2021}. For each star, \textit{Gaia} provides several useful parameters that enable a more detailed analysis, some of which are extensively used as binarity indicators. Additionally, to ensure a more comprehensive selection, we augmented the target list with complementary catalogs, resulting in a sample of 2047 stars. In Section \ref{sec:Sample}, we provide a detailed description of our sample selection and the cross-correlation with other catalogs. In Section \ref{sec:param}, we analyze the reliability of some \textit{Gaia} parameters as binarity indicators. The observations for each hemisphere, along with the observational constraints and strategy, are described in Section \ref{sec:Obs}. Data processing, astrometric calibrations, and preliminary results of the pilot observations carried out in July 2022 with the 2.1-m telescope at OAN-SPM are presented in Section \ref{sec:Pilot}; the full results of the complete sample will be discussed in detail in a series of forthcoming papers. Finally, Section \ref{sec:conclusions} provides the main conclusions of this manuscript.

This is Paper I in a series of forthcoming papers. Paper II will present the complete results of the northern hemisphere observations; Paper III will present the complete results of the southern hemisphere observations; and Paper IV will present the comparison of the overlapping observations from both hemispheres, as well as the analysis and the empirical comparison between \textit{Gaia} binarity indicators for each observed star and the observational results of the complete survey. This will allow us to assess the reliability of these indicators against the control sample presented in Section \ref{sec:param} of this paper. 

\section{SAMPLE SELECTION }
\label{sec:Sample}

The goal of this study is to conduct an unbiased, volume-limited survey of stars within 18 pc of the Sun. To achieve this, we build a comprehensive sample using \textit{Gaia} as baseline and then supplement it with additional catalogs to enhance completeness. This section details the construction of this target list, which comprises 2047 stars.

\subsection{\textit{Gaia} Sample}

The starting point for our sample selection was the {\it Gaia} Early Data Realease 3 (eDR3) \citep{Gaia2021}, using a parallax cutoff as the sole selection criterion. We included all stars with parallaxes $\varpi >$ 55.55 mas, corresponding to distances $d \leq18$ pc. This selection yielded 1,900 sources. For each object, \textit{Gaia} eDR3 provides 99 parameters, including astrometry (equatorial coordinates, parallax, proper motions and their uncertainties), broad-band photometry, and several diagnostic quantities related to the quality of the astrometric and image parameter determination (IPD) solutions. A complete description of these parameters is available in Section 13.1.1 of the \textit{Gaia} eDR3 documentation\footnote{\url{gea.esac.esa.int/archive/documentation/GEDR3/Gaia_archive/chap_datamodel/sec_dm_main_tables/ssec_dm_gaia_source.html}}. 

We note that the third \textit{Gaia} data release (\textit{Gaia} DR3) was published in 2023--after this work had already begun. The astrometric solution and broad-band photometry in \textit{Gaia} DR3 are identical to those in eDR3, with additional parameters, including the new $G_{\rm RVS}$ photometry band \citep{Gaia2023}. In this work, we adopt eDR3 astrometric solutions and parameter definitions, while using the DR3 source identifier for consistency with the current catalog.

To ensure the reliability of our initial target list, we cross-matched it with the \textit{Gaia} Catalog of Nearby Stars \citep[GCNS,][]{GCNS2021}, which provides a well-characterized sample of objects within 100 pc of the Sun with reliable astrometric and photometric solutions based on \textit{Gaia} eDR3. To select the objects comprising the GCNS, \citet{GCNS2021} used a random forest classifier \citep{Breiman2001} to identify and remove sources with spurious or poor astrometric solutions, including objects affected by photometric contamination, stellar crowding or insufficient scan coverage. The model was trained using predictor variables selected from a set of 41 features listed in Table A.1 of \citet{GCNS2021}; for more details on the rejected sources, see Section 2.1 of that paper. After applying the cross-match, 1.875 stars remained in our list. The 25 objects rejected by the GCNS were removed from our sample, as they could introduce systematic errors in both distance and photometry, compromising the integrity of the sample,  even though some may correspond to real stars with unreliable astrometric solutions rather than purely spurious detection.

\subsection{Cross-Matches with other Catalogs}
\label{subsec:x-match}

Although {\it Gaia} represents the most accurate survey of the solar neighborhood, its completeness has limitations at both the bright end ($G \lesssim 3$~mag)\footnote{Magnitudes $G$, $G_{\rm BP}$ and $G_{\rm RP}$ are given in the {\it Gaia} photometric system.} and the faint end ($G \approx 21$ mag). For $G \lesssim 6$ mag, saturation effects degrade the astrometric and photometric quality, affecting the detection and characterization of the brightest stars \citep{Lindegren2021}. At the faint end, {\it Gaia} loses sensitivity to sub-stellar objects and ultracool dwarfs, whose low luminosity places them beyond the survey's detection limits \citep{GCNS2021,Lindegren2021}. Furthermore, due to stricter quality filters applied in {\it Gaia} eDR3, some sources observed by the mission lack an astrometric solution in this release, either because of insufficient observational data, degraded solutions, or inconsistencies detected during processing \citep{Lindegren2021}. Consequently, some sources with a published parallax measurement in DR2 do not have one in eDR3. Additionally, secondary components of close binary systems resolved by \textit{Gaia} eDR3 may lack a reported parallax in that release.

To address the limitations described above, we supplemented our \textit{Gaia} sample with stars from three additional catalogs: 1) the Hipparcos catalog \citep{VanLeeuwen2007}, which allow us to recover saturated stars missing from the  {\it Gaia} catalog; 2) the compiled catalog of \citet{Kirkpatrick2024}, which is itself composed of other comprehensive astronomical catalogs and incorporated ultracool objects and very low-mass dwarfs that lie below \textit{Gaia}'s sensitivity limits; and 3) the SIMBAD astronomical database \citep{Simbad2000}, used to further improve completeness. For each catalog, the selection criteria were the same: we included all stars with a parallax greater than 55.55 mas (corresponding to a distance of 18 pc), and with reported optical photometry. Each star was included only once, following catalog-specific priorities: 1) from Hipparcos, we included all the stars not present in \textit{Gaia} eDR3, and with validated parallax within our range; 2) from \citet{Kirkpatrick2024} we included all stars that are not present in the combined \textit{Gaia} eDR3 and Hipparcos sample that have reported optical magnitudes--this criterion automatically excludes the majority of L,T, and Y type objects, which lack optical photometry; and 3) from SIMBAD we included all the stars not present in the combined sample from \textit{Gaia} eDR3, Hipparcos and \citet{Kirkpatrick2024}, excluding objects classified as non-stellar, such as brown dwarfs and exoplanets.

Although perfect completeness is impossible to achieve, integrating these catalogs is essential for constructing a sample as complete as possible of stars within 18~pc of the Sun. Our final sample, after including these catalogs, consists of 2047 objects with reported optical photometry. Table \ref{tab:columns} lists the columns of Table \ref{tab:list}, which presents a fragment of our target list\footnote{The complete table is available at \url{https://share.obspm.fr/s/B3gAwq9xfdb4Fff}.}. Our comprehensive catalog is a valuable product of our survey and will be made available for the community. Details of the adopted cross-matching procedures are described in the following subsections. 

\begin{table}[ht]
    \caption{Description of the columns in Table \ref{tab:list}.}
    \label{tab:columns}
    \centering
    \small
    \begin{tabular}{l@{\hspace*{9pt}}l}
    \toprule
    \toprule
    Column & Description \\
    \midrule
    Main ID & Main ID in SIMBAD \\
    Gaia ID & Gaia designation \\
    WDS ID & WDS designation \\
    TIC ID & TESS Input Catalog designation \\
    TOI ID & TESS Object of Interest designation \\
    RA Dec Ref & Reference for RA and Dec coordinates$^a$ \\
    RA J2000 & Decimal J2000 RA coordinates\\
    Dec J2000 & Decimal J2000 Dec coordinates \\
    RUWE & Gaia eDR3 Renormalised Unit Weight Error \\
    IPDfmp & Gaia eDR3 {\tt IPD\_frac\_multi\_peak} \\
    Plx-PM Ref & Reference for parallax and proper motion values \\
    Plx & Parallax (mas) \\
    PM RA & Proper Motion in RA (mas yr$^{-1}$) \\
    PM Dec & Proper Motion in Dec (mas yr$^{-1}$) \\
    G & G-band magnitude from Gaia eDR3  \\
    $G_{\rm BP}$ & $G_{\rm BP}$-band magnitude from Gaia eDR3 \\
    $G_{\rm RP}$ & $G_{\rm RP}$-band magnitude from Gaia eDR3 \\
    $B$ & $B$-band magnitude$^b$ \\
    $V$ & $V$-band magnitude$^b$  \\
    $I$ & $I$-band magnitude$^b$ \\
    SP Type SIMBAD & Spectral type from SIMBAD \\
    RV Gaia & Radial Velocity from Gaia (km s$^{-1}$) \\
    RV SIMBAD & Radial Velocity from SIMBAD (km s$^{-1}$) \\
    Catalog & Catalog reference \\
    Observed & Observation status: \\
    & ON: observed only in the Northern hemisphere \\
    & OS: observed only in the Southern hemisphere \\
    & ON/OS: observed in both hemispheres \\
    & --: No observed \\
    \bottomrule
    \bottomrule
    \multicolumn{2}{p{0.95\columnwidth}}{$a$: J2000 coordinates for \textit{Gaia} sources were retrieved from the VizieR, since \textit{Gaia} natively provides coordinates in epoch J2016. For all other sources, the "RA Dec Ref" column indicates the specific reference for their coordinates.} \\
    \multicolumn{2}{p{0.95\columnwidth}}{$b$: $V$, $I$ and $B$ magnitudes for \textit{Gaia} sources were computed using the transformation equations from the \textit{Gaia} photometric system to the Johnson-Cousins system provided by \citet{Riello2021}.}
    \end{tabular}
\end{table}

\subsubsection{Hipparcos}
\label{sssec:Hip}

The Hipparcos catalog, published by \citet{ESA1997}, represents the first space mission devoted specifically to high-precision astrometry, providing positions, parallaxes, and proper motions for more than 118,000 stars. In this work, we used the revised version of the Hipparcos catalog by \citet{VanLeeuwen2007}, which significantly improves the astrometric precision through a complete re-reduction of the original data. This version is available in the VizieR database of astronomical catalogs \citep{Vizier} under the identifier I/311\footnote{VizieR allows users to download astronomical catalogs, either complete or in part, using different selection criteria depending on the parameters available in each catalog. Additionally, VizieR computes coordinates in different epochs and systems (e.g., galactic coordinates, coordinates in epoch J2000, among others) using the catalogs information.}. We downloaded the Hipparcos sources using the parallax as the sole selection criterion, obtaining a total of 765 objects within 18 pc. 

We incorporated Hipparcos data through a cross-match between the Hipparcos catalog and {\it Gaia} eDR3, using the software TOPCAT\footnote{\url{https://www.star.bris.ac.uk/~mbt/topcat/}}, which employs equatorial coordinates (RA, Dec) in epoch J2000 for both catalogs\footnote{The coordinates in epoch J2000 for both catalogs were computed by VizieR}. The cross-match was performed using a search radius of 5 arcsec. We obtain 631 stars from Gaia and Hipparcos, and 134 stars only in Hipparcos. 

For each new star identified in Hipparcos, we validated the parallax using the SIMBAD Astronomical Database\footnote{\url{https://simbad.cds.unistra.fr/simbad/}} to ensure that the object is truly located within 18~pc. This procedure allowed us to identify 60 of 134 stars, that are truly located within 18 pc and are either not present in {\it Gaia} or lack astrometric solutions in eDR3. 

\subsubsection{\citet{Kirkpatrick2024}}
\label{sssec:Kirk}

To complement the stellar census with low-mass object, we included sources from the catalog of stars within 20 pc published by \citet{Kirkpatrick2024}. This compilation provides comprehensive coverage of low-mass and extremely cool objects, including spectral types L, T and Y. 

\citet{Kirkpatrick2024} assembled this census using multiple catalogs and surveys, such as the Preliminary Version of the Third Catalog of Nearby Stars \citep{Gliese1991}, {\it Gaia} eDR3 \citep{Gaia2021}, {\it Gaia} DR3 \citep{Gaia2023},  the Wide-field Infrared Survey Explorer (WISE) mission \citep{WISE2010}, among others. We note that \citet{Kirkpatrick2024} did not include the Hipparcos catalog in their compilation.  

We integrated these objects through a cross-match with our updated list (combining {\it Gaia} eDR3 and Hipparcos sources) using J2000 equatorial coordinates, following the same procedure employed with the Hipparcos catalog. This process identified 302 additional stars, 253 of which are very faint (intrinsically and apparently), with spectral types L, T and Y. Due to a lack of available optical photometric measurements for these faint objects, they were not included in our final list. Consequently, we added only 49 stars from this catalog to our sample. 

We note that the catalog in \citet{Kirkpatrick2024} identifies and specifies the components of all known binaries and multiple systems within 20 pc of the Sun. For this reason, when performing the cross-match, we used the coordinates of the primary star in close binaries or multiple systems.

\subsubsection{SIMBAD Astronomical Database}

SIMBAD (Set of Identifications, Measurements, and Bibliography for Astronomical Data) is one of the most comprehensive astronomical database, produced and maintained by the Strasbourg Astronomical Data Center (CDS). It provides bibliographic references, fundamental data (e.g. astrometry, photometry, velocities, proper motions), and a wide range of additional measurements depending on the object type \citep{Simbad2000}. 

Given its extensive coverage, SIMBAD was used to further improve the completeness of our sample. We performed a query with a single criterion--parallax (plx $>$ 55.55 mas)--obtaining a total of 2361 objects \footnote{This query was performed on December 4, 2025}. We cross-matched this list with our current target list (constructed combining \textit{Gaia} eDR3, Hipparcos and \citet{Kirkpatrick2024}) using J2000 equatorial coordinates, following the same process applied to the previous catalogs. We then discarded sources classified as brown dwarfs, as well as objects identified as extrasolar planets. As a result, 63 additional stars were added into our final sample.

\subsection{The Final Compiled Sample}

\begin{table*}[ht!]
\centering
\small
\caption{Fragment of the final target list.}
\label{tab:list}
\begin{tabular}{llllll}
\toprule
\toprule
Main ID & Gaia ID & WDS & TIC ID & TOI & RA Dec Ref \\
 &  &  &  &  & \\
\midrule
2MASS J00001354+2554180 & -- & -- & TIC 117927630 & -- & SIMBAD \\
G 158-27 & Gaia DR3 2441630500517079808 & -- & TIC 176287658 & -- & DR3 \\
BD+66 34A & Gaia DR3 527956488340229632 & WDS J00321+6715A & TIC 420470815 & -- & DR3 \\
G 272-10 & Gaia DR3 2457862762476854656 & -- & TIC 32550424 & -- & DR3 \\
* bet Ari & -- & WDS J01546+2049AB & TIC 91356756 & -- & SIMBAD \\
SSSPM J0306-3648 & Gaia DR3 5047423236725995136 & -- & TIC 308243298 & -- & DR3 \\
L 736-30 & Gaia DR3 2979590513145784192 & -- & TIC 246902275 & -- & DR3 \\
G 222-11 & Gaia DR3 1143321672882274560 & -- & TIC 138421058 & -- & DR3 \\
G 51-15 & Gaia DR3 703790044252850688 & -- & TIC 3664898 & -- & DR3 \\
HD 101581 & Gaia DR3 5378886891123024512 & -- & TIC 397362481 & TOI 6276 & DR3 \\
G 180-18 & Gaia DR3 1372215976327300480 & -- & TIC 188589164 & TOI 2013 & DR3 \\
G 23-24 & Gaia DR3 4248817876711932416 & -- & TIC 345472607 & -- & DR3 \\
HD 239960A & Gaia DR3 2007876324466455040 & WDS J22280+5742A & TIC 471012608 & -- & SIMBAD \\
G 267-11 & Gaia DR3 2313022171603701888 & -- & TIC 183532804 & -- & DR3\\ 
\bottomrule
\bottomrule
\end{tabular}

\medskip
\medskip 
\noindent\rule{\textwidth}{0.1pt} 
\vspace{2pt}

\begin{tabular}{cccccccccc}
\toprule
\toprule
RA J2000 & Dec J2000 & RUWE & IPDfmp & Plx-PM Ref & Plx & PM RA & PM Dec & $G$ & $G_{BP}$ \\
deg & deg &  &  &  & mas & mas yr$^{-1}$ & mas yr$^{-1}$ & mag & mag \\
\midrule
0.056 & 25.905 & -- & -- & SIMBAD & 70.8 & 6 & 130 & 21.38 & -- \\
1.680 & -7.538 & 1.337 & 4 & DR3 & 206.35 & -811.57 & -1893.25 & 11.77 & 14.07 \\
8.123 & 67.236 & 5.886 & 44 & DR3 & 101.09 & 1748.32 & -347.59 & 9.57 & 10.73 \\
20.825 & -12.940 & 1.794 & 4 & DR3 & 63.79 & -12.51 & 331.27 & 11.86 & 13.23 \\
28.660 & 20.808 & -- & -- & SIMBAD & 55.6 & 98.74 & -110.41 & -- & -- \\
46.548 & -36.798 & 1.016 & 0 & DR3 & 75.42 & -172.17 & -669.24 & 15.99 & 19.53 \\
73.458 & -17.773 & 1.106 & 0 & DR3 & 83.69 & 408.57 & -644.46 & 9.94 & 11.14 \\
92.583 & 82.107 & 1.075 & 2 & DR3 & 105.69 & 50.26 & -1336.25 & 9.47 & 10.72 \\
127.456 & 26.776 & 1.155 & 1 & DR3 & 279.25 & -1113.69 & -612.19 & 12.17 & 15.10 \\
175.260 & -44.405 & 0.916 & 0 & DR3 & 78.23 & -660.63 & 242.10 & 7.39 & 7.98 \\
239.578 & 35.407 & 1.185 & 0 & DR3 & 63.17 & -143.06 & 318.12 & 11.51 & 12.97 \\
300.962 & 5.996 & 1.316 & 3 & DR3 & 75.25 & -495.48 & -799.62 & 11.13 & 12.35 \\
336.998 & 57.695 & -- & -- & SIMBAD & 249.39 & -725.23 & -223.46 & 8.67 & -- \\
359.964 & -34.112 & 1.613 & 0 & DR3 & 63.76 & 698.29 & -643.74 & 11.50 & 13.06 \\ 
\bottomrule
\bottomrule
\end{tabular}

\medskip
\medskip 
\noindent\rule{\textwidth}{0.1pt} 
\vspace{2pt}

\begin{tabular}{ccccccccc}
\toprule
\toprule
$G_{RP}$ & $B$ & $V$ & $I$ & SP Type Simbad & RV Gaia & RV Simbad & Cat & Observed \\
mag & mag & mag & mag &  & km s$^{-1}$ & km s$^{-1}$ &  &  \\
\midrule
19.65 & -- & -- & -- & T4.5 & -- & -- & Kirkpatrick & -- \\
10.40 & 14.65 & 13.95 & 10.30 & M5.5V & -40.46 & -33.7 & Gaia & ON/OS \\
8.42 & 10.97 & 10.53 & 8.32 & M2V & -3.37 & -4.02 & Gaia & -- \\
10.69 & 13.49 & 13.00 & 10.55 & dM6.5 & -1.26 & -0.788 & Gaia & OS \\
-- & 2.78 & 2.65 & 2.43 & kA4hA5mA5Va & -- & -3.1 & Hipparcos & ON \\
14.46 & 20.60 & 19.20 & -- & M8.7V & -- & 11.44 & Gaia & -- \\
8.85 & 11.33 & 10.89 & 8.70 & M3V & -14.87 & -14.478 & Gaia & OS \\
8.36 & 10.93 & 10.48 & 8.21 & M2.5V & -2.02 & -1.5 & Gaia & -- \\
10.71 & 15.98 & 15.09 & 10.72 & M6.5Ve & -- & 10.62 & Gaia & ON \\
6.67 & 7.96 & 7.74 & 6.55 & K4.5Vk: & 13.55 & 13.736 & Gaia & OS \\
10.32 & 13.25 & 12.74 & 10.17 & M3.5V & 10.01 & 10.14 & Gaia & ON \\
10.04 & 12.55 & 12.10 & 9.89 & M2.5V & 30.09 & 30.5 & Gaia & ON/OS \\
-- & 11.82 & 11.41 & -- & M3V & -- & -24 & SIMBAD & -- \\
10.27 & 13.39 & 12.85 & 10.13 & M3V & -12.86 & -- & Gaia & OS \\
\bottomrule
\bottomrule
\end{tabular}

\medskip
\medskip 
\noindent\rule{\textwidth}{0.1pt}

\begin{tabular}{p{0.95\textwidth}}
 NOTE --- Table 2 can be downloaded in csv format from \url{https://share.obspm.fr/s/B3gAwq9xfdb4Fff}.\\
\end{tabular}
\end{table*}

By combining the refined \textit{Gaia} sample (1875 stars) with the objects identified in the Hipparcos (60), \citet{Kirkpatrick2024} (49), and SIMBAD (63) catalogs, we obtained a final, comprehensive and as complete as possible sample of 2047 stars with reported optical photometry within 18 pc of the Sun. To further verify the completeness of our sample, we performed a cross-match with the Fifth Catalog of Nearby Stars \citep[CNS5;][]{Golovin2023}. This cross-match confirmed that our catalog is comprehensive, as contains 85 more stars than CNS5. This compiled list serves as the master input catalog for selecting the observable target list for our high-angular-resolution, volume-limited survey. The observational program is described in Section \ref{sec:Obs}.


\section{THE UNBIASED SURVEY: ASSESSING THE RELIABILITY OF \textit{GAIA} BINARITY INDICATORS }
\label{sec:param}

The results of the analysis described in what follows provide a quantitative justification for selecting observable targets independently of \textit{Gaia} binarity indicators, as described in Section \ref{subsec:SB}, leading - in this sense - to an unbiased survey. Our analysis shows that an important fraction of close binary stars may be classified as single stars based on \textit{Gaia} binarity indicators. Therefore, restricting observations to objects flagged as binary stars by these indicators would introduce an important selection bias in the survey results. 

The {\it Gaia} mission provides a wide range of valuable parameters for each of its sources, some of which are especially useful for identifying potential close binary systems. Nevertheless, these parameters must be interpreted with caution: they are statistical in nature and can be affected by observational limitations, and therefore cannot provide a definitive classification of binarity on their own. Having defined a final, as complete as possible, volume-limited sample of 2047 stars in the previous section, an important question arises: should we prioritize for high-resolution imaging targets based on their potential binary status, according to their \textit{Gaia} parameters? In this section, we investigate the reliability of these parameters, and show that a significant fraction of confirmed binaries have \textit{Gaia} parameters indistinguishable from those of single stars.

We introduce and briefly describe three key {\it Gaia} indicators--the Renormalized Unit Weight Error (RUWE), the fraction of windows with multiple peaks ({\tt ipd\_frac\_multi\_peak}, or IPDfmp), and the fraction of windows with odd/conflict windows ({\tt ipd\_frac\_odd\_win}, or IPDfow)--in order to evaluate how useful they are for identifying the binary nature of nearby stars. To do so, we use a sample of stars confirmed as bona fide close binaries from the Washington Double Star Catalog \citep[WDS,][]{WDSCat2001}. Additionally, we briefly present two related parameters--{\tt astrometric\_excess\_noise} and IPD goodness-of-fit harmonic amplitude ({\tt ipd\_gof\_harmonic\_amplitude}, or IPDgofha)--and discuss why they are not used in our analysis. 

In Paper IV of this series, we will compare the findings from this analysis with the actual results from our survey, allowing us to draw more robust conclusions about the true binary fraction among nearby stars and its relationship to the \textit{Gaia} binarity indicators. 

\subsection{A Control Sample of Confirmed Close Binaries}

To evaluate the reliability of \textit{Gaia} binarity indicators, we constructed a control sample of stars that are confirmed as binary systems in the WDS\footnote{Available at \url{https://www.astro.gsu.edu/wds/}}--the most comprehensive catalog of known binary systems, which is updated regularly. We first chose all stars from our \textit{Gaia} sample that have a WDS identifier reported in SIMBAD, i.e., known binaries. Subsequently, using the WDS information of these, we selected those that met the following selection criteria:

\begin{enumerate}
    \item Separation $\rho <$ \SI{1}{\arcsecond}. This criterion is important because \textit{Gaia}'s effective angular resolution is $\sim$\SI{0.4}{\arcsecond} in the along-scan direction but significantly worse in the across direction \citep{Fabricius2021,Gaia2016}. Therefore, binaries with $\rho <$ \SI{1}{\arcsecond} may remain completely unresolved in DR3 and are treated as single sources, so they represent the most difficult test for the binarity indicators.

    \item More than one observation is listed in WDS. Multiple observations confirm the bona-fide nature of these binary systems, reducing the possibility of including spurious or optical doubles as they usually have only one observation. In our final control sample of 70 binary stars, only seven have exactly two observations, while the remaining 63 have three or more. 

    \item Binary system not resolved by \textit{Gaia}, to ensure that the binarity indicators we are evaluating correspond to the combined light of the components.
\end{enumerate}

This selection yielded a clean control sample of 70 confirmed close binary systems within 18 pc, across the full magnitude range, with magnitude differences ($\Delta$m) ranging from 0 to 9.8 mag. Since all the stars in this sample are confirmed binaries, they allow us to examine how precisely the \textit{Gaia} parameters identify them as a binary system. 

The analysis described in the following subsections is based mainly on the separations between the components of the binary systems, due to incomplete magnitude measurements for secondary components in the WDS catalog, which does not contain data for all binary stars in the control sample. Additionally, the catalog does not specify the photometric bands in which all magnitudes are reported. Consequently, a comprehensive analysis of our control sample results as a function of magnitude difference is not feasible with the available data. Moreover, the goal of this analysis is to demonstrate empirically that the \textit{Gaia} parameters can miss a significant fraction of close binaries, rather than to investigate the physical or photometric factors underlying their limitations. Analyzing how photometric and astrometric properties affect the \textit{Gaia} parameters is beyond the scope of this work.

\subsection{Individual \textit{Gaia} Parameters as Binarity Indicators}

To evaluate the reliability of the \textit{Gaia} parameters as individual binarity indicators, we begin by presenting the distribution of each in our main \textit{Gaia} sample of 1875 stars and, most importantly, their distribution in our control sample of 70 confirmed binaries. The color-magnitude diagrams (CMDs) presented in this section were produced using the raw {\it Gaia} photometry for illustrative purposes; we have not applied the magnitude corrections proposed by \citet{Riello2021}.  

\subsubsection{Renormalized Unit Weight Error (RUWE)}

This parameter quantifies the goodness-of-fit of the single-star astrometric model applied to each source, normalized by its expected value as a function of color and magnitude. The RUWE is expected to be $\sim$1.0 for well-behaved sources--those for which the single-star model provides a good fit to astrometric observations \citep{Gaia2021,Lindegren2021}. A RUWE value significantly greater than 1.0 (typically $>$1.4) suggest a degraded fit, possibly caused by issues in the astrometric solutions or by the presence of a non-single star \citep{Lindegren2021}. A threshold of RUWE $>1.4$ is commonly used as a first-order indicator of potential unresolved binary stars \citep{Belokurov2020,Castro-G2024}. \citet{Tokovinin2023} suggested a threshold of RUWE $> 2$ to infer inner subsystems in wide binaries. 

\begin{figure}[ht!]
\centering
\includegraphics[width=0.45\textwidth]{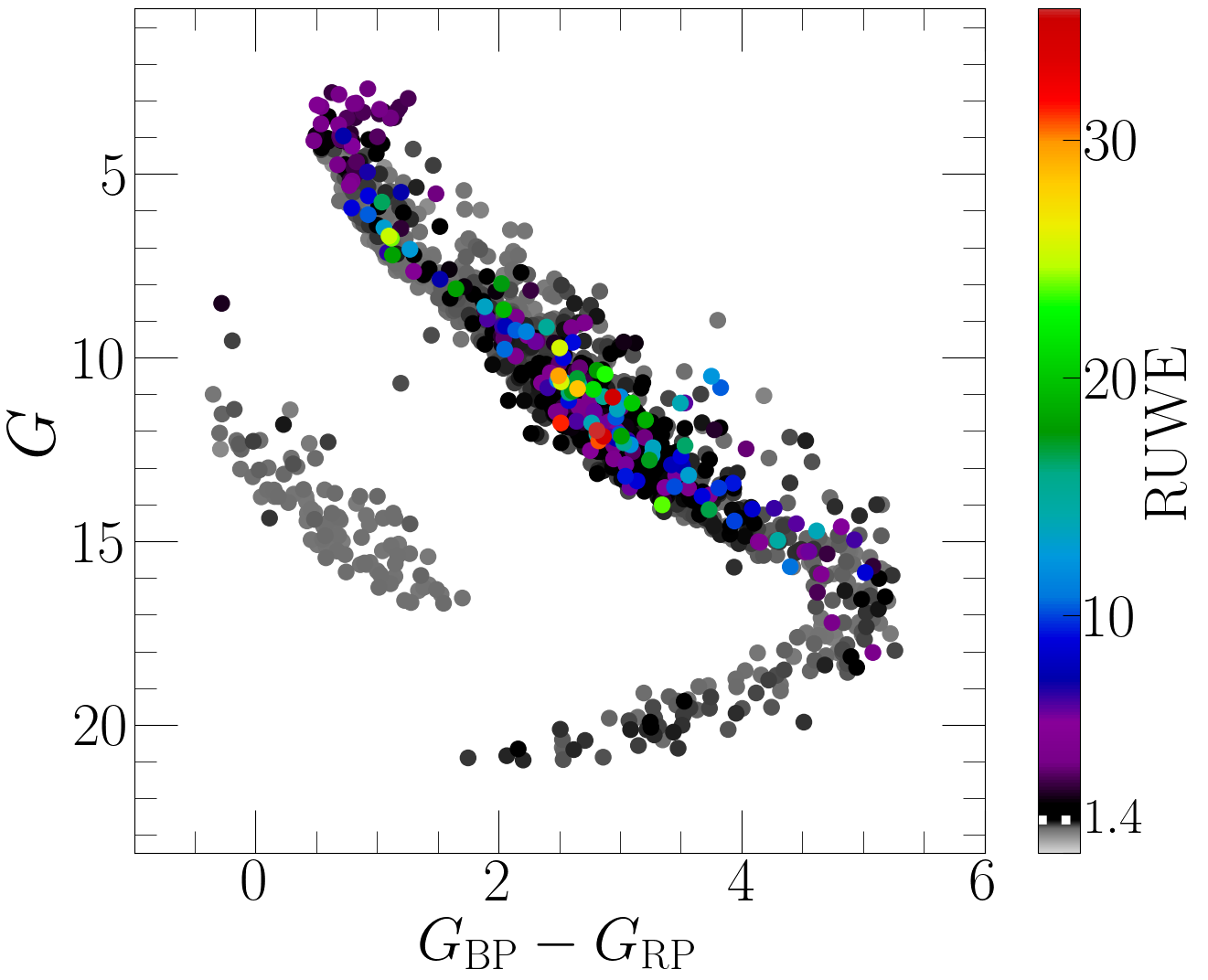}
\includegraphics[width=0.45\textwidth]{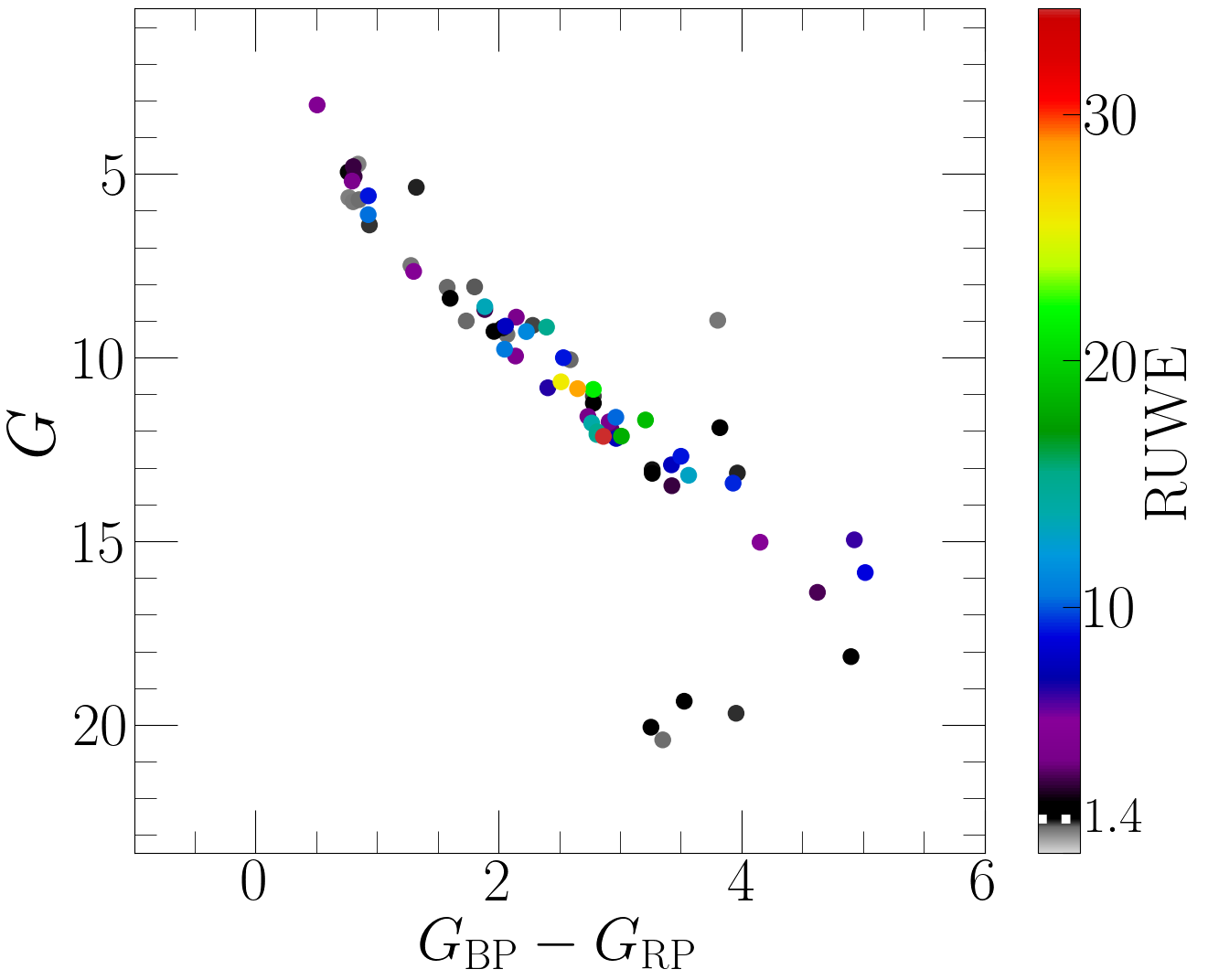}
\captionsetup{justification=justified}
\caption{Top: CMD of the 1,875-star {\it Gaia} sample. Bottom: CMD of the 70-star control sample of confirmed close binaries ($\rho <$ \SI{1}{\arcsecond}) from the WDS. In both panels, the color scale represents RUWE, the gray dots represent stars with RUWE $< 1.4$, and the white dashed line marks RUWE $= 1.4$. A significant fraction (31\%) of these known binaries have RUWE $< 1.4$, demonstrating that RUWE alone is not a reliable indicator of binarity.  }
\label{fig:cmd-ruwe}
\end{figure}

Figure \ref{fig:cmd-ruwe} (top) shows the CMD of our 1,875-star \textit{Gaia} sample, with the color scale representing RUWE. Applying the condition RUWE $> 1.4$, we find 463 stars (24.7\%) that are potential unresolved binary systems.

It is important to emphasize that RUWE is statistical in nature and does not show a one-to-one correspondence with binarity. This is clearly illustrated in Figure~\ref{fig:cmd-ruwe} (bottom), which shows the CMD of our 70-star control sample of confirmed binaries from the WDS. If the RUWE were a reliable indicator, all 70 stars would have high values. However, we find that 22 of these stars (31\%) have RUWE $< 1.4$, and would be misclassified as single stars. Even if we adopt a more stringent threshold of RUWE = 1, there are 5 stars (7\%) below this limit. These results underscore the need for high-resolution imaging follow-up of stars with RUWE $< 1.4$. Consistently, \citet{Ziegler2020} selected TESS planet candidate hosts for speckle observations based on RUWE $> 1.4$ and null RUWE values. Their selection criteria imply that RUWE alone is not sufficient to identify all candidate binaries, which is consistent with our findings. Our analysis reveals a limitation in the sensitivity of RUWE and suggests that using RUWE alone can omit a significant fraction of close binary stars. 

On the other hand, a high RUWE is not necessarily caused by binarity. Another explanation for an elevated RUWE value is photometric variability, which can introduce astrometric scatter that mimics an unresolved companion, leading to RUWE $> 1.4$ in a single variable star \citep{Golovin2023}. Moreover, the astrometric solution can be degraded by calibration issues, particularly for bright stars near saturation, leading to a high RUWE \citep{Lindegren2021}. Conversely, unresolved systems do not necessarily produce a poor single-star astrometric fit. The photocenter can remain stable for some binary systems (e.g. those with long orbital periods, nearly equal-brightness pairs, very faint companions, or very tight orbits), yielding a fit to the single-astrometric model and, consequently, a RUWE close to unity \citep{Penoyre2020, ElBadry2024}. This can explain the false-negatives observed in our control sample.

\subsubsection{{\tt astrometric\_excess\_noise}}

Another parameter noted in the literature as an indicator of binarity is the {\tt astrometric\_excess\_noise} \citep{Gaia2023multiplicity, Gandhi2022}. This parameter quantifies the extra noise per observation of a source that explains the scatter in its astrometric solution. The {\tt astrometric\_excess\_noise} is considered statistically significant when {\tt astrometric\_excess\_noise\_sig} $> 2$.

Although this parameter is correlated with RUWE, the two are not identical and may react differently to specific system configurations. Nevertheless, both parameters measure deviations from the single-star model, and therefore encode essentially the same information \citep{Lindegren2021}. For this reason, we do not include {\tt astrometric\_excess\_noise} in our analysis and instead focus on RUWE, which is the more commonly adopted binarity indicator.

\subsubsection{IPD Goodness-of-Fit Harmonic Amplitude (IPDgofha)}

The {\tt ipd\_gof\_harmonic\_amplitude} or IPDgofha is a statistical parameter that measures the amplitude of the natural logarithm of the variations in the image parameter determination (IPD) goodness-of-fit (GoF) as a function of the scan position angle \citep{Gaia2021, Holl2023, Lindegren2021}. This value is normally small ($\le 0.1$); however, if the source exhibits an elongated image along a particular orientation, the parameter can become larger, which may indicate the presence of an unresolved binary \citep{Gaia2021, Halbwachs2023, Penoyre2022}. Additionally, sources with this parameter above 0.1 and with a RUWE $\ge 1.4$ are indicative of unresolved binaries, or resolved binaries which have not been correctly handled yet \citep{Fabricius2021}. 

However, \textit{Gaia} DR3 reports that this parameter was erroneously computed using the \textit{G}-band flux rather than being determined by the GoF of the IPD for all individual observations of a given source. While \textit{G}-band flux and IPDgofha are typically correlated--except for certain source types--this parameter might still appear useful for data analysis and filtering. Nevertheless, due to its incorrect calculation, its use is officially discouraged\footnote{See \url{https://www.cosmos.esa.int/web/gaia/dr3-known-issues}}. The correctly computed version will only be available in DR4. Consequently, we do not include IPDgofha in our analysis.

\subsubsection{IPD Fraction of Multiple Peaks (IPDfmp)}

The {\tt ipd\_frac\_multi\_peak}, or IPDfmp, specifies the fraction of windows with a successful IPD process in which more than one peak is identified in the flux distributions, rather than a single symmetric profile. This can be explained by the presence of two nearby stars whose light overlaps, possibly corresponding to a visually resolved double star, which could be either an optical pair or a physical binary \citep{Penoyre2022, Gaia2021}. This parameter ranges from 0 to 100 (as percentage); a value close to zero indicates an isolated, single-peaked source, while a high value suggest the presence of a resolved companion. The typical threshold used as binarity indicator is IPDfmp $> 2$ \citep{Halbwachs2023}. Figure~2 of \citet{Tokovinin2023} shows elevated IPDfmp values for close pairs ($\rho <$ \SI{2.5}{\arcsecond}). He adopts a threshold of IPDfmp $> 10$ as another indicator of inner subsystems in hierarchical binaries.

\begin{figure}[ht!]
\centering
\includegraphics[width=0.45\textwidth]{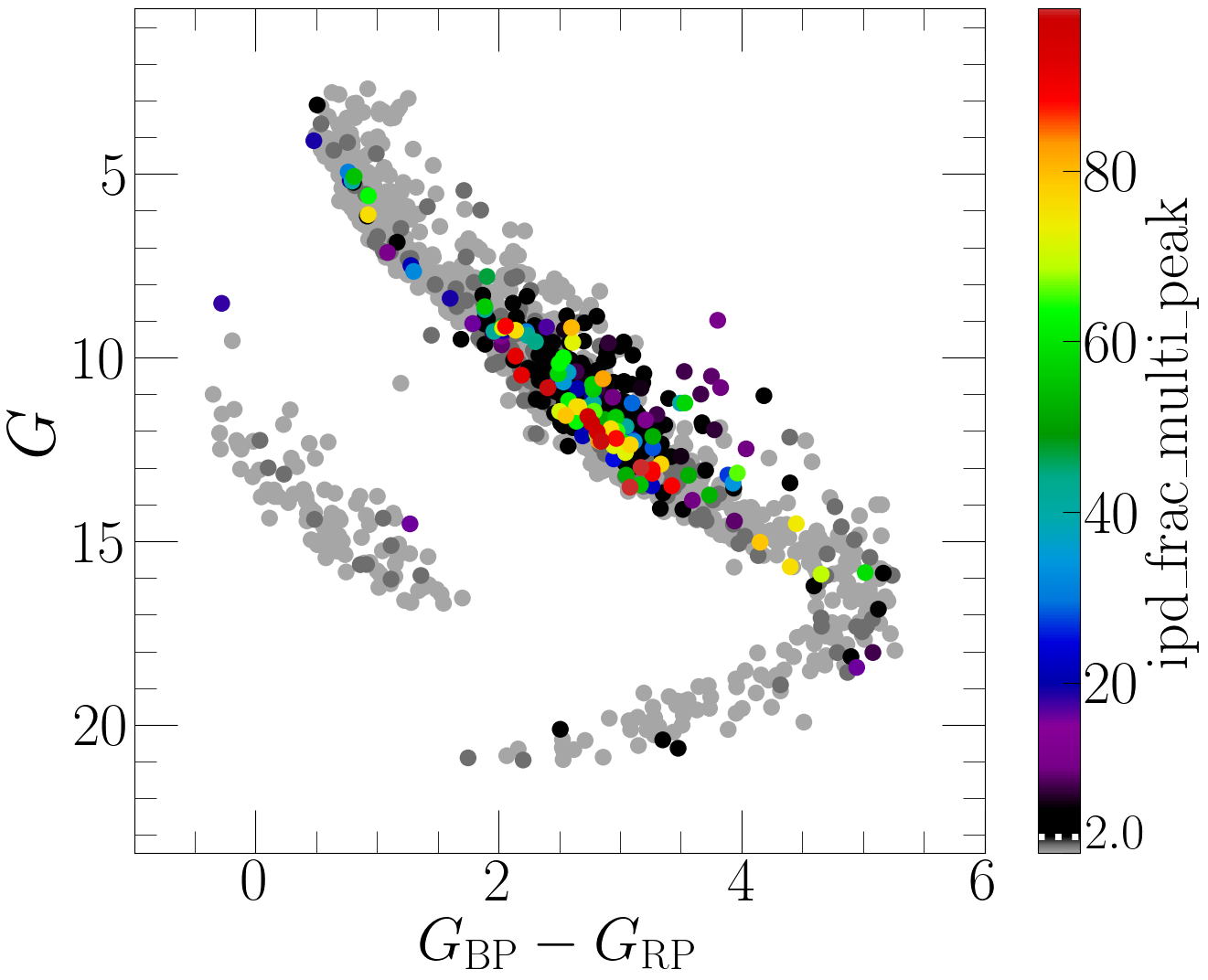}
\includegraphics[width=0.45\textwidth]{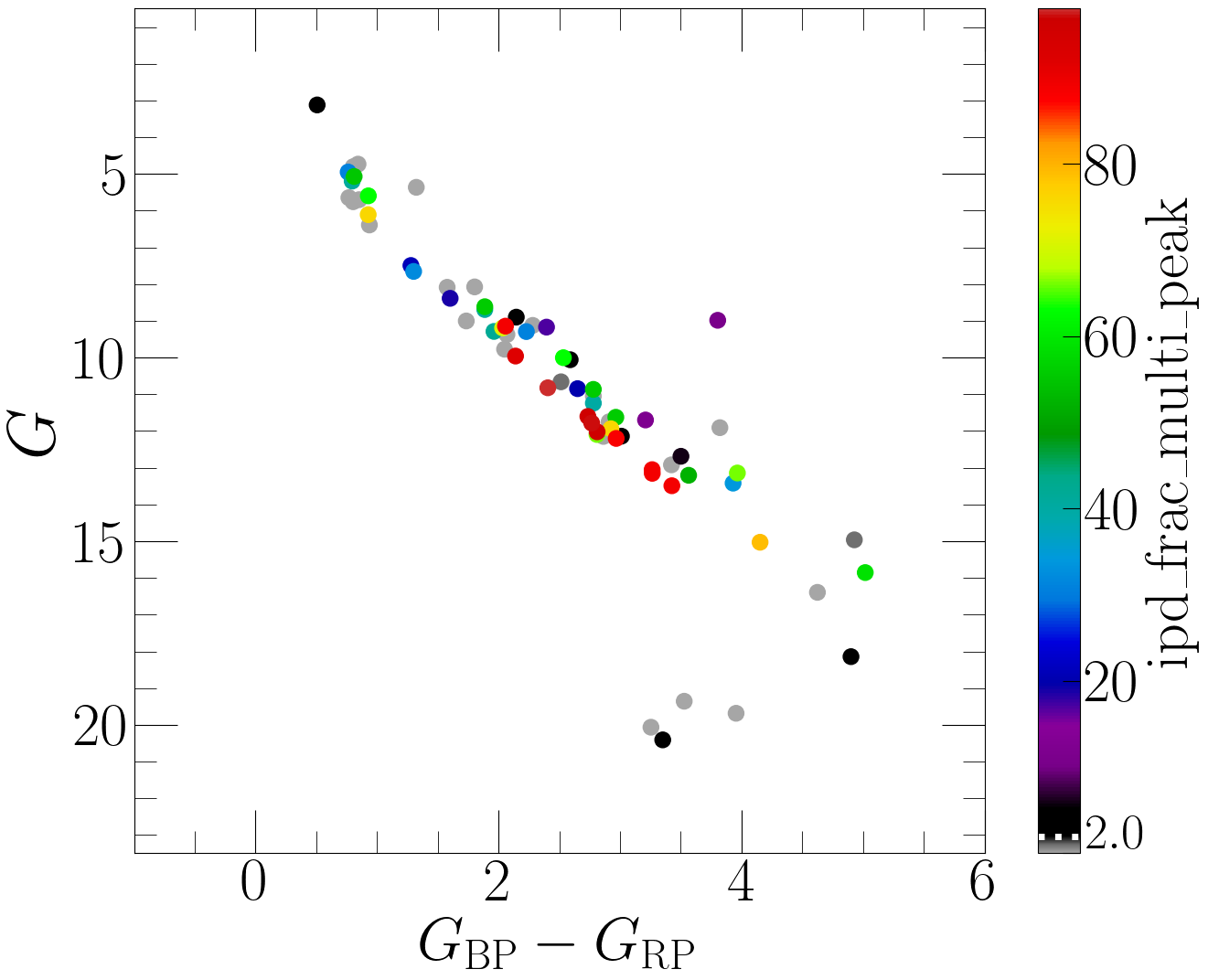}
\captionsetup{justification=justified}
\caption{Top: CMD of 1,875-star {\it Gaia} sample. Bottom: CMD of 70-star control sample of confirmed close binaries from the WDS. In both panels, the color scale represents IPDfmp, the gray dots represent stars with IPDfmp $< 2$, and the white dashed line marks IPDfmp = 2. Approximately 36\% of these known binaries have IPDfmp $\leq 2$, i.e., they are detected as isolated, single-peaked sources in the IPD process.}
\label{fig:ipdfmp}
\end{figure}
    
Figure \ref{fig:ipdfmp} (top) shows the CMD of our 1,875-star \textit{Gaia} sample, with the color scale representing IPDfmp. Using the threshold IPDfmp $> 2$, we find 236 stars that could be double. Figure \ref{fig:ipdfmp} (bottom) shows the CMD of our 70-star control sample of confirmed close binaries. We find that 25 stars (36\%) have IPDfmp $\le 2$, meaning that only one peak was detected in the IPD process despite the star being a known binary. 

These results suggest that IPDfmp, while useful, is not a fully reliable indicator of binarity on its own. Moreover, spurious multiple peaks can occur in the flux distribution of a single star due to image artifacts, background galaxies or cosmic ray hits, yielding false-positive. Conversely, a binary star does not always appear as a multiple peak in the IPD process. For example, a binary system with a large magnitude difference between its components may appear as a single-peaked source, as the fainter component is lost in the wing of the brighter star's PSF \citep{ElBadry2024}. Other cases occur when the components are too close to be spatially resolved  or when the binary axis is not optimally aligned with the scan direction; in such situations, the IPD process may identify only a single peak \citep{Fabricius2021,Tokovinin2023}.

\subsubsection{IPD Fraction of Odd Windows (IPDfow)}

The {\tt ipd\_frac\_odd\_win} or IPDfow, indicates the fraction of transits with conflict windows--that is, cases where the flux distribution appears asymmetric, as identified by the IPD process \citep{Fabricius2021,Lindegren2021}. Such a distortion can be caused by nearby sources, which might be attributed to close binaries or unrelated field stars. This parameter tends to zero when the source is symmetric, and is greater than 0.1 when the source is distorted \citep{Gaia2021, Lehtinen2023}.

In our {\it Gaia} sample, only 27 stars have IPDfow $> 0.1$. Among these, only 14 are reported as binaries in the WDS catalog, with separations between \SI{0.1}{\arcsecond} and \SI{2.6}{\arcsecond}, and only seven of these have $\rho < $ \SI{1}{\arcsecond}. However, we know that our sample contains many more binary stars with $\rho < $ \SI{1}{\arcsecond} than just those seven objects. This is clearly reflected in our control sample: most of the true binaries (66 stars, i.e., 94.3\%) have IPDfow = 0: that is, they are identified as symmetric sources by the IPD process. This result suggests that IPDfow is not a sensitive indicator of binarity; for this reason we did not include it in the analysis of parameters in combination. 

\subsubsection{Separation Dependence of the Individual Parameters}

\begin{figure}[ht!]
    \centering
    \includegraphics[width=0.45\textwidth]{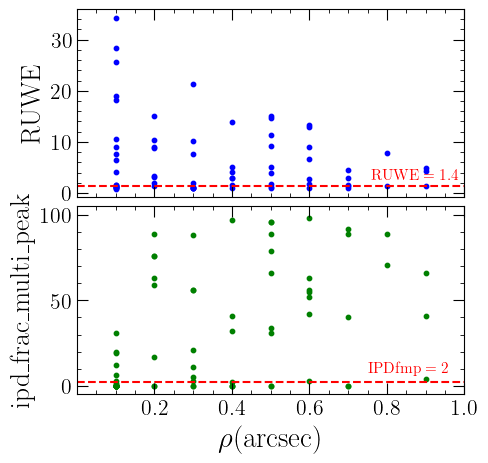}
    \captionsetup{justification=justified}
    \caption{Separation ($\rho$) versus binarity indicators for our control sample of 70 confirmed close binaries. Top: RUWE tends to increase at small separations ($\rho <$ \SI{0.5}{\arcsecond}). Bottom: IPDfmp shows no strong correlation with separation; high values occur across the entire range of separations ($\rho <$ \SI{1}{\arcsecond}). Red dashed line mark the thresholds: RUWE = 1.4, and IPDfmp = 2. The intervals along the x-axis are discrete due to the fact that WDS provides only one decimal place in the separation value.}
    \label{fig:paramsvsrho}
\end{figure}

To evaluate the behavior of the two relevant parameters in our analysis, Figure \ref{fig:paramsvsrho} shows the separation versus each binarity indicator for our 70-star control sample of confirmed close binaries. For RUWE (top panel), we can clearly see that values tend to be higher at smaller separations, indicating its sensitivity to close binaries. In the case of IPDfmp (bottom panel), we observe that high values can occur across all separations. Taking into account that our analysis is in the regime of $\rho <$ \SI{1}{\arcsecond}, this behavior is consistent with the finding of \citet{Tokovinin2023}, who notes that IPDfmp tends to be higher at small separations ($\rho <$ \SI{2.5}{\arcsecond}).

\subsection{Comparison of \textit{Gaia} Parameters Combinations}
\label{sssec:comparison}

Although the individual {\it Gaia} parameters discussed above can help detect unresolved binaries, we have found that individually, they show limited completeness when they are applied to the control sample. However, in combination they can become more powerful--but also more restrictive. In this section we examine the combined performance of RUWE and IPDfmp. For this analysis, we adopted the typical threshold for each binarity indicator, resulting in the following conditions:

\begin{enumerate}
    \item RUWE $> 1.4$,
    \item IPDfmp $> 2$
\end{enumerate}

From these conditions, we can identify three cases: \textbf{(1)} stars that do not meet any of the criteria, which likely indicates they are single; \textbf{(2)} stars that meet at least one condition; and \textbf{(3)} stars that meet both conditions simultaneously.

\begin{figure}[ht!]
\centering
\includegraphics[width=0.40\textwidth]{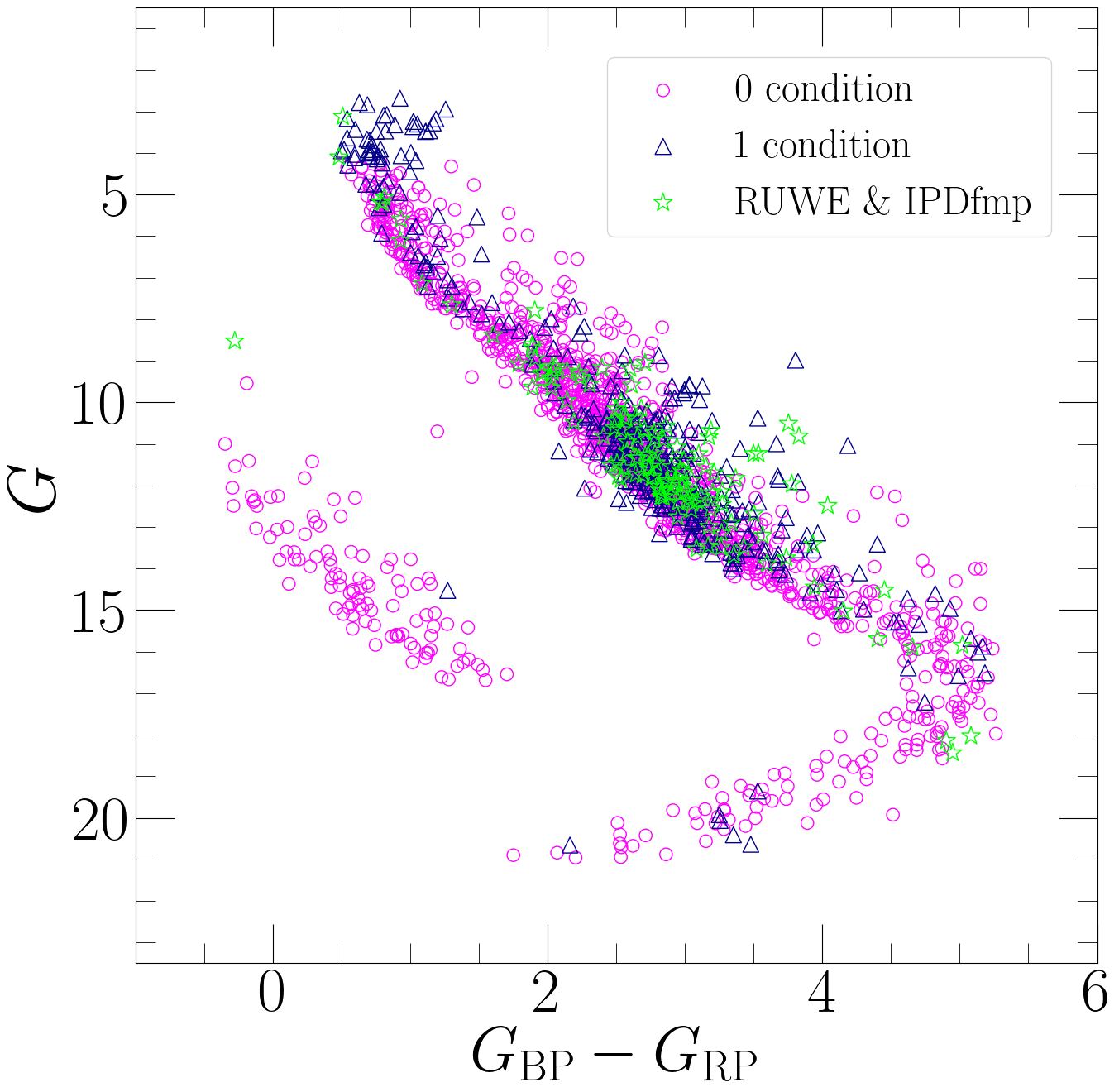}
\includegraphics[width=0.40\textwidth]{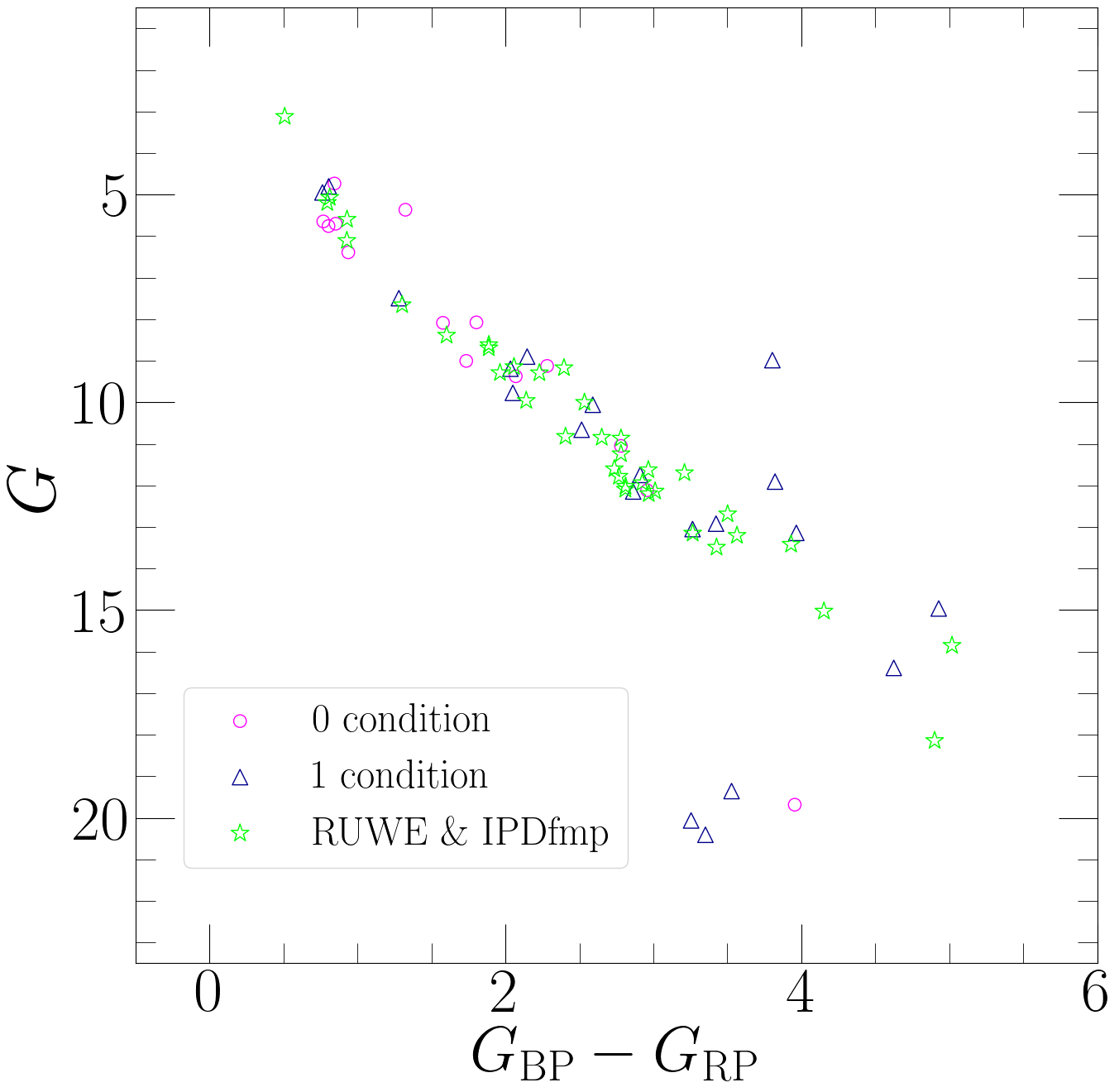}
\captionsetup{justification=justified}
\caption{CMD showing the distribution of stars according to binarity indicators. Top: 1,875-star {\it Gaia} sample. Bottom: Control sample of 70 confirmed close binaries. In both plots, magenta circles mark stars that meet no conditions; blue triangles indicate stars meeting only one condition; green stars correspond to stars meeting both RUWE $> 1.4$ and IPDfmp $> 2$. In the control sample, only 36 stars ($\sim$51.4\%) satisfy both conditions, highlighting the incompleteness of binary indicators, even when used in combination.}
\label{fig:comparison} 
\end{figure}

For our main sample of 1,875 {\it Gaia} stars, we find that 1,322 stars ($\sim$70.5\%) do not meet any of the binarity conditions, suggesting they are likely single stars. A total of 407 stars ($\sim$21.7\%) met one condition. Among these, 317 stars meet only the RUWE $> 1.4$ condition, and 90 stars meet only IPDfmp $> 2$. Although these stars could be unresolved binaries, our results from the control sample suggest that meeting only one condition is not sufficient to reliably indicate binarity. Regarding the combination of the two indicators, 146 stars ($\sim$7.8\%) meet both conditions simultaneously. These results are shown in the CMD of our {\it Gaia} sample in Figure \ref{fig:comparison} (top). On the other hand, it is interesting to note that in this figure, most of the stars meeting one or both conditions are concentrated at $G_{\rm BP} - G_{\rm RP} \ge 2$, which corresponds to M dwarfs, i.e., low-mass stars. This may be explained by the fact that, for a low-mass binary system, the secondary component is likely to have a mass similar to that of the primary, which is itself a low-mass star. Consequently, the mass ratio $q = M_2 / M_1$ will be close to 1, meaning that the photocentric motion is larger. In contrast, for a massive primary star with a low-mass secondary component, $q$ will be small, and the photocenter will hardly move. Therefore, it is expected that RUWE and IPDfmp are more sensitive to binarity in low-mass stars.  

For our control sample of 70 confirmed close binaries, we find the following distribution. Only 14 stars ($\sim$20\%) do not meet any of the binarity conditions, that is, they appear single according to both indicators. A total of 20 stars ($\sim$28.6\%) meet only one condition. Among these, 12 stars meet only RUWE $> 1.4$ and 8 stars meet only IPDfmp $> 2$. Finally, 36 stars ($\sim$51.4\%) satisfy both RUWE and IPDfmp simultaneously. These results are shown in Figure \ref{fig:comparison} (bottom). These results underscore that although using the combination of both parameters yields high-confidence binary detections, it fails to identify almost half of the confirmed close binaries in the control sample. This illustrates the trade-off between reliability and completeness: increasing confidence in detections requires multiple binarity indicators; however, this approach missed a substantial number of true binary stars.

\begin{figure}[ht!]
    \centering
    \includegraphics[width=0.45\textwidth]{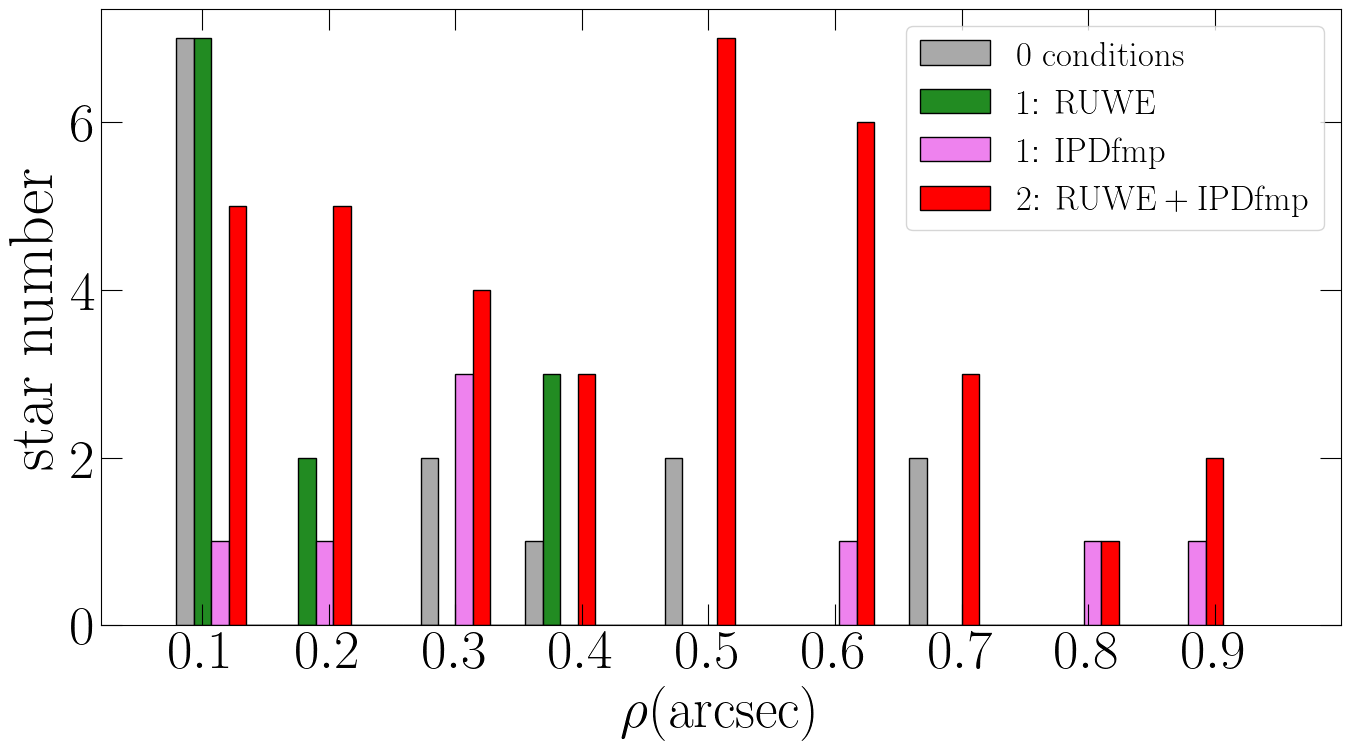}
    \captionsetup{justification=justified}
    \caption{Histogram of the distribution of stars in our control sample as a function of separation. Each bar is colored according to the binarity condition(s) satisfied: grey indicates stars meeting no conditions; green, stars with RUWE $> 1.4$ only; magenta, stars with IPDfmp $> 2$ only; red, stars that meet both indicators. }
    \label{fig:hist}
\end{figure}

Figure \ref{fig:hist} shows a histogram for our control sample as a function of separation. Each bar indicates how many stars at a given separation satisfy a given indicator or combination. Of the objects that meet one condition, most satisfy the RUWE criterion, which tends to be higher at the smallest separations ($\rho <$ \SI{0.5}{\arcsecond}), while those that meet only IPDfmp are distributed across all separations. On the other hand, most of the stars that satisfy both conditions (RUWE and IPDfmp) span the entire separation range; however, almost half of the true binaries are missed when this stricter condition is applied. 

\subsection{The Necessity of an Unbiased Survey}

Our analysis of the control sample of 70 confirmed close binaries within 18 pc shows that relying on binarity indicators--either individually or together--carries the risk of missing an important fraction of true double stars: RUWE alone misclassifies 31\% of confirmed close binaries as single, while IPDfmp alone misclassifies 36\%. The combination of the parameters  increases the confidence of detecting binaries but decreases substantially the fraction of detected true binaries. Additionally, the fact that the value of these parameters are influenced by other factors (e.g. orbital configuration, magnitude difference of components, calibration issues or photometric variability of a single star) that can leads false-positives or false-negatives further reduces their reliability.

These results are consistent with previous studies that have independently demonstrated the limitations of \textit{Gaia} astrometric quality parameters as binarity indicators. For example, \citet{Kervella2019} showed that an approach based on proper motion anomalies can detect companions missed by RUWE in \textit{Gaia} DR2. Similarly, \citet{Vrijmoet2020}, using a sample of M dwarfs within 25 pc, showed that a fraction of confirmed unresolved binaries can be missed by RUWE, and established cutoffs for several \textit{Gaia} parameters which can help to detect potential multiples in \textit{Gaia} DR2. Nevertheless, RUWE remains a useful parameter for selecting potential unresolved binaries; but as we show in this analysis, it is not infallible and can miss a significant fraction of close binaries. 

This conclusion underscores the necessity of conducting an unbiased, high-angular-resolution survey, independent of these indicators, in order to obtain a more complete census of binarity among stars within 18 pc of the Sun. In the following section we describe our observational program to carry out the described blind survey. In Paper IV of this series, we will compare the findings from this analysis with the actual results from our survey, providing an empirical measure of how many stars classified as single according to \textit{Gaia} parameters are detected as binaries (and vice versa) using high-resolution imaging. 


\section{OBSERVATIONS}
\label{sec:Obs}

To carry out an unbiased survey of stars within 18~pc of the Sun, we have conducted observations in both hemispheres using high-speed cameras that acquire short-exposure frames. This observational approach allows the application of high-angular-resolution techniques, including speckle interferometry \citep{Labeyrie1970}, a widely used high-resolution technique for studying binary and multiple systems at optical wavelengths \citep{Hartkopf2015,Balega2017,Tokovinin2020,Howell2021,Orlov2021,Mendez2025}. 

High-speed imaging efficiently removes atmospheric effects by taking a large number of very short-exposure images, during which the atmospheric turbulence is effectively frozen, generating a fixed pattern of phase variations--called a speckle pattern--in each frame. This approach allows the telescope to reach its diffraction limit and enables precise astrometry (angular separation, $\rho$ and position angle, PA) and photometry (brightness contrast, $\Delta m$) through computational data processing.  

\subsection{Observational Constraints and Sample Selection Biases}
\label{subsec:SB}

In the context of our high-angular-resolution survey, the term "unbiased" refers specifically to our observable target selection strategy: we apply only two criteria to select observable targets--the magnitude limits inherent to the observational techniques and the declination ranges imposed by the geographic locations of the telescopes. These are physical constraints that cannot be modified or avoided. Every star in the volume-limited sample that satisfies these conditions is included in the observable target list, regardless of its \textit{Gaia} binarity indicators, spectral type, known binary status, etc. This distinguishes our approach from surveys that preferentially observe stars flagged as binary candidates by \textit{Gaia} indicators--which would introduce a selection bias against systems undetectable by those indicators, as demonstrated in Section \ref{sec:param}--or known binary stars selected for follow-up \citep[e.g.,][]{Mason2023,Tokovinin2023,Tokovinin2024}, or samples restricted to a specific spectral type \citep[e.g.,][]{Raghavan2010,Roberts2017,Balega2017,Clark2024}.

In practice, the two selection criteria described above translate into the following observational constraints. For targets to be observed with the 2.1~m telescope at OAN-SPM we applied an apparent magnitude cut of $V \le 12$~mag, while at the NTT the brightness cut was $I < 16$~mag. At 18~pc, these limits imply a minimum main-sequence luminosity and mass of $\sim$0.029 L$_{\odot}$ and $\sim$0.44 M$_{\odot}$ for the northern targets, and $\sim$3 $\times 10^{-4}$ L$_{\odot}$ and $\sim$0.079 M$_{\odot}$ for the southern targets, respectively, using the relations from the EEM dwarf stellar parameters table ("A Modern Mean Dwarf Stellar Color and Effective Temperature Sequence", version 2022.04.16)\footnote{\url{https://www.pas.rochester.edu/~emamajek/EEM_dwarf_UBVIJHK_colors_Teff.txt}}, based on \citet{Pecaut2013}. After further imposing the declination limits for each telescope, as indicated in Sections~\ref{subsec:NH} and \ref{subsec:SH}, the number of observable targets is 648 and 1446 for the Northern and Southern samples, respectively. The sum of these values does not equal the total sample size for two reasons: first, several targets are observable from both hemispheres due to the overlapping declination range of the observatories; second, some stars remain unobservable from either location due to the magnitude and geographic constraints. 

In addition to the observable target list, in the north we also observed several stars fainter than $V=12$ during very good seeing conditions. In the Southern sample, we focused on stars without WDS entries in SIMBAD, leaving 894 stars. However, a few stars with WDS entries were still observed, as their WDS information was not reported in SIMBAD at the time of observation. Moreover, several stars were observed in more than one observing run. The observation status of each star is indicated in the Observed column of Table \ref{tab:list}.

Beyond the magnitude and declination limits, there are additional factors that affect the detection of stellar companions, including the angular separation $\rho$, magnitude difference $\Delta m$ between the components, primary star spectral type (color), filter choice and atmospheric conditions at the time of observation. A companion is detectable if its magnitude difference $\Delta m$ is below a threshold that depends on separation $\rho$. It is well established that a threshold of $5\sigma$ provides reliable detections \citep{Tokovinin2010}, although the achievable $\Delta m$ at the $5\sigma$ level depends strongly on the stellar magnitude, spectral type, separation, filter, seeing, and instrument. As preliminary examples, three representative contrast curves for stars of K and M spectral types are presented in Section \ref{subsec:TOI}. For instance, TOI 2013 (\textit{Gaia} DR3 1372215976327300480, $I \sim 10.2$, M3.5V, $d \sim 15.8$ pc), observed in the $I$-band with the OAN-SPM 2.1-m telescope, achieves a detection limit of $\Delta m = 6.06$ at $\rho =$ \SI{0.8} {\arcsecond}. This means that at separations smaller than $\rho =$ \SI{0.8}{\arcsecond} (corresponding to $\sim$10.1 AU), only companions with $\Delta m < 6.06$ are detectable; fainter or closer companions (with $\Delta m \ge 6.06$) remain undetectable. A full characterization of the detection limits and the complete contrast curves for the observed sample will be presented in Paper II and III for northern and southern observations, respectively. 

\subsection{Northern Hemisphere}
\label{subsec:NH}

In the northern hemisphere, we conducted observations over three years using a 2.1-m Ritchey-Chretien telescope on an equatorial mount, located at the OAN-SPM in the Sierra de San Pedro Martir, Baja California, in the northeast of Mexico. The telescope is equipped with the "Berkut" instrument, which uses an EMCCD iXon Ultra 888 camera from Andor Technology. This is a highly sensitive camera equipped with an Electron-Multiplying (EM) CCD sensor, with a pixel array of $1024 \times 1024$, and a pixel size of 13 $\mu$m. Its design provides high quantum efficiency--over 80\% across the wavelength range from 450 to 750 nm, and peaking at 95\% around 550 nm. The low readout noise enabled by on-chip electron multiplying allows for very short exposures, which is ideal for speckle interferometry. Additionally, the camera can be thermoelectrically cooled down to $-95^{\circ}$C, which effectively eliminates dark noise, making it well suited for detecting stars as faint as $V = 12$ mag in a single exposure with 30 ms integration time\footnote{For observations of close binaries and multiple systems, which often appear as single stars, it is essential to have high spatial and temporal resolution. For that reason, the choice of an EMCCD camera is particularly advantageous. Its low readout noise and high sensitivity make it possible to detect faint companions near bright primaries, improving contrast in crowded fields. Moreover, its fast frame rate allows us to freeze atmospheric turbulence and achieve diffraction-limited resolution.}. 

Berkut features a filter wheel equipped with a Johnson-Cousins $UBVRI$ set. To achieve Nyquist sampling of the specklegrams in the $I$ filter, Berkut used the $f/15$ secondary mirror combined with a $\times4$ microscope objective lens, that provides a pixel scale of about 35 mas/pixel, yielding a $35 \times 35$ arcsec field of view (FoV) in the full frame mode. Since longer wavelengths are less affected by atmospheric turbulence, the $I$-band offers better performance under typical atmospheric conditions than the other available filters, improving the quality of the speckle pattern and increasing the probability of resolving close companions.

At the 2.1-m telescope at OAN-SPM, under good seeing conditions, the resolution is between \SI{0.7}{\arcsecond} to \SI{1}{\arcsecond}, which is about ten to fifteen times worse than the diffraction-limit for the telescope's aperture, estimated to be $\sim$ 70 mas at a wavelength of 550 nm. In other words, at a distance of 18 pc, all double stars (visual or a bona-fide binary) with angular separation $\rho > 70$~mas (corresponding to a physical separation of approximately 1.25 AU) could be detected\footnote{See Section~\ref{subsec:SB} for the effective absolute magnitude and mass limit though.}.

Due to the geographic location of the observatory, we can only observe targets within a declination range of $-40^{\circ}$ to $+60^{\circ}$. Additionally, the speckle interferometry technique can only be applied to stars brighter than $V = 12$ mag. Faint objects can be observed, but only under the most optimal whether conditions.

The observations were performed using the full Johnson-Cousins filter set ($U$, $B$, $V$, $R$ and $I$) available in the instrument's filter wheel, with the choice of filter depending on the magnitude of the target star to optimize the speckle pattern quality, not for photometric analysis. The number of speckle frames and the exposure time were both selected according to the instantaneous sky conditions (seeing) and the brightness of the target. We recorded between 1000 to 4000 frames per data cubes per target, using exposure times ranging from 10~ms to~30 ms\footnote{The exposure time values and the number of frames were chosen after analyzing the pilot observations where the exposure time was higher and the number of frames was lower. See Section \ref{sec:Pilot}}. The EM gain was adjusted according to the magnitude of the target star; however, the majority of the observations were taken using an EM gain of 300. Frames of 256 $\times$ 256, 512 $\times$ 512 and 1024 $\times$ 1024 pixels were used, depending on the number of stars in the FoV and the seeing conditions. The dates, number of total nights, number of observable nights, and number of data cubes per filter of each observing runs are listed in Table \ref{tab:ObsSPM}. In total, 1863 data cubes were obtained, corresponding to observations of 703 stars throughout the observing runs. The complete observational results for northern hemisphere will be presented in Paper II of this series.

\begin{table}[ht]
    \captionsetup{justification=justified}
    \caption{Summary of observing runs at the 2.1-m telescope in OAN-SPM.}
    \label{tab:ObsSPM}
    \centering
    \small
    \begin{tabular}{l@{\hspace*{5pt}}c@{\hspace*{7pt}}c@{\hspace*{7pt}}c@{\hspace*{7pt}}c@{\hspace*{7pt}}c@{\hspace*{7pt}}c@{\hspace*{7pt}}c}
    \toprule
    \toprule
    Run & N Total & N Observable & N$_U$ & N$_B$ & N$_V$ & N$_R$ & N$_I$ \\
    Date & Nights & Nights &  &  &  &  & \\
    \midrule
    July 2022 & 4 & 4 & 5 & 18 & 10 & 20 & 76 \\
    November 2022 & 5 & 5 & -- & -- & 3 & -- & 298 \\
    June 2023 & 7 & 7 & -- & -- & 270 & -- & 307 \\
    September 2023 & 8 & 5 & -- & -- & 24 & 110 & 1 \\
    February 2024 & 10 & 5 & 2 & -- & -- & 357 & -- \\
    September 2024 & 5 & 5 & 1 & 1 & 14 & 81 & 39 \\
    November 2024 & 8 & 6 & 7 & 2 & -- & 3 & 76 \\  
    June 2025 & 4 & 2 & 2 & 2 & 1 & 1 & 132 \\
    \midrule
    Total & 51 & 39 & 17 & 23 & 322 & 572 & 929 \\
    \bottomrule
    \bottomrule
    \multicolumn{8}{p{0.95\columnwidth}}{NOTE --- In July 2022, five faint stars were observed without a filter. } \\
    \end{tabular}
\end{table}

\subsection{Southern Hemisphere}
\label{subsec:SH}

In the southern hemisphere, we conducted the observations over one year using the New Technology Telescope (NTT), a 3.58-meter Richey-Chretien telescope with an Alt-Az mount, located at the ESO/La Silla Observatory in the southern part of the Atacama Desert, Chile. The telescope was equipped with the AstraLux Sur (ALS) - the twin instrument of AstraLux North \citep{Hormuth2008, Hippler2009} - a visitor instrument.

The geographic location of the observatory allows us to observe targets within the declination range of $-90^{\circ}$ to $+29.5^{\circ}$. Due to the configuration of the telescope and instrument, we are able to observe stars brighter than $I < 16$, giving us the possibility to obtain a more complete observational sample. AstraLux Sur is designed to implement the Lucky Imaging technique; for this reason, each data cube typically contains 10,000 frames. In total, 875 data cubes were obtained, corresponding to observations of 645 stars throughout the observing runs.

Although AstraLux Sur is specifically designed for Lucky Imaging, the individual frames of the data cubes contain specklegrams of the observed stars. Therefore, the data acquired with AstraLux Sur can be processed using other high-angular-resolution techniques. For this reason, we also processed the southern data with the pipeline developed for the northern hemisphere (see Section \ref{ssec:DPN}), which includes classical speckle interferometry, speckle holography, self-calibrated shift-and-add, the autocorrelation function in polar coordinates, and the computation of contrast curve to determine the detection limits for a secondary components. This enables a direct comparison between the two data processing pipelines. In paper III of this series, we will present for the southern hemisphere: a detailed description of the instrumentation; a summary of its observing runs; the total number of stars observed and their observational results; and a comparison of the results obtained with both data reduction pipelines applied to each observed southern star. Paper III will also include a descriptive comparison of the telescopes and instruments between both hemispheres.

\subsection{Observational strategy}
\label{subsec:OS}

The observations were conducted in northern and southern hemisphere to access the majority of the celestial vault, giving us access to declinations from $-90^{\circ}$ to $+60^{\circ}$, according to the declination constraints of each hemisphere. The observational strategy was the same for both hemispheres. The target selection for each run depends on the right ascension (RA) at the local sidereal time (LST). We considered targets within one hour of the initial and final LST of the night, respectively, taking into account that the observation should preferably be performed near the meridian, within a maximum hour angle of $\pm 2$ hour, so that the airmass is as low as possible to avoid refraction-induced distortions (neither telescope are equipped with an ADC).

To maximize the number of targets observed per night and to increase observing efficiency, we organized the list not only by RA but also, within each RA hour, by Dec, such that the telescope movement from one target to the next was minimized. This strategy allowed us to observe as many as 200 targets per night in the north, and about 90 in the south. The difference in the number of targets between the hemispheres is due to the number of frames per data cube taken with each telescope (see Sections \ref{subsec:NH} and \ref{subsec:SH}). These target numbers represent the maximum possible under ideal conditions. The actual number of targets observed per night depends on weather, the specific RA/Dec distribution of targets available during each run, the length of the night, and technical factors such as dome, telescope, and instrument performance. Table \ref{tab:ObsSPM} includes only data cubes of sufficient quality to yield scientifically meaningful results; observations affected by poor conditions or technical issues were excluded from further analysis.


\section{Pilot Observations Conducted with the 2.1 m Telescope at OAN-SPM in July 2022}
\label{sec:Pilot}

Pilot observations for our survey were conducted over five nights in July 2022 using the 2.1-m telescope at the Observatorio Astronómico Nacional, San Pedro Mártir (OAN-SPM). The main objectives were: (1) to validate the high-angular-resolution observing strategy; and (2) to quantify the achievable angular resolution and establish detection limits.

The seeing, estimated from contemporaneous long-exposure images, ranged from $0.7''$ to $1.2''$, with a median value of $0.9''$. Atmospheric transparency was photometric on three nights and variable on two.

A total of 134 data cubes were acquired for 116 stars. Each cube consists of 1000 short-exposure frames obtained with an exposure time of 46 ms and an EM gain of 1000. In several cases, multiple stars were recorded within a single field, and some targets were observed in more than one cube. All datasets were bias-subtracted and processed using standard speckle reconstruction techniques.

\subsection{Astrometric calibrations}
\label{ssec:AC}

Astrometric calibrations for speckle interferometry aim to determine two quantities that are independent of the computational technique used for data processing. The first is the position angle offset, which corresponds to the orientation of the instrument relative to the (instantaneous) celestial north, and the second is the pixel scale, which relates the size of a pixel to an angular distance on the sky. 

To estimate the position angle offset and pixel scale of the speckle interferometer Berkut, we observed 63 double stars. The targets were selected to have components with very slow relative motion and separations in the range of \SI{4}{\arcsecond}–\SI{6}{\arcsecond}. For each system, we adopted the most recent values of separation ($\rho$) and position angle (PA) from the WDS catalog. From these observations, we derived the following calibration parameters: a position angle offset of $\theta_0 = 0.1^{\circ} \pm 0.5^{\circ}$ and a pixel scale of $s = 31.46 \pm 0.16$ mas/px.

\subsection{Data Processing Pipeline}
\label{ssec:DPN}

The first step in data processing is a dark correction of the raw image $I'_n(\mathbf{x})$:
\begin{equation}
I_n(\mathbf{x}) = I'_n(\mathbf{x}) - Dark(\mathbf{x}),
\end{equation}

where $\mathbf{x}$ represents the 2D spatial coordinate, $I_n(\mathbf{x})$ is the corrected image, and $Dark(\mathbf{x})$ is the median dark frame obtained with the shutter closed. In addition, all pixel value less than a pre-defined threshold - set at the level where the signal amplifier begins to contribute spurious counts - are set to zero, thereby effectively suppressing the contribution of read noise. 

The true gain in an EMCCD can be obtained from a linear fit to the exponential tail (in logarithm scale) that appears in the histogram of a dark cube, corresponding to -1/slope of the fit \citep{Harpsoe2012}. In this way, to accurately estimate the number of detected photons, the corrected image $I_n(\mathbf{x})$ is divided by the real gain. 

\begin{figure}[ht!]
\centering
\includegraphics[width=0.35\textwidth]{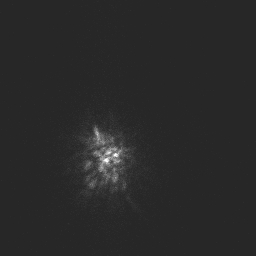}
\includegraphics[width=0.35\textwidth]{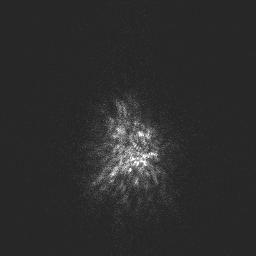}
\captionsetup{justification=justified}
\caption{Data processing. Best (top frame) and worst (bottom frame) speckle pattern in a series of 1000 specklegrams of Gaia DR3 1333760419707327744, a non-binary star.}
\label{fig:BestWorst}
\end{figure}

The second step consists of assessing the quality of the specklegrams. Several parameters can be used to evaluate image quality \citep{McCarthy1986}. In our case, we use the sharpness parameter (Sh1) for each frame, defined as:

\begin{equation}
    {\rm Sh1} = \frac{[\sum I_i(x,y)]^2}{\sum [I_i^2(x,y)]}, 
    \label{eq:sh1}
\end{equation}

where $I(x,y)$ is the intensity on the pixel $(x,y)$. Within each data cube, the images are sorted from the best (lowest Sh1) to the worst (largest Sh1) values, allowing us to discard a percentage of low-quality images and to improve the recovered information (see Figure \ref{fig:BestWorst}). 

\begin{figure}[ht!]
\centering
\includegraphics[width=0.35\textwidth]{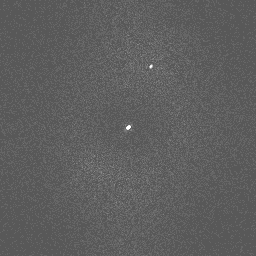}
\includegraphics[width=0.35\textwidth]{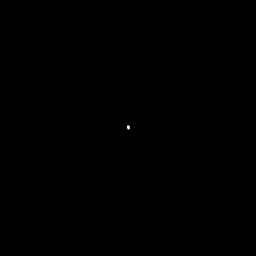}
\captionsetup{justification=justified}
\caption{Data processing. High-resolution images reconstructed using Speckle Holography. A companion detected at a separation greater than \SI{1.7}{\arcsecond} in the star Gaia DR3 1586968533015434624 (top). No distant companion detected in the star Gaia DR3 4389844948935164544 (bottom). }
\label{fig:WideClose}
\end{figure}

The next step is to apply Speckle Holography (SH) to all observed stars, following the methodology described in \citet{Luna2020}, to obtain reconstructed high-resolution images. If a companion is found at a separation greater than 1.7~arcseconds (see Figure \ref{fig:WideClose} top), the data cube is split into two separate cubes, one for each star.

\begin{figure}[ht!]
\centering
\includegraphics[width=0.35\textwidth]{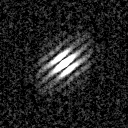}
\includegraphics[width=0.35\textwidth]{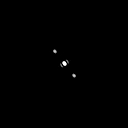}
\captionsetup{justification=justified}
\caption{Data processing. Power spectrum (top) and ACF (bottom) of binary star Gaia DR3 6322245224682035456 (J15157-0725 Aa Bb) with a separation of \SI{0.6}{\arcsecond}.}
\label{fig:PWS}
\end{figure}

The following step is the classic speckle interferometry technique developed by \citet{Labeyrie1970}, which consists of calculating the averaged power spectrum (PS) for each data cube:

\begin{equation}
    PS(\mathbf{f}) = \langle |FT\{ I_n(\mathbf{x})\}|^2 \rangle,
\end{equation}

where $\mathbf{f}$ is the spatial frequency, $FT\{\cdots\}$ denotes the Fourier transform and $\langle \cdots \rangle$ indicates the average over all images. To obtain the photon-bias–corrected power spectrum (see Figure \ref{fig:PWS} top), we follow the procedure described in Section~3.1 of \citep{Orlov2021}. 
By dividing the power spectrum of the object by that of the reference star and applying the inverse Fourier transform, we reconstruct the autocorrelation function (ACF) of the object (see Figure \ref{fig:PWS} bottom).

\begin{figure}[ht!]
\centering
\includegraphics[width=0.45\textwidth]{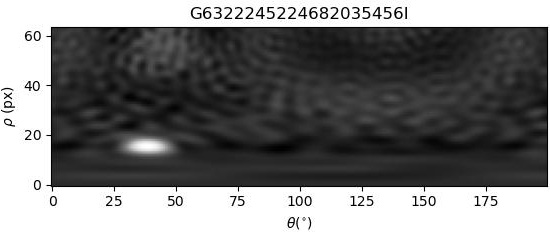}
\includegraphics[width=0.35\textwidth]{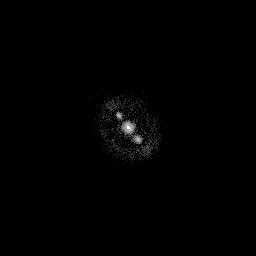}
\captionsetup{justification=justified}
\caption{Data processing. The autocorrelation function calculated in polar coordinates of the same binary star shown in Figure~\ref{fig:PWS}, with a separation of \SI{0.6}{\arcsecond} (top). High-resolution image reconstructed using Self-Calibration Shift-And-Add (bottom). Because the magnitude difference between the stellar components is very small, the reconstructed image is nearly indistinguishable from the ACF (and, hence, it presents the typical quadrant ambiguity).}
\label{fig:HrPolar}
\end{figure}

If a companion is detected, we derive the astrometric parameters from the autocorrelation function calculated in polar coordinates (see Figure~\ref{fig:HrPolar}, top). To resolve the 180° ambiguity, we apply the Self-Calibration Shift-and-Add (SCSAA) image reconstruction technique (see Figure~\ref{fig:HrPolar}, bottom) developed by \citet{Christou1986}. This method enables high-resolution imaging without the need for a reference star. When the component magnitudes are similar, however, the reconstructed image closely resembles the diffraction-limited autocorrelation function, and the ambiguity cannot be resolved. The technique, described in detail by its author in \citet{Christou1986}, has been incorporated into our data-processing pipeline.

Sometimes, a close companion is not detected, and it is therefore necessary to estimate the contrast curves and, from them, the detection limits. Currently, the high-resolution imaging community has established that results must be accompanied by contrast curves. These are essential to confirm the absence of companions brighter than the detection limits defined by the observations. The detection limit describes how faint a companion can be while still remaining detectable in the data. It is defined by the minimum signal-to-noise ratio required to distinguish a real signal from random noise. In practice, a threshold of $5\sigma$ is commonly adopted, which reduces the probability of confusing noise with real detection to much below $1\%$. A lower threshold of $3\sigma$ is sometimes used, although this increases the risk of spurious detections caused by noise \citep{Tokovinin2010}.

At large separations from the primary star (more than about three times the seeing disk) the noise reaches a steady level, and the detection limit becomes roughly constant. 
At smaller separations, however, it strongly depends on the angular distance from the primary star. This behavior is summarized with a contrast curve (Figure \ref{fig:CurvTOI}). The contrast curve plots the faintest detectable companion (in terms of brightness difference, $\Delta m$) as a function of separation from the primary star. In other words, it tells: If there had been a companion this bright at this distance, we would have seen it.
Contrast curves were computed as a function of angular separation by evaluating the standard deviation of intensity fluctuations within annular regions centered on the primary star.

\begin{table*}[ht!]
    \centering
    \caption{Astrometry of binary stars.}
    \label{tab:ObsSPMBin}
    \begin{adjustbox}{max width=\linewidth}
    \begin{tabular}{ccccccccccc}
    \toprule
    \toprule
    WDS & Disc Desig & Simbad ID & JY & $\rho$ & $\Delta \rho$ & $\Theta$ & $\Delta \Theta$ & Filter & GAIA DR3 Component A & GAIA DR3 Component B \\
    &  &  &  & $arcsec$ & $arcsec$ & $grad$ & $grad$ &  &  &  \\
    \midrule
    15009+4526 & HDS2118 & BD+45 2247 & 2022.5426 & 2.047 & 0.010 & 159.6 & 0.6 & R & Gaia DR3 1586968533015434624 & Gaia DR3 1586968533013974528	\\
    15157-0725	& RAO 532 &	G 152-1B & 2022.5454 & 0.497 & 0.003 & 38.1 & 0.6 & I & Gaia DR3 6322245224682035456 & Gaia DR3 6322245327760092800	\\
    15400+4330 & VBS 25 & VB 24 A & 2022.5426 & 4.403 & 0.023 & 156.6 & 0.6 & I & Gaia DR3 1391241478857408768 & Gaia DR3 1391241478857408640	\\
    15496+3449 & BWL 41 & G 179-57 & 2022.5426 & 0.627 & 0.003 & 114.2 & 0.6 & I & Gaia DR3 1372546413931606016 & N/A	\\
    16579+4722 & A 1874 & HD 153557B & 2022.5427 & 4.194 & 0.021 & 123.8 & 0.6 & V & Gaia DR3 1408029509584967168 & Gaia DR3 1408029509583934464	\\
    17465+2743 & AC 7 &	* mu.02 Her B & 2022.551 & 1.058 & 0.006 & 32.0 & 0.6 & I & Gaia DR3 4594497838483177984 & Gaia DR3 4594497834189213184	\\
    19464+3201 & KAM 3 & HD 331161A & 2022.5430 & 5.517 & 0.028 & 135.1 & 0.6 & I & Gaia DR3 2034581091118259328 & Gaia DR3 2034581091125010048	\\
    21000+4004 & KUI 103 & BD+40 883A & 2022.5431 & 1.013 & 0.005 & 82.9 & 0.6 & I & Gaia DR3 1873073175244118784 & Gaia DR3 1873073175249217408	\\
    21072-1355 & STF2752 & HD 200968 & 2022.5432 & 3.450 & 0.018 & 7.4 & 0.6 & I & Gaia DR3 6885342959337741184 & Gaia DR3 6885342955043185664	\\
    22234+3228 & WOR 11 & Wolf 1225 A & 2022.5432 & 1.035 & 0.005 & 100.4 & 0.6 & I & Gaia DR3 1901773448080491392 & Gaia DR3 1901773452377248512	\\
    23317+1956 & WIR 1 & BD+19 5116A & 2022.5433 & 3.494 & 0.018 & 119.1 & 0.6 & I & Gaia DR3 2824770686019003904 & Gaia DR3 2824770686019004032	\\
    23577+2318 & CAC 1 & G 129-47 & 2022.5433 & 0.118 & 0.001 & 85.1 & 0.6 & I & Gaia DR3 2848303052351577344 & N/A \\
    \bottomrule
    \bottomrule
    \end{tabular}%
    \end{adjustbox}
\end{table*}

\subsection{Observation Results}
\subsubsection{Astrometry of Binary Stars}

Table \ref{tab:ObsSPMBin} presents current epoch astrometric measurements of 12 binary stars. For the triple system J15157-0725, we resolve the  components Gaia DR3 6322245224682035456 and Gaia DR3 6322245323464909440. Unfortunately, we were unable to resolve the third component, Gaia DR3 6322245327760092800, in this system, at an expected separation of about 0.3~arcsec. The binary stars J15496+3449 and J23577+2318 are not resolved in Gaia.

\subsubsection{TOI}
\label{subsec:TOI}

\begin{figure*}
\centering
\includegraphics[width=0.33\textwidth]{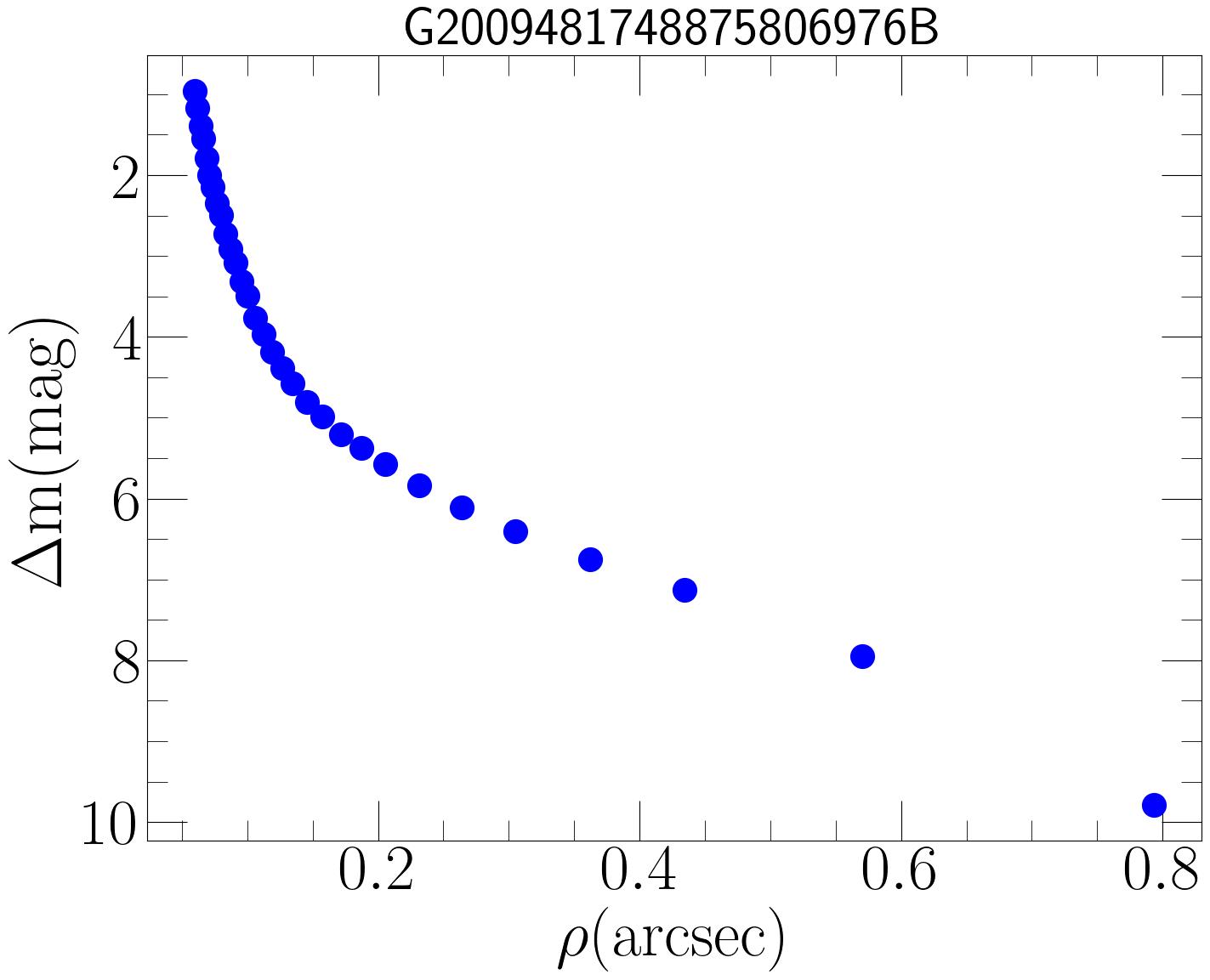}
\includegraphics[width=0.324\textwidth]{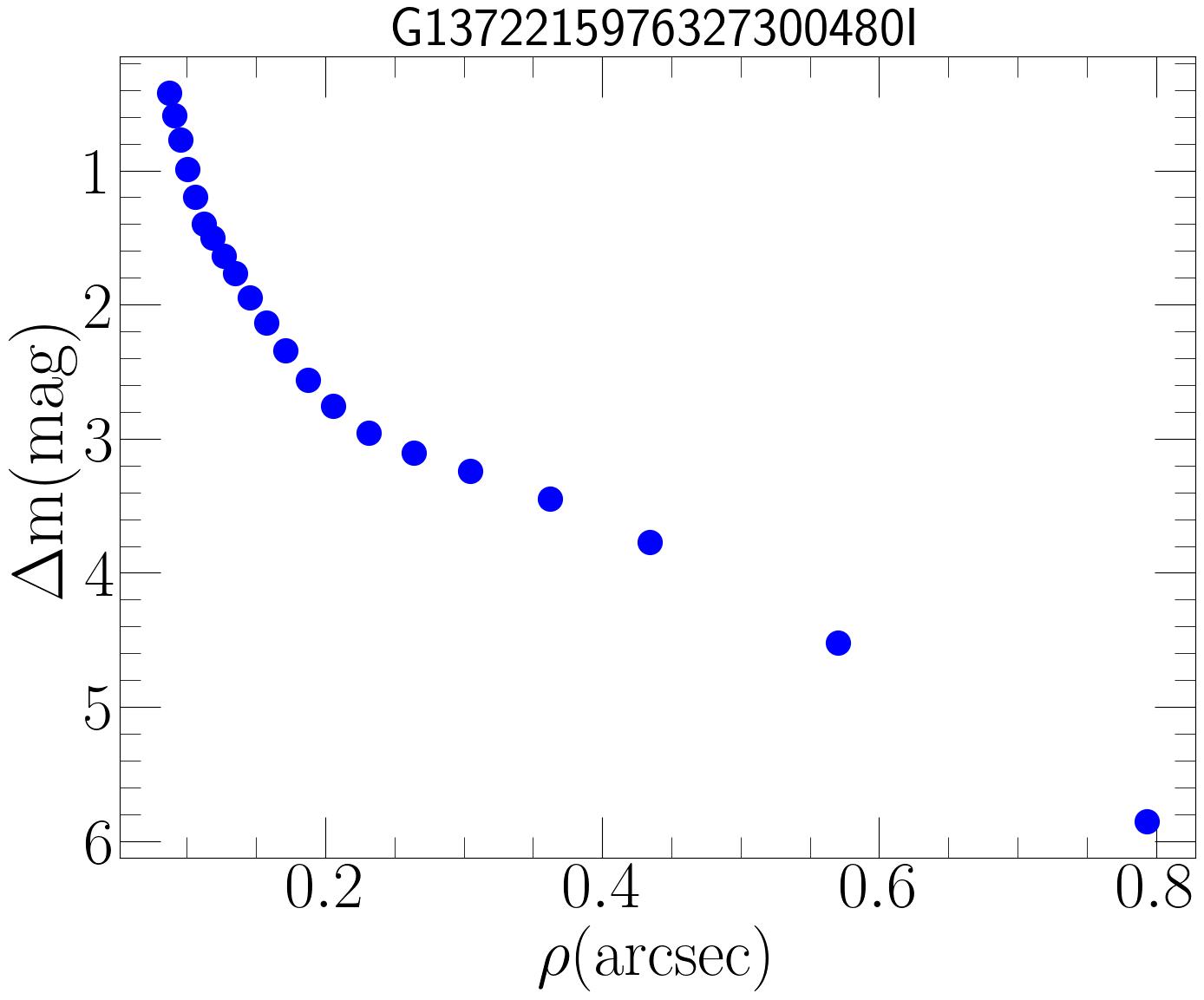}
\includegraphics[width=0.325\textwidth]{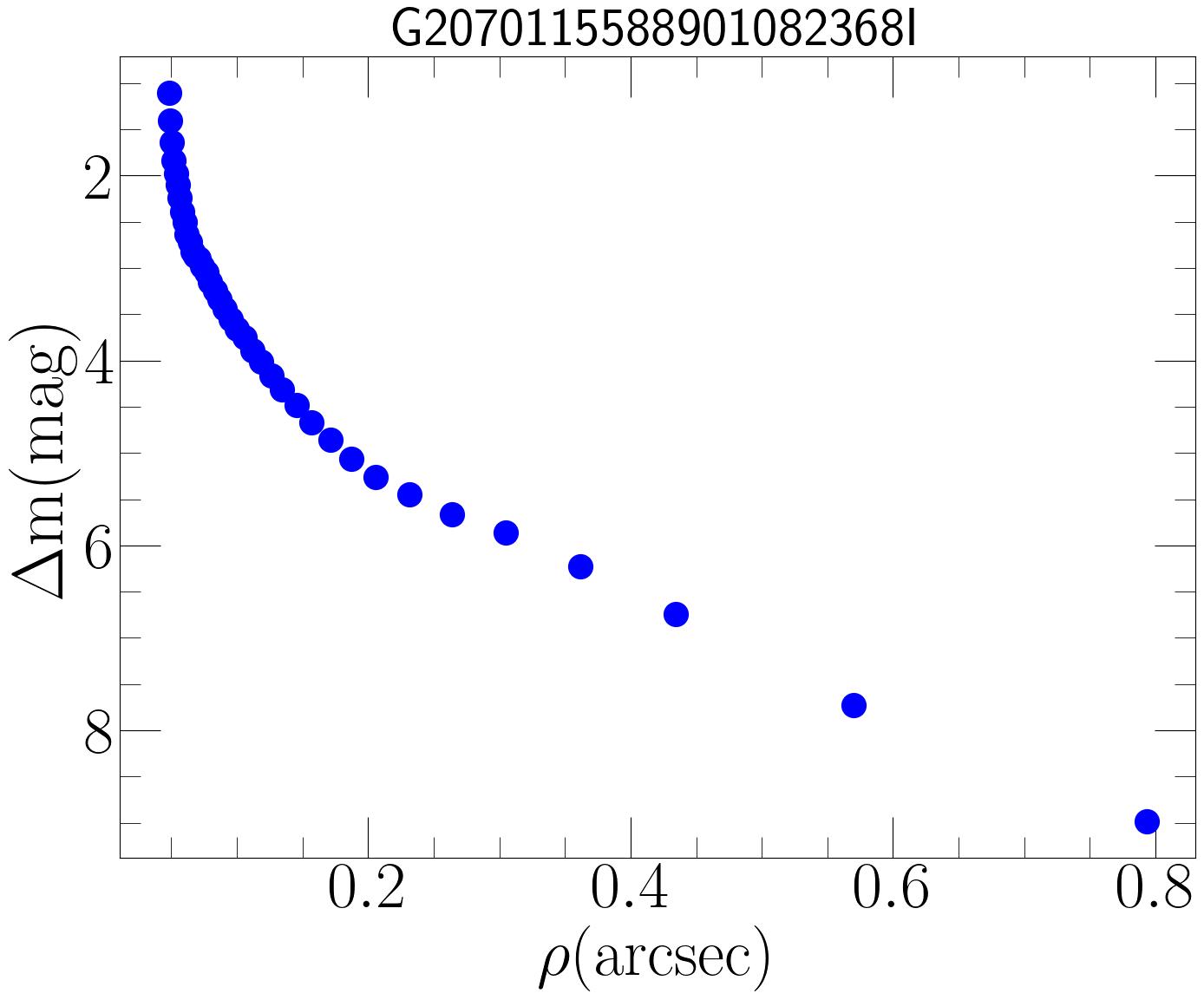}
\caption{Contrast curves for the three TOI targets (from left to right): TOI 1469, TOI 2013, and TOI 4481. No companions were detected. }
\label{fig:CurvTOI}
\end{figure*}

We also observed three TOI targets. Observations of TOI 1469, TOI 2013, and TOI 4481 confirm the absence of close companions that exceed the detection thresholds $5\sigma$ derived from the contrast curves (see Figure~\ref{fig:CurvTOI}).

\subsubsection{Stars within 18 pc of the Sun without Detected Close Companions}

No close companions were detected for 102 stars in our sample. Table~\ref{Tab:ContCurv5p} lists five representative points from the contrast curves, corresponding to magnitude differences at angular separations of $0.1'', 0.2'', 0.3'', 0.5'',$ and $0.8''$. Most targets were observed in one of the Johnson–Cousins filters (U, B, V, R, or I), while the faintest objects were observed without a filter (N). A subset of the sample was observed in three bands (V, R, and I). Comparison of these contrast curves shows that the detection limits are generally best in the I band, occasionally comparable in the R band, and consistently worse in the V band.

\section{Conclusions}
\label{sec:conclusions}

We have compiled a new, well-characterized catalog of 2047 stars within 18~pc of the Sun, constructed from Gaia EDR3 and complemented by the Hipparcos catalog, the catalog of nearby stars by \citet{Kirkpatrick2024}, and the SIMBAD astronomical database. Table \ref{tab:columns} provides the columns that are included in our compiled catalog. 

We have performed a detailed analysis on the effectiveness of three \textit{Gaia} parameters commonly used as binarity flags, namely, RUWE, IPDfmp and IPDfow. For this purpose, we used a control sample of 70 stars confirmed as bona-fide close binaries in the WDS, with separations $\rho <$ \SI{1}{\arcsecond} and with more than one observation, excluding those resolved by \textit{Gaia}. Our key findings are:

\begin{itemize}
    \item Individual parameters are unreliable: a significant fraction of true binaries can be missed by single-parameter thresholds. We found that $\sim20\%$ of our control sample did not trigger any of the binarity flags described in Section \ref{sssec:comparison}. Furthermore, IPDfow shows little to no correlation with binarity in the close-separation regime and is therefore poor indicators on their own.
    
    \item The combination of RUWE and IPDfmp increases detection confidence but remains incomplete: only 36 of the 70 close binaries in our control sample ($\sim51.4\%$) satisfy both parameters simultaneously. Thus, although the combination increases reliability relative to single-parameter thresholds, these indicators still miss a large fraction of true binaries.

    \item \textit{Gaia} indicators do not provide a complete census: relying on \textit{Gaia} binarity indicators, whether individually or in combination, risks missing an important fraction of true double stars. In our control sample, the fraction of missed systems ranges from $\sim20\%$ (stars that do not meet any criteria) to $\sim 48.6\%$ when applying the stricter condition of combining RUWE and IPDfmp.
    
\end{itemize}

These findings underscore a critical limitation of using the \textit{Gaia} parameters as binarity flags to identify potential binaries or to estimate multiplicity fractions, highlighting the necessity of conducting an unbiased observational survey, independent of these indicators. 

The pilot observing campaign conducted in July 2022 achieved its two primary objectives: 

\begin{enumerate}
    \item Validation of the high-angular-resolution observing strategy: the campaign enabled us to refine the observing strategy, establish a robust data-reduction pipeline, and standardize the format for presenting the results. We found that integrations of 1,000 frames yield insufficient signal-to-noise ratios for sources fainter than $m = 10.5$ in the selected filter. Accordingly, all subsequent observations were performed using 3,000–4,000 speckle frames to ensure adequate sensitivity.

    \item Quantification of the achievable angular resolution and establishing detection limits: we determined that the $I$ filter provides superior data quality compared to the other filters--which is consistent with the fact that longer wavelengths are less affected by atmospheric turbulence. Therefore, we prioritized this band for all subsequent observations whenever feasible. Given the anticipated lack of sub-arcsecond companions for most targets, we report detection limits using five representative points sampled from the contrast curves (see Table \ref{Tab:ContCurv5p}). This format is particularly useful for TOIs (see Figure \ref{fig:CurvTOI}) from our sample, as they require confirmation of the absence of close, faint companions.
\end{enumerate}

Although the TOIs in our sample have been observed in multiple epochs, continued follow-up is important, as stellar companions may appear in future epochs that were not detected in earlier observations. This can occur because a binary system can become unresolvable (or conversely, become resolvable after being unresolved). For example, when the secondary component goes through periastron, it can become too close to the primary, near conjunction or behind it, at the time of previous observations. If the orbital period is short, the secondary component is expected to reappear after a few years or less. Another case is if the secondary component is a faint companion that becomes detectable only at a more favorable orbital separation or position angle. Additionally, improvements in instrument sensitivity or observing conditions over time may also reveal previously undetected companions.

\section{FACILITIES}

The observations for this work were acquired using the 2.1-m telescope at the Observatorio Astronómico Nacional San Pedro Mártir, equipped with the Berkut instrument, at the Northern Hemisphere; and the 3.58-m New Technology Telescope at ESO's La Silla Observatory, equipped with the visit instrument AstraLux, at the Southern Hemisphere.


\renewcommand{\refname}{REFERENCES}
\bibliography{RMxAA}

@ARTICLE{Gaia2021,
       author = {{Gaia Collaboration} and {Brown}, A.~G.~A. and {Vallenari}, A. and {Prusti}, T. and {de Bruijne}, J.~H.~J. and {Babusiaux}, C. and {Biermann}, M. and {Creevey}, O.~L. and {Evans}, D.~W. and {Eyer}, L. and {Hutton}, A. and {Jansen}, F. and {Jordi}, C. and {Klioner}, S.~A. and {Lammers}, U. and {Lindegren}, L. and {Luri}, X. and {Mignard}, F. and {Panem}, C. and {Pourbaix}, D. and {Randich}, S. and {Sartoretti}, P. and {Soubiran}, C. and {Walton}, N.~A. and {Arenou}, F. and {Bailer-Jones}, C.~A.~L. and {Bastian}, U. and {Cropper}, M. and {Drimmel}, R. and {Katz}, D. and {Lattanzi}, M.~G. and {van Leeuwen}, F. and {Bakker}, J. and {Cacciari}, C. and {Casta{\~n}eda}, J. and {De Angeli}, F. and {Ducourant}, C. and {Fabricius}, C. and {Fouesneau}, M. and {Fr{\'e}mat}, Y. and {Guerra}, R. and {Guerrier}, A. and {Guiraud}, J. and {Jean-Antoine Piccolo}, A. and {Masana}, E. and {Messineo}, R. and {Mowlavi}, N. and {Nicolas}, C. and {Nienartowicz}, K. and {Pailler}, F. and {Panuzzo}, P. and {Riclet}, F. and {Roux}, W. and {Seabroke}, G.~M. and {Sordo}, R. and {Tanga}, P. and {Th{\'e}venin}, F. and {Gracia-Abril}, G. and {Portell}, J. and {Teyssier}, D. and {Altmann}, M. and {Andrae}, R. and {Bellas-Velidis}, I. and {Benson}, K. and {Berthier}, J. and {Blomme}, R. and {Brugaletta}, E. and {Burgess}, P.~W. and {Busso}, G. and {Carry}, B. and {Cellino}, A. and {Cheek}, N. and {Clementini}, G. and {Damerdji}, Y. and {Davidson}, M. and {Delchambre}, L. and {Dell'Oro}, A. and {Fern{\'a}ndez-Hern{\'a}ndez}, J. and {Galluccio}, L. and {Garc{\'\i}a-Lario}, P. and {Garcia-Reinaldos}, M. and {Gonz{\'a}lez-N{\'u}{\~n}ez}, J. and {Gosset}, E. and {Haigron}, R. and {Halbwachs}, J. -L. and {Hambly}, N.~C. and {Harrison}, D.~L. and {Hatzidimitriou}, D. and {Heiter}, U. and {Hern{\'a}ndez}, J. and {Hestroffer}, D. and {Hodgkin}, S.~T. and {Holl}, B. and {Jan{\ss}en}, K. and {Jevardat de Fombelle}, G. and {Jordan}, S. and {Krone-Martins}, A. and {Lanzafame}, A.~C. and {L{\"o}ffler}, W. and {Lorca}, A. and {Manteiga}, M. and {Marchal}, O. and {Marrese}, P.~M. and {Moitinho}, A. and {Mora}, A. and {Muinonen}, K. and {Osborne}, P. and {Pancino}, E. and {Pauwels}, T. and {Petit}, J. -M. and {Recio-Blanco}, A. and {Richards}, P.~J. and {Riello}, M. and {Rimoldini}, L. and {Robin}, A.~C. and {Roegiers}, T. and {Rybizki}, J. and {Sarro}, L.~M. and {Siopis}, C. and {Smith}, M. and {Sozzetti}, A. and {Ulla}, A. and {Utrilla}, E. and {van Leeuwen}, M. and {van Reeven}, W. and {Abbas}, U. and {Abreu Aramburu}, A. and {Accart}, S. and {Aerts}, C. and {Aguado}, J.~J. and {Ajaj}, M. and {Altavilla}, G. and {{\'A}lvarez}, M.~A. and {{\'A}lvarez Cid-Fuentes}, J. and {Alves}, J. and {Anderson}, R.~I. and {Anglada Varela}, E. and {Antoja}, T. and {Audard}, M. and {Baines}, D. and {Baker}, S.~G. and {Balaguer-N{\'u}{\~n}ez}, L. and {Balbinot}, E. and {Balog}, Z. and {Barache}, C. and {Barbato}, D. and {Barros}, M. and {Barstow}, M.~A. and {Bartolom{\'e}}, S. and {Bassilana}, J. -L. and {Bauchet}, N. and {Baudesson-Stella}, A. and {Becciani}, U. and {Bellazzini}, M. and {Bernet}, M. and {Bertone}, S. and {Bianchi}, L. and {Blanco-Cuaresma}, S. and {Boch}, T. and {Bombrun}, A. and {Bossini}, D. and {Bouquillon}, S. and {Bragaglia}, A. and {Bramante}, L. and {Breedt}, E. and {Bressan}, A. and {Brouillet}, N. and {Bucciarelli}, B. and {Burlacu}, A. and {Busonero}, D. and {Butkevich}, A.~G. and {Buzzi}, R. and {Caffau}, E. and {Cancelliere}, R. and {C{\'a}novas}, H. and {Cantat-Gaudin}, T. and {Carballo}, R. and {Carlucci}, T. and {Carnerero}, M.~I. and {Carrasco}, J.~M. and {Casamiquela}, L. and {Castellani}, M. and {Castro-Ginard}, A. and {Castro Sampol}, P. and {Chaoul}, L. and {Charlot}, P. and {Chemin}, L. and {Chiavassa}, A. and {Cioni}, M. -R.~L. and {Comoretto}, G. and {Cooper}, W.~J. and {Cornez}, T. and {Cowell}, S. and {Crifo}, F. and {Crosta}, M. and {Crowley}, C. and {Dafonte}, C. and {Dapergolas}, A. and {David}, M. and {David}, P.},
        title = {Gaia Early Data Release 3. Summary of the contents and survey properties},
      journal = {\aap},
         year = 2021,
        month = may,
       volume = {649},
          eid = {A1},
        pages = {A1},
          doi = {10.1051/0004-6361/202039657},
archivePrefix = {arXiv},
       eprint = {2012.01533},
 primaryClass = {astro-ph.GA},
       adsurl = {https://ui.adsabs.harvard.edu/abs/2021A&A...649A...1G}
}

@ARTICLE{GCNS2021,
       author = {{Gaia Collaboration} and {Smart}, R.~L. and {Sarro}, L.~M. and {Rybizki}, J. and {Reyl{\'e}}, C. and {Robin}, A.~C. and {Hambly}, N.~C. and {Abbas}, U. and {Barstow}, M.~A. and {de Bruijne}, J.~H.~J. and {Bucciarelli}, B. and {Carrasco}, J.~M. and {Cooper}, W.~J. and {Hodgkin}, S.~T. and {Masana}, E. and {Michalik}, D. and {Sahlmann}, J. and {Sozzetti}, A. and {Brown}, A.~G.~A. and {Vallenari}, A. and {Prusti}, T. and {Babusiaux}, C. and {Biermann}, M. and {Creevey}, O.~L. and {Evans}, D.~W. and {Eyer}, L. and {Hutton}, A. and {Jansen}, F. and {Jordi}, C. and {Klioner}, S.~A. and {Lammers}, U. and {Lindegren}, L. and {Luri}, X. and {Mignard}, F. and {Panem}, C. and {Pourbaix}, D. and {Randich}, S. and {Sartoretti}, P. and {Soubiran}, C. and {Walton}, N.~A. and {Arenou}, F. and {Bailer-Jones}, C.~A.~L. and {Bastian}, U. and {Cropper}, M. and {Drimmel}, R. and {Katz}, D. and {Lattanzi}, M.~G. and {van Leeuwen}, F. and {Bakker}, J. and {Casta{\~n}eda}, J. and {De Angeli}, F. and {Ducourant}, C. and {Fabricius}, C. and {Fouesneau}, M. and {Fr{\'e}mat}, Y. and {Guerra}, R. and {Guerrier}, A. and {Guiraud}, J. and {Jean-Antoine Piccolo}, A. and {Messineo}, R. and {Mowlavi}, N. and {Nicolas}, C. and {Nienartowicz}, K. and {Pailler}, F. and {Panuzzo}, P. and {Riclet}, F. and {Roux}, W. and {Seabroke}, G.~M. and {Sordo}, R. and {Tanga}, P. and {Th{\'e}venin}, F. and {Gracia-Abril}, G. and {Portell}, J. and {Teyssier}, D. and {Altmann}, M. and {Andrae}, R. and {Bellas-Velidis}, I. and {Benson}, K. and {Berthier}, J. and {Blomme}, R. and {Brugaletta}, E. and {Burgess}, P.~W. and {Busso}, G. and {Carry}, B. and {Cellino}, A. and {Cheek}, N. and {Clementini}, G. and {Damerdji}, Y. and {Davidson}, M. and {Delchambre}, L. and {Dell'Oro}, A. and {Fern{\'a}ndez-Hern{\'a}ndez}, J. and {Galluccio}, L. and {Garc{\'\i}a-Lario}, P. and {Garcia-Reinaldos}, M. and {Gonz{\'a}lez-N{\'u}{\~n}ez}, J. and {Gosset}, E. and {Haigron}, R. and {Halbwachs}, J. -L. and {Harrison}, D.~L. and {Hatzidimitriou}, D. and {Heiter}, U. and {Hern{\'a}ndez}, J. and {Hestroffer}, D. and {Holl}, B. and {Jan{\ss}en}, K. and {Jevardat de Fombelle}, G. and {Jordan}, S. and {Krone-Martins}, A. and {Lanzafame}, A.~C. and {L{\"o}ffler}, W. and {Lorca}, A. and {Manteiga}, M. and {Marchal}, O. and {Marrese}, P.~M. and {Moitinho}, A. and {Mora}, A. and {Muinonen}, K. and {Osborne}, P. and {Pancino}, E. and {Pauwels}, T. and {Recio-Blanco}, A. and {Richards}, P.~J. and {Riello}, M. and {Rimoldini}, L. and {Roegiers}, T. and {Siopis}, C. and {Smith}, M. and {Ulla}, A. and {Utrilla}, E. and {van Leeuwen}, M. and {van Reeven}, W. and {Abreu Aramburu}, A. and {Accart}, S. and {Aerts}, C. and {Aguado}, J.~J. and {Ajaj}, M. and {Altavilla}, G. and {{\'A}lvarez}, M.~A. and {{\'A}lvarez Cid-Fuentes}, J. and {Alves}, J. and {Anderson}, R.~I. and {Anglada Varela}, E. and {Antoja}, T. and {Audard}, M. and {Baines}, D. and {Baker}, S.~G. and {Balaguer-N{\'u}{\~n}ez}, L. and {Balbinot}, E. and {Balog}, Z. and {Barache}, C. and {Barbato}, D. and {Barros}, M. and {Bartolom{\'e}}, S. and {Bassilana}, J. -L. and {Bauchet}, N. and {Baudesson-Stella}, A. and {Becciani}, U. and {Bellazzini}, M. and {Bernet}, M. and {Bertone}, S. and {Bianchi}, L. and {Blanco-Cuaresma}, S. and {Boch}, T. and {Bombrun}, A. and {Bossini}, D. and {Bouquillon}, S. and {Bragaglia}, A. and {Bramante}, L. and {Breedt}, E. and {Bressan}, A. and {Brouillet}, N. and {Burlacu}, A. and {Busonero}, D. and {Butkevich}, A.~G. and {Buzzi}, R. and {Caffau}, E. and {Cancelliere}, R. and {C{\'a}novas}, H. and {Cantat-Gaudin}, T. and {Carballo}, R. and {Carlucci}, T. and {Carnerero}, M.~I. and {Casamiquela}, L. and {Castellani}, M. and {Castro-Ginard}, A. and {Castro Sampol}, P. and {Chaoul}, L. and {Charlot}, P. and {Chemin}, L. and {Chiavassa}, A. and {Cioni}, M. -R.~L. and {Comoretto}, G. and {Cornez}, T. and {Cowell}, S. and {Crifo}, F. and {Crosta}, M. and {Crowley}, C. and {Dafonte}, C. and {Dapergolas}, A.},
        title = {Gaia Early Data Release 3. The Gaia Catalogue of Nearby Stars},
      journal = {\aap},
         year = 2021,
        month = may,
       volume = {649},
          eid = {A6},
        pages = {A6},
          doi = {10.1051/0004-6361/202039498},
archivePrefix = {arXiv},
       eprint = {2012.02061},
 primaryClass = {astro-ph.SR},
       adsurl = {https://ui.adsabs.harvard.edu/abs/2021A&A...649A...6G}
}

@ARTICLE{Lindegren2021,
       author = {{Lindegren}, L. and {Klioner}, S.~A. and {Hern{\'a}ndez}, J. and {Bombrun}, A. and {Ramos-Lerate}, M. and {Steidelm{\"u}ller}, H. and {Bastian}, U. and {Biermann}, M. and {de Torres}, A. and {Gerlach}, E. and {Geyer}, R. and {Hilger}, T. and {Hobbs}, D. and {Lammers}, U. and {McMillan}, P.~J. and {Stephenson}, C.~A. and {Casta{\~n}eda}, J. and {Davidson}, M. and {Fabricius}, C. and {Gracia-Abril}, G. and {Portell}, J. and {Rowell}, N. and {Teyssier}, D. and {Torra}, F. and {Bartolom{\'e}}, S. and {Clotet}, M. and {Garralda}, N. and {Gonz{\'a}lez-Vidal}, J.~J. and {Torra}, J. and {Abbas}, U. and {Altmann}, M. and {Anglada Varela}, E. and {Balaguer-N{\'u}{\~n}ez}, L. and {Balog}, Z. and {Barache}, C. and {Becciani}, U. and {Bernet}, M. and {Bertone}, S. and {Bianchi}, L. and {Bouquillon}, S. and {Brown}, A.~G.~A. and {Bucciarelli}, B. and {Busonero}, D. and {Butkevich}, A.~G. and {Buzzi}, R. and {Cancelliere}, R. and {Carlucci}, T. and {Charlot}, P. and {Cioni}, M. -R.~L. and {Crosta}, M. and {Crowley}, C. and {del Peloso}, E.~F. and {del Pozo}, E. and {Drimmel}, R. and {Esquej}, P. and {Fienga}, A. and {Fraile}, E. and {Gai}, M. and {Garcia-Reinaldos}, M. and {Guerra}, R. and {Hambly}, N.~C. and {Hauser}, M. and {Jan{\ss}en}, K. and {Jordan}, S. and {Kostrzewa-Rutkowska}, Z. and {Lattanzi}, M.~G. and {Liao}, S. and {Licata}, E. and {Lister}, T.~A. and {L{\"o}ffler}, W. and {Marchant}, J.~M. and {Masip}, A. and {Mignard}, F. and {Mints}, A. and {Molina}, D. and {Mora}, A. and {Morbidelli}, R. and {Murphy}, C.~P. and {Pagani}, C. and {Panuzzo}, P. and {Pe{\~n}alosa Esteller}, X. and {Poggio}, E. and {Re Fiorentin}, P. and {Riva}, A. and {Sagrist{\`a} Sell{\'e}s}, A. and {Sanchez Gimenez}, V. and {Sarasso}, M. and {Sciacca}, E. and {Siddiqui}, H.~I. and {Smart}, R.~L. and {Souami}, D. and {Spagna}, A. and {Steele}, I.~A. and {Taris}, F. and {Utrilla}, E. and {van Reeven}, W. and {Vecchiato}, A.},
        title = "{Gaia Early Data Release 3. The astrometric solution}",
      journal = {\aap},
         year = 2021,
        month = may,
       volume = {649},
          eid = {A2},
        pages = {A2},
          doi = {10.1051/0004-6361/202039709},
archivePrefix = {arXiv},
       eprint = {2012.03380},
 primaryClass = {astro-ph.IM},
       adsurl = {https://ui.adsabs.harvard.edu/abs/2021A&A...649A...2L}
}

@ARTICLE{VanLeeuwen2007,
       author = {{van Leeuwen}, F.},
        title = "{Validation of the new Hipparcos reduction}",
      journal = {\aap},
         year = 2007,
        month = nov,
       volume = {474},
       number = {2},
        pages = {653-664},
          doi = {10.1051/0004-6361:20078357},
archivePrefix = {arXiv},
       eprint = {0708.1752},
 primaryClass = {astro-ph},
       adsurl = {https://ui.adsabs.harvard.edu/abs/2007A&A...474..653V}
}

@ARTICLE{Kirkpatrick2024,
       author = {{Kirkpatrick}, J. Davy and {Marocco}, Federico and {Gelino}, Christopher R. and {Raghu}, Yadukrishna and {Faherty}, Jacqueline K. and {Bardalez Gagliuffi}, Daniella C. and {Schurr}, Steven D. and {Apps}, Kevin and {Schneider}, Adam C. and {Meisner}, Aaron M. and {Kuchner}, Marc J. and {Caselden}, Dan and {Smart}, R.~L. and {Casewell}, S.~L. and {Raddi}, Roberto and {Kesseli}, Aurora and {Stevnbak Andersen}, Nikolaj and {Antonini}, Edoardo and {Beaulieu}, Paul and {Bickle}, Thomas P. and {Bilsing}, Martin and {Chieng}, Raymond and {Colin}, Guillaume and {Deen}, Sam and {Dereveanco}, Alexandru and {Doll}, Katharina and {Durantini Luca}, Hugo A. and {Frazer}, Anya and {Gantier}, Jean Marc and {Gramaize}, L{\'e}opold and {Grant}, Kristin and {Hamlet}, Leslie K. and {Higashimura}, Hiro and {Hyogo}, Michiharu and {Ja{\l}owiczor}, Peter A. and {Jonkeren}, Alexander and {Kabatnik}, Martin and {Kiwy}, Frank and {Martin}, David W. and {Michaels}, Marianne N. and {Pendrill}, William and {Pessanha Machado}, Celso and {Pumphrey}, Benjamin and {Rothermich}, Austin and {Russwurm}, Rebekah and {Sainio}, Arttu and {Sanchez}, John and {Sapelkin-Tambling}, Fyodor Theo and {Sch{\"u}mann}, J{\"o}rg and {Selg-Mann}, Karl and {Singh}, Harshdeep and {Stenner}, Andres and {Sun}, Guoyou and {Tanner}, Christopher and {Th{\'e}venot}, Melina and {Ventura}, Maurizio and {Voloshin}, Nikita V. and {Walla}, Jim and {W{\k{e}}dracki}, Zbigniew and {Adorno}, Jose I. and {Aganze}, Christian and {Allers}, Katelyn N. and {Brooks}, Hunter and {Burgasser}, Adam J. and {Calamari}, Emily and {Connor}, Thomas and {Costa}, Edgardo and {Eisenhardt}, Peter R. and {Gagn{\'e}}, Jonathan and {Gerasimov}, Roman and {Gonzales}, Eileen C. and {Hsu}, Chih-Chun and {Kiman}, Rocio and {Li}, Guodong and {Low}, Ryan and {Mamajek}, Eric and {Pantoja}, Blake M. and {Popinchalk}, Mark and {Rees}, Jon M. and {Stern}, Daniel and {Su{\'a}rez}, Genaro and {Theissen}, Christopher and {Tsai}, Chao-Wei and {Vos}, Johanna M. and {Zurek}, David and {The Backyard Worlds: Planet 9 Collaboration}},
        title = "{The Initial Mass Function Based on the Full-sky 20 pc Census of {\ensuremath{\sim}}3600 Stars and Brown Dwarfs}",
      journal = {\apjs},
         year = 2024,
        month = apr,
       volume = {271},
       number = {2},
          eid = {55},
        pages = {55},
          doi = {10.3847/1538-4365/ad24e2},
archivePrefix = {arXiv},
       eprint = {2312.03639},
 primaryClass = {astro-ph.SR},
       adsurl = {https://ui.adsabs.harvard.edu/abs/2024ApJS..271...55K}
}

@PROCEEDINGS{ESA1997,
        title = "{The HIPPARCOS and TYCHO catalogues. Astrometric and photometric star catalogues derived from the ESA HIPPARCOS Space Astrometry Mission}",
    booktitle = {ESA Special Publication},
         year = 1997,
       editor = {{ESA}},
       series = {ESA Special Publication},
       volume = {1200},
        month = jan,
       adsurl = {https://ui.adsabs.harvard.edu/abs/1997ESASP1200.....E}
}

@MISC{Gliese1991,
       author = {{Gliese}, W. and {Jahrei{\ss}}, H.},
        title = "{Preliminary Version of the Third Catalogue of Nearby Stars}",
 howpublished = {On: The Astronomical Data Center CD-ROM: Selected Astronomical Catalogs, Vol. I; L.E. Brotzmann, S.E. Gesser (eds.), NASA/Astronomical Data Center, Goddard Space Flight Center, Greenbelt, MD},
         year = 1991,
        month = jan,
       adsurl = {https://ui.adsabs.harvard.edu/abs/1991adc..rept.....G}
}

@ARTICLE{Gaia2023,
       author = {{Gaia Collaboration} and {Vallenari}, A. and {Brown}, A.~G.~A. and {Prusti}, T. and {de Bruijne}, J.~H.~J. and {Arenou}, F. and {Babusiaux}, C. and {Biermann}, M. and {Creevey}, O.~L. and {Ducourant}, C. and {Evans}, D.~W. and {Eyer}, L. and {Guerra}, R. and {Hutton}, A. and {Jordi}, C. and {Klioner}, S.~A. and {Lammers}, U.~L. and {Lindegren}, L. and {Luri}, X. and {Mignard}, F. and {Panem}, C. and {Pourbaix}, D. and {Randich}, S. and {Sartoretti}, P. and {Soubiran}, C. and {Tanga}, P. and {Walton}, N.~A. and {Bailer-Jones}, C.~A.~L. and {Bastian}, U. and {Drimmel}, R. and {Jansen}, F. and {Katz}, D. and {Lattanzi}, M.~G. and {van Leeuwen}, F. and {Bakker}, J. and {Cacciari}, C. and {Casta{\~n}eda}, J. and {De Angeli}, F. and {Fabricius}, C. and {Fouesneau}, M. and {Fr{\'e}mat}, Y. and {Galluccio}, L. and {Guerrier}, A. and {Heiter}, U. and {Masana}, E. and {Messineo}, R. and {Mowlavi}, N. and {Nicolas}, C. and {Nienartowicz}, K. and {Pailler}, F. and {Panuzzo}, P. and {Riclet}, F. and {Roux}, W. and {Seabroke}, G.~M. and {Sordo}, R. and {Th{\'e}venin}, F. and {Gracia-Abril}, G. and {Portell}, J. and {Teyssier}, D. and {Altmann}, M. and {Andrae}, R. and {Audard}, M. and {Bellas-Velidis}, I. and {Benson}, K. and {Berthier}, J. and {Blomme}, R. and {Burgess}, P.~W. and {Busonero}, D. and {Busso}, G. and {C{\'a}novas}, H. and {Carry}, B. and {Cellino}, A. and {Cheek}, N. and {Clementini}, G. and {Damerdji}, Y. and {Davidson}, M. and {de Teodoro}, P. and {Nu{\~n}ez Campos}, M. and {Delchambre}, L. and {Dell'Oro}, A. and {Esquej}, P. and {Fern{\'a}ndez-Hern{\'a}ndez}, J. and {Fraile}, E. and {Garabato}, D. and {Garc{\'\i}a-Lario}, P. and {Gosset}, E. and {Haigron}, R. and {Halbwachs}, J. -L. and {Hambly}, N.~C. and {Harrison}, D.~L. and {Hern{\'a}ndez}, J. and {Hestroffer}, D. and {Hodgkin}, S.~T. and {Holl}, B. and {Jan{\ss}en}, K. and {Jevardat de Fombelle}, G. and {Jordan}, S. and {Krone-Martins}, A. and {Lanzafame}, A.~C. and {L{\"o}ffler}, W. and {Marchal}, O. and {Marrese}, P.~M. and {Moitinho}, A. and {Muinonen}, K. and {Osborne}, P. and {Pancino}, E. and {Pauwels}, T. and {Recio-Blanco}, A. and {Reyl{\'e}}, C. and {Riello}, M. and {Rimoldini}, L. and {Roegiers}, T. and {Rybizki}, J. and {Sarro}, L.~M. and {Siopis}, C. and {Smith}, M. and {Sozzetti}, A. and {Utrilla}, E. and {van Leeuwen}, M. and {Abbas}, U. and {{\'A}brah{\'a}m}, P. and {Abreu Aramburu}, A. and {Aerts}, C. and {Aguado}, J.~J. and {Ajaj}, M. and {Aldea-Montero}, F. and {Altavilla}, G. and {{\'A}lvarez}, M.~A. and {Alves}, J. and {Anders}, F. and {Anderson}, R.~I. and {Anglada Varela}, E. and {Antoja}, T. and {Baines}, D. and {Baker}, S.~G. and {Balaguer-N{\'u}{\~n}ez}, L. and {Balbinot}, E. and {Balog}, Z. and {Barache}, C. and {Barbato}, D. and {Barros}, M. and {Barstow}, M.~A. and {Bartolom{\'e}}, S. and {Bassilana}, J. -L. and {Bauchet}, N. and {Becciani}, U. and {Bellazzini}, M. and {Berihuete}, A. and {Bernet}, M. and {Bertone}, S. and {Bianchi}, L. and {Binnenfeld}, A. and {Blanco-Cuaresma}, S. and {Blazere}, A. and {Boch}, T. and {Bombrun}, A. and {Bossini}, D. and {Bouquillon}, S. and {Bragaglia}, A. and {Bramante}, L. and {Breedt}, E. and {Bressan}, A. and {Brouillet}, N. and {Brugaletta}, E. and {Bucciarelli}, B. and {Burlacu}, A. and {Butkevich}, A.~G. and {Buzzi}, R. and {Caffau}, E. and {Cancelliere}, R. and {Cantat-Gaudin}, T. and {Carballo}, R. and {Carlucci}, T. and {Carnerero}, M.~I. and {Carrasco}, J.~M. and {Casamiquela}, L. and {Castellani}, M. and {Castro-Ginard}, A. and {Chaoul}, L. and {Charlot}, P. and {Chemin}, L. and {Chiaramida}, V. and {Chiavassa}, A. and {Chornay}, N. and {Comoretto}, G. and {Contursi}, G. and {Cooper}, W.~J. and {Cornez}, T. and {Cowell}, S. and {Crifo}, F. and {Cropper}, M. and {Crosta}, M. and {Crowley}, C. and {Dafonte}, C. and {Dapergolas}, A. and {David}, M. and {David}, P. and {de Laverny}, P. and {De Luise}, F. and {De March}, R.},
        title = "{Gaia Data Release 3. Summary of the content and survey properties}",
      journal = {\aap},
         year = 2023,
        month = jun,
       volume = {674},
          eid = {A1},
        pages = {A1},
          doi = {10.1051/0004-6361/202243940},
archivePrefix = {arXiv},
       eprint = {2208.00211},
 primaryClass = {astro-ph.GA},
       adsurl = {https://ui.adsabs.harvard.edu/abs/2023A&A...674A...1G}
}

@ARTICLE{WISE2010,
       author = {{Wright}, Edward L. and {Eisenhardt}, Peter R.~M. and {Mainzer}, Amy K. and {Ressler}, Michael E. and {Cutri}, Roc M. and {Jarrett}, Thomas and {Kirkpatrick}, J. Davy and {Padgett}, Deborah and {McMillan}, Robert S. and {Skrutskie}, Michael and {Stanford}, S.~A. and {Cohen}, Martin and {Walker}, Russell G. and {Mather}, John C. and {Leisawitz}, David and {Gautier}, III, Thomas N. and {McLean}, Ian and {Benford}, Dominic and {Lonsdale}, Carol J. and {Blain}, Andrew and {Mendez}, Bryan and {Irace}, William R. and {Duval}, Valerie and {Liu}, Fengchuan and {Royer}, Don and {Heinrichsen}, Ingolf and {Howard}, Joan and {Shannon}, Mark and {Kendall}, Martha and {Walsh}, Amy L. and {Larsen}, Mark and {Cardon}, Joel G. and {Schick}, Scott and {Schwalm}, Mark and {Abid}, Mohamed and {Fabinsky}, Beth and {Naes}, Larry and {Tsai}, Chao-Wei},
        title = "{The Wide-field Infrared Survey Explorer (WISE): Mission Description and Initial On-orbit Performance}",
      journal = {\aj},
         year = 2010,
        month = dec,
       volume = {140},
       number = {6},
        pages = {1868-1881},
          doi = {10.1088/0004-6256/140/6/1868},
archivePrefix = {arXiv},
       eprint = {1008.0031},
 primaryClass = {astro-ph.IM},
       adsurl = {https://ui.adsabs.harvard.edu/abs/2010AJ....140.1868W}
}

@ARTICLE{Holl2023,
       author = {{Holl}, B. and {Fabricius}, C. and {Portell}, J. and {Lindegren}, L. and {Panuzzo}, P. and {Bernet}, M. and {Casta{\~n}eda}, J. and {Jevardat de Fombelle}, G. and {Audard}, M. and {Ducourant}, C. and {Harrison}, D.~L. and {Evans}, D.~W. and {Busso}, G. and {Sozzetti}, A. and {Gosset}, E. and {Arenou}, F. and {De Angeli}, F. and {Riello}, M. and {Eyer}, L. and {Rimoldini}, L. and {Gavras}, P. and {Mowlavi}, N. and {Nienartowicz}, K. and {Lecoeur-Ta{\"\i}bi}, I. and {Garc{\'\i}a-Lario}, P. and {Pourbaix}, D.},
        title = "{Gaia Data Release 3. Gaia scan-angle-dependent signals and spurious periods}",
      journal = {\aap},
         year = 2023,
        month = jun,
       volume = {674},
          eid = {A25},
        pages = {A25},
          doi = {10.1051/0004-6361/202245353},
archivePrefix = {arXiv},
       eprint = {2212.11971},
 primaryClass = {astro-ph.IM},
       adsurl = {https://ui.adsabs.harvard.edu/abs/2023A&A...674A..25H}
}

@ARTICLE{Halbwachs2023,
       author = {{Halbwachs}, Jean-Louis and {Pourbaix}, Dimitri and {Arenou}, Fr{\'e}d{\'e}ric and {Galluccio}, Laurent and {Guillout}, Patrick and {Bauchet}, Nathalie and {Marchal}, Olivier and {Sadowski}, Gilles and {Teyssier}, David},
        title = "{Gaia Data Release 3. Astrometric binary star processing}",
      journal = {\aap},
         year = 2023,
        month = jun,
       volume = {674},
          eid = {A9},
        pages = {A9},
          doi = {10.1051/0004-6361/202243969},
archivePrefix = {arXiv},
       eprint = {2206.05726},
 primaryClass = {astro-ph.SR},
       adsurl = {https://ui.adsabs.harvard.edu/abs/2023A&A...674A...9H}
}

@ARTICLE{Penoyre2022,
       author = {{Penoyre}, Zephyr and {Belokurov}, Vasily and {Evans}, N. Wyn},
        title = "{Astrometric identification of nearby binary stars - II. Astrometric binaries in the Gaia Catalogue of Nearby Stars}",
      journal = {\mnras},
         year = 2022,
        month = jul,
       volume = {513},
       number = {4},
        pages = {5270-5289},
          doi = {10.1093/mnras/stac1147},
archivePrefix = {arXiv},
       eprint = {2202.06963},
 primaryClass = {astro-ph.SR},
       adsurl = {https://ui.adsabs.harvard.edu/abs/2022MNRAS.513.5270P}
}

@ARTICLE{Fabricius2021,
       author = {{Fabricius}, C. and {Luri}, X. and {Arenou}, F. and {Babusiaux}, C. and {Helmi}, A. and {Muraveva}, T. and {Reyl{\'e}}, C. and {Spoto}, F. and {Vallenari}, A. and {Antoja}, T. and {Balbinot}, E. and {Barache}, C. and {Bauchet}, N. and {Bragaglia}, A. and {Busonero}, D. and {Cantat-Gaudin}, T. and {Carrasco}, J.~M. and {Diakit{\'e}}, S. and {Fabrizio}, M. and {Figueras}, F. and {Garcia-Gutierrez}, A. and {Garofalo}, A. and {Jordi}, C. and {Kervella}, P. and {Khanna}, S. and {Leclerc}, N. and {Licata}, E. and {Lambert}, S. and {Marrese}, P.~M. and {Masip}, A. and {Ramos}, P. and {Robichon}, N. and {Robin}, A.~C. and {Romero-G{\'o}mez}, M. and {Rubele}, S. and {Weiler}, M.},
        title = "{Gaia Early Data Release 3. Catalogue validation}",
      journal = {\aap},
         year = 2021,
        month = may,
       volume = {649},
          eid = {A5},
        pages = {A5},
          doi = {10.1051/0004-6361/202039834},
archivePrefix = {arXiv},
       eprint = {2012.06242},
 primaryClass = {astro-ph.GA},
       adsurl = {https://ui.adsabs.harvard.edu/abs/2021A&A...649A...5F}
}

@ARTICLE{Lehtinen2023,
       author = {{Lehtinen}, K. and {Prusti}, T. and {de Bruijne}, J. and {Lammers}, U. and {Manara}, C.~F. and {Ness}, J. -U. and {Siddiqui}, H. and {Poutanen}, M. and {Muinonen}, K. and {Morrison}, O.},
        title = "{Carte du Ciel and Gaia. I. Astrometry}",
      journal = {\aap},
         year = 2023,
        month = mar,
       volume = {671},
          eid = {A16},
        pages = {A16},
          doi = {10.1051/0004-6361/202142929},
       adsurl = {https://ui.adsabs.harvard.edu/abs/2023A&A...671A..16L}
}

@ARTICLE{Tokovinin2023,
       author = {{Tokovinin}, Andrei},
        title = "{Exploring Thousands of Nearby Hierarchical Systems with Gaia and Speckle Interferometry}",
      journal = {\aj},
         year = 2023,
        month = apr,
       volume = {165},
       number = {4},
          eid = {180},
        pages = {180},
          doi = {10.3847/1538-3881/acc464},
       adsurl = {https://ui.adsabs.harvard.edu/abs/2023AJ....165..180T}
}

@ARTICLE{Gaia2016,
       author = {{Gaia Collaboration} and {Brown}, A.~G.~A. and {Vallenari}, A. and {Prusti}, T. and {de Bruijne}, J.~H.~J. and {Mignard}, F. and {Drimmel}, R. and {Babusiaux}, C. and {Bailer-Jones}, C.~A.~L. and {Bastian}, U. and {Biermann}, M. and {Evans}, D.~W. and {Eyer}, L. and {Jansen}, F. and {Jordi}, C. and {Katz}, D. and {Klioner}, S.~A. and {Lammers}, U. and {Lindegren}, L. and {Luri}, X. and {O'Mullane}, W. and {Panem}, C. and {Pourbaix}, D. and {Randich}, S. and {Sartoretti}, P. and {Siddiqui}, H.~I. and {Soubiran}, C. and {Valette}, V. and {van Leeuwen}, F. and {Walton}, N.~A. and {Aerts}, C. and {Arenou}, F. and {Cropper}, M. and {H{\o}g}, E. and {Lattanzi}, M.~G. and {Grebel}, E.~K. and {Holland}, A.~D. and {Huc}, C. and {Passot}, X. and {Perryman}, M. and {Bramante}, L. and {Cacciari}, C. and {Casta{\~n}eda}, J. and {Chaoul}, L. and {Cheek}, N. and {De Angeli}, F. and {Fabricius}, C. and {Guerra}, R. and {Hern{\'a}ndez}, J. and {Jean-Antoine-Piccolo}, A. and {Masana}, E. and {Messineo}, R. and {Mowlavi}, N. and {Nienartowicz}, K. and {Ord{\'o}{\~n}ez-Blanco}, D. and {Panuzzo}, P. and {Portell}, J. and {Richards}, P.~J. and {Riello}, M. and {Seabroke}, G.~M. and {Tanga}, P. and {Th{\'e}venin}, F. and {Torra}, J. and {Els}, S.~G. and {Gracia-Abril}, G. and {Comoretto}, G. and {Garcia-Reinaldos}, M. and {Lock}, T. and {Mercier}, E. and {Altmann}, M. and {Andrae}, R. and {Astraatmadja}, T.~L. and {Bellas-Velidis}, I. and {Benson}, K. and {Berthier}, J. and {Blomme}, R. and {Busso}, G. and {Carry}, B. and {Cellino}, A. and {Clementini}, G. and {Cowell}, S. and {Creevey}, O. and {Cuypers}, J. and {Davidson}, M. and {De Ridder}, J. and {de Torres}, A. and {Delchambre}, L. and {Dell'Oro}, A. and {Ducourant}, C. and {Fr{\'e}mat}, Y. and {Garc{\'\i}a-Torres}, M. and {Gosset}, E. and {Halbwachs}, J. -L. and {Hambly}, N.~C. and {Harrison}, D.~L. and {Hauser}, M. and {Hestroffer}, D. and {Hodgkin}, S.~T. and {Huckle}, H.~E. and {Hutton}, A. and {Jasniewicz}, G. and {Jordan}, S. and {Kontizas}, M. and {Korn}, A.~J. and {Lanzafame}, A.~C. and {Manteiga}, M. and {Moitinho}, A. and {Muinonen}, K. and {Osinde}, J. and {Pancino}, E. and {Pauwels}, T. and {Petit}, J. -M. and {Recio-Blanco}, A. and {Robin}, A.~C. and {Sarro}, L.~M. and {Siopis}, C. and {Smith}, M. and {Smith}, K.~W. and {Sozzetti}, A. and {Thuillot}, W. and {van Reeven}, W. and {Viala}, Y. and {Abbas}, U. and {Abreu Aramburu}, A. and {Accart}, S. and {Aguado}, J.~J. and {Allan}, P.~M. and {Allasia}, W. and {Altavilla}, G. and {{\'A}lvarez}, M.~A. and {Alves}, J. and {Anderson}, R.~I. and {Andrei}, A.~H. and {Anglada Varela}, E. and {Antiche}, E. and {Antoja}, T. and {Ant{\'o}n}, S. and {Arcay}, B. and {Bach}, N. and {Baker}, S.~G. and {Balaguer-N{\'u}{\~n}ez}, L. and {Barache}, C. and {Barata}, C. and {Barbier}, A. and {Barblan}, F. and {Barrado y Navascu{\'e}s}, D. and {Barros}, M. and {Barstow}, M.~A. and {Becciani}, U. and {Bellazzini}, M. and {Bello Garc{\'\i}a}, A. and {Belokurov}, V. and {Bendjoya}, P. and {Berihuete}, A. and {Bianchi}, L. and {Bienaym{\'e}}, O. and {Billebaud}, F. and {Blagorodnova}, N. and {Blanco-Cuaresma}, S. and {Boch}, T. and {Bombrun}, A. and {Borrachero}, R. and {Bouquillon}, S. and {Bourda}, G. and {Bouy}, H. and {Bragaglia}, A. and {Breddels}, M.~A. and {Brouillet}, N. and {Br{\"u}semeister}, T. and {Bucciarelli}, B. and {Burgess}, P. and {Burgon}, R. and {Burlacu}, A. and {Busonero}, D. and {Buzzi}, R. and {Caffau}, E. and {Cambras}, J. and {Campbell}, H. and {Cancelliere}, R. and {Cantat-Gaudin}, T. and {Carlucci}, T. and {Carrasco}, J.~M. and {Castellani}, M. and {Charlot}, P. and {Charnas}, J. and {Chiavassa}, A. and {Clotet}, M. and {Cocozza}, G. and {Collins}, R.~S. and {Costigan}, G. and {Crifo}, F. and {Cross}, N.~J.~G. and {Crosta}, M. and {Crowley}, C. and {Dafonte}, C. and {Damerdji}, Y. and {Dapergolas}, A. and {David}, P. and {David}, M. and {De Cat}, P.},
        title = "{Gaia Data Release 1. Summary of the astrometric, photometric, and survey properties}",
      journal = {\aap},
         year = 2016,
        month = nov,
       volume = {595},
          eid = {A2},
        pages = {A2},
          doi = {10.1051/0004-6361/201629512},
archivePrefix = {arXiv},
       eprint = {1609.04172},
 primaryClass = {astro-ph.IM},
       adsurl = {https://ui.adsabs.harvard.edu/abs/2016A&A...595A...2G}
}

@ARTICLE{Labeyrie1970,
       author = {{Labeyrie}, A.},
        title = "{Attainment of Diffraction Limited Resolution in Large Telescopes by Fourier Analysing Speckle Patterns in Star Images}",
      journal = {\aap},
         year = 1970,
        month = may,
       volume = {6},
        pages = {85},
       adsurl = {https://ui.adsabs.harvard.edu/abs/1970A&A.....6...85L}
}

@ARTICLE{Orlov2021,
       author = {{Orlov}, V.~G.},
        title = "{Speckle Interferometry at the Observatorio Astron{\'o}mico Nacional. VII}",
      journal = {\RMxAA},
         year = 2021,
        month = apr,
       volume = {57},
        pages = {67-80},
          doi = {10.22201/ia.01851101p.2021.57.01.04},
       adsurl = {https://ui.adsabs.harvard.edu/abs/2021RMxAA..57...67O}
}

@ARTICLE{Tokovinin2020,
       author = {{Tokovinin}, Andrei and {Mason}, Brian D. and {Mendez}, Rene A. and {Costa}, Edgardo and {Horch}, Elliott P.},
        title = "{Speckle Interferometry at SOAR in 2019}",
      journal = {\aj},
         year = 2020,
        month = jul,
       volume = {160},
       number = {1},
          eid = {7},
        pages = {7},
          doi = {10.3847/1538-3881/ab91c1},
archivePrefix = {arXiv},
       eprint = {2005.05305},
 primaryClass = {astro-ph.SR},
       adsurl = {https://ui.adsabs.harvard.edu/abs/2020AJ....160....7T}
}

@INPROCEEDINGS{Balega2017,
       author = {{Balega}, Yu. Yu. and {Dyachenko}, V.~V. and {Maksimov}, A.~F. and {Malogolovets}, E.~V. and {Rastegaev}, D.~A.},
        title = "{Multiplicity Survey of Bright Stars in the Nearby OB Associations}",
    booktitle = {Stars: From Collapse to Collapse},
         year = 2017,
       editor = {{Balega}, Yu. Yu. and {Kudryavtsev}, D.~O. and {Romanyuk}, I.~I. and {Yakunin}, I.~A.},
       series = {Astronomical Society of the Pacific Conference Series},
       volume = {510},
        month = jun,
        pages = {325},
       adsurl = {https://ui.adsabs.harvard.edu/abs/2017ASPC..510..325B}
}

@INPROCEEDINGS{Hormuth2008,
       author = {{Hormuth}, Felix and {Hippler}, Stefan and {Brandner}, Wolfgang and {Wagner}, Karl and {Henning}, Thomas},
        title = "{AstraLux: the Calar Alto lucky imaging camera}",
    booktitle = {Ground-based and Airborne Instrumentation for Astronomy II},
         year = 2008,
       editor = {{McLean}, Ian S. and {Casali}, Mark M.},
       series = {Society of Photo-Optical Instrumentation Engineers (SPIE) Conference Series},
       volume = {7014},
        month = jul,
          eid = {701448},
        pages = {701448},
          doi = {10.1117/12.787384},
archivePrefix = {arXiv},
       eprint = {0807.0497},
 primaryClass = {astro-ph},
       adsurl = {https://ui.adsabs.harvard.edu/abs/2008SPIE.7014E..48H}
}

@ARTICLE{Hippler2009,
       author = {{Hippler}, Stefan and {Bergfors}, Carolina and {Brandner Wolfgang} and {Daemgen}, Sebastian and {Henning}, Thomas and {Hormuth}, Felix and {Huber}, Armin and {Janson}, Markus and {Rochau}, Boyke and {Rohloff}, Ralf-Rainer and {Wagner}, Karl},
        title = "{The AstraLux Sur Lucky Imaging Instrument at the NTT}",
      journal = {The Messenger},
         year = 2009,
        month = sep,
       volume = {137},
        pages = {14-17},
       adsurl = {https://ui.adsabs.harvard.edu/abs/2009Msngr.137...14H}
}

@ARTICLE{Hartkopf2015,
       author = {{Hartkopf}, William I. and {Mason}, Brian D.},
        title = "{Speckle Interferometry at the U.S. Naval Observatory. XX}",
      journal = {\aj},
         year = 2015,
        month = oct,
       volume = {150},
       number = {4},
          eid = {136},
        pages = {136},
          doi = {10.1088/0004-6256/150/4/136},
       adsurl = {https://ui.adsabs.harvard.edu/abs/2015AJ....150..136H}
}

@ARTICLE{Howell2021,
       author = {{Howell}, Steve B. and {Scott}, Nicholas J. and {Matson}, Rachel A. and {Everett}, Mark E. and {Furlan}, Elise and {Gnilka}, Crystal L. and {Ciardi}, David R. and {Lester}, Kathryn V.},
        title = "{The NASA High-Resolution Speckle Interferometric Imaging Program: Validation and Characterization of Exoplanets and Their Stellar Hosts}",
      journal = {\frass},
         year = 2021,
        month = mar,
       volume = {8},
          eid = {10},
        pages = {10},
          doi = {10.3389/fspas.2021.635864},
archivePrefix = {arXiv},
       eprint = {2101.08378},
 primaryClass = {astro-ph.EP},
       adsurl = {https://ui.adsabs.harvard.edu/abs/2021FrASS...8...10H}
}

@ARTICLE{WDSCat2001,
       author = {{Mason}, Brian D. and {Wycoff}, Gary L. and {Hartkopf}, William I. and
         {Douglass}, Geoffrey G. and {Worley}, Charles E.},
        title = "{The 2001 US Naval Observatory Double Star CD-ROM. I. The Washington Double Star Catalog}",
      journal = {\aj},
         year = 2001,
        month = dec,
       volume = {122},
       number = {6},
        pages = {3466-3471},
          doi = {10.1086/323920},
       adsurl = {https://ui.adsabs.harvard.edu/abs/2001AJ....122.3466M}
}

@ARTICLE{Gaia2023multiplicity,
       author = {{Gaia Collaboration} and {Arenou}, F. and {Babusiaux}, C. and {Barstow}, M.~A. and {Faigler}, S. and {Jorissen}, A. and {Kervella}, P. and {Mazeh}, T. and {Mowlavi}, N. and {Panuzzo}, P. and {Sahlmann}, J. and {Shahaf}, S. and {Sozzetti}, A. and {Bauchet}, N. and {Damerdji}, Y. and {Gavras}, P. and {Giacobbe}, P. and {Gosset}, E. and {Halbwachs}, J. -L. and {Holl}, B. and {Lattanzi}, M.~G. and {Leclerc}, N. and {Morel}, T. and {Pourbaix}, D. and {Re Fiorentin}, P. and {Sadowski}, G. and {S{\'e}gransan}, D. and {Siopis}, C. and {Teyssier}, D. and {Zwitter}, T. and {Planquart}, L. and {Brown}, A.~G.~A. and {Vallenari}, A. and {Prusti}, T. and {de Bruijne}, J.~H.~J. and {Biermann}, M. and {Creevey}, O.~L. and {Ducourant}, C. and {Evans}, D.~W. and {Eyer}, L. and {Guerra}, R. and {Hutton}, A. and {Jordi}, C. and {Klioner}, S.~A. and {Lammers}, U.~L. and {Lindegren}, L. and {Luri}, X. and {Mignard}, F. and {Panem}, C. and {Randich}, S. and {Sartoretti}, P. and {Soubiran}, C. and {Tanga}, P. and {Walton}, N.~A. and {Bailer-Jones}, C.~A.~L. and {Bastian}, U. and {Drimmel}, R. and {Jansen}, F. and {Katz}, D. and {van Leeuwen}, F. and {Bakker}, J. and {Cacciari}, C. and {Casta{\~n}eda}, J. and {De Angeli}, F. and {Fabricius}, C. and {Fouesneau}, M. and {Fr{\'e}mat}, Y. and {Galluccio}, L. and {Guerrier}, A. and {Heiter}, U. and {Masana}, E. and {Messineo}, R. and {Nicolas}, C. and {Nienartowicz}, K. and {Pailler}, F. and {Riclet}, F. and {Roux}, W. and {Seabroke}, G.~M. and {Sordo}, R. and {Th{\'e}venin}, F. and {Gracia-Abril}, G. and {Portell}, J. and {Altmann}, M. and {Andrae}, R. and {Audard}, M. and {Bellas-Velidis}, I. and {Benson}, K. and {Berthier}, J. and {Blomme}, R. and {Burgess}, P.~W. and {Busonero}, D. and {Busso}, G. and {C{\'a}novas}, H. and {Carry}, B. and {Cellino}, A. and {Cheek}, N. and {Clementini}, G. and {Davidson}, M. and {de Teodoro}, P. and {Nu{\~n}ez Campos}, M. and {Delchambre}, L. and {Dell'Oro}, A. and {Esquej}, P. and {Fern{\'a}ndez-Hern{\'a}ndez}, J. and {Fraile}, E. and {Garabato}, D. and {Garc{\'\i}a-Lario}, P. and {Haigron}, R. and {Hambly}, N.~C. and {Harrison}, D.~L. and {Hern{\'a}ndez}, J. and {Hestroffer}, D. and {Hodgkin}, S.~T. and {Jan{\ss}en}, K. and {Jevardat de Fombelle}, G. and {Jordan}, S. and {Krone-Martins}, A. and {Lanzafame}, A.~C. and {L{\"o}ffler}, W. and {Marchal}, O. and {Marrese}, P.~M. and {Moitinho}, A. and {Muinonen}, K. and {Osborne}, P. and {Pancino}, E. and {Pauwels}, T. and {Recio-Blanco}, A. and {Reyl{\'e}}, C. and {Riello}, M. and {Rimoldini}, L. and {Roegiers}, T. and {Rybizki}, J. and {Sarro}, L.~M. and {Smith}, M. and {Utrilla}, E. and {van Leeuwen}, M. and {Abbas}, U. and {{\'A}brah{\'a}m}, P. and {Abreu Aramburu}, A. and {Aerts}, C. and {Aguado}, J.~J. and {Ajaj}, M. and {Aldea-Montero}, F. and {Altavilla}, G. and {{\'A}lvarez}, M.~A. and {Alves}, J. and {Anders}, F. and {Anderson}, R.~I. and {Anglada Varela}, E. and {Antoja}, T. and {Baines}, D. and {Baker}, S.~G. and {Balaguer-N{\'u}{\~n}ez}, L. and {Balbinot}, E. and {Balog}, Z. and {Barache}, C. and {Barbato}, D. and {Barros}, M. and {Bartolom{\'e}}, S. and {Bassilana}, J. -L. and {Becciani}, U. and {Bellazzini}, M. and {Berihuete}, A. and {Bernet}, M. and {Bertone}, S. and {Bianchi}, L. and {Binnenfeld}, A. and {Blanco-Cuaresma}, S. and {Blazere}, A. and {Boch}, T. and {Bombrun}, A. and {Bossini}, D. and {Bouquillon}, S. and {Bragaglia}, A. and {Bramante}, L. and {Breedt}, E. and {Bressan}, A. and {Brouillet}, N. and {Brugaletta}, E. and {Bucciarelli}, B. and {Burlacu}, A. and {Butkevich}, A.~G. and {Buzzi}, R. and {Caffau}, E. and {Cancelliere}, R. and {Cantat-Gaudin}, T. and {Carballo}, R. and {Carlucci}, T. and {Carnerero}, M.~I. and {Carrasco}, J.~M. and {Casamiquela}, L. and {Castellani}, M. and {Castro-Ginard}, A. and {Chaoul}, L. and {Charlot}, P. and {Chemin}, L. and {Chiaramida}, V. and {Chiavassa}, A. and {Chornay}, N. and {Comoretto}, G.},
        title = "{Gaia Data Release 3. Stellar multiplicity, a teaser for the hidden treasure}",
      journal = {\aap},
         year = 2023,
        month = jun,
       volume = {674},
          eid = {A34},
        pages = {A34},
          doi = {10.1051/0004-6361/202243782},
archivePrefix = {arXiv},
       eprint = {2206.05595},
 primaryClass = {astro-ph.SR},
       adsurl = {https://ui.adsabs.harvard.edu/abs/2023A&A...674A..34G}
}

@ARTICLE{Gandhi2022,
       author = {{Gandhi}, P. and {Buckley}, D.~A.~H. and {Charles}, P.~A. and {Hodgkin}, S. and {Scaringi}, S. and {Knigge}, C. and {Rao}, A. and {Paice}, J.~A. and {Zhao}, Y.},
        title = "{Astrometric excess noise in Gaia EDR3 and the search for X-ray binaries}",
      journal = {\mnras},
         year = 2022,
        month = mar,
       volume = {510},
       number = {3},
        pages = {3885-3895},
          doi = {10.1093/mnras/stab3771},
archivePrefix = {arXiv},
       eprint = {2201.00833},
 primaryClass = {astro-ph.HE},
       adsurl = {https://ui.adsabs.harvard.edu/abs/2022MNRAS.510.3885G}
}

@ARTICLE{Pecaut2013,
       author = {{Pecaut}, Mark J. and {Mamajek}, Eric E.},
        title = "{Intrinsic Colors, Temperatures, and Bolometric Corrections of Pre-main-sequence Stars}",
      journal = {\apjs},
         year = 2013,
        month = sep,
       volume = {208},
       number = {1},
          eid = {9},
        pages = {9},
          doi = {10.1088/0067-0049/208/1/9},
archivePrefix = {arXiv},
       eprint = {1307.2657},
 primaryClass = {astro-ph.SR},
       adsurl = {https://ui.adsabs.harvard.edu/abs/2013ApJS..208....9P}
}

@ARTICLE{Breiman2001,
       author = {{Breiman}, Leo},
        title = "{Random Forests.}",
      journal = {Machine Learning},
         year = 2001,
        month = jan,
       volume = {45},
        pages = {5-32},
          doi = {10.1023/A:1010933404324},
       adsurl = {https://ui.adsabs.harvard.edu/abs/2001MachL..45....5B}
}

@INPROCEEDINGS{McCarthy1986,
       author = {{McCarthy}, Jr., D.~W. and {Cobb}, M.~L.},
        title = "{Application of image sharpness criteria in infrared speckle interferometry.}",
    booktitle = {Instrumentation in astronomy VI},
         year = 1986,
       editor = {{Crawford}, David L.},
       series = {Society of Photo-Optical Instrumentation Engineers (SPIE) Conference Series},
       volume = {627},
        month = jan,
        pages = {797-804},
          doi = {10.1117/12.968161},
       adsurl = {https://ui.adsabs.harvard.edu/abs/1986SPIE..627..797M}
}

@ARTICLE{Tokovinin2010,
       author = {{Tokovinin}, Andrei and {Mason}, Brian D. and {Hartkopf}, William I.},
        title = "{Speckle Interferometry at the Blanco and SOAR Telescopes in 2008 and 2009}",
      journal = {\aj},
         year = 2010,
        month = feb,
       volume = {139},
       number = {2},
        pages = {743-756},
          doi = {10.1088/0004-6256/139/2/743},
archivePrefix = {arXiv},
       eprint = {0911.5718},
 primaryClass = {astro-ph.SR},
       adsurl = {https://ui.adsabs.harvard.edu/abs/2010AJ....139..743T}
}

@ARTICLE{Luna2020,
       author = {{Luna}, Alonso and {Orlov}, Valeri G.},
        title = "{Speckle Holography of Visual Double Stars at the 2.1 m Telescope of OAN-SPM: First Results}",
      journal = {\aj},
         year = 2020,
        month = jul,
       volume = {160},
       number = {1},
          eid = {9},
        pages = {9},
          doi = {10.3847/1538-3881/ab9120},
       adsurl = {https://ui.adsabs.harvard.edu/abs/2020AJ....160....9L}
}

@ARTICLE{Christou1986,
       author = {{Christou}, J.~C. and {Hege}, E.~K. and {Freeman}, J.~D. and {Ribak}, E.},
        title = "{Self-calibrating shift-and-add technique for speckle imaging}",
      journal = {Journal of the Optical Society of America A},
         year = 1986,
        month = feb,
       volume = {3},
        pages = {204-209},
          doi = {10.1364/JOSAA.3.000204},
       adsurl = {https://ui.adsabs.harvard.edu/abs/1986JOSAA...3..204C}
}

@ARTICLE{Riello2021,
       author = {{Riello}, M. and {De Angeli}, F. and {Evans}, D.~W. and {Montegriffo}, P. and {Carrasco}, J.~M. and {Busso}, G. and {Palaversa}, L. and {Burgess}, P.~W. and {Diener}, C. and {Davidson}, M. and {Rowell}, N. and {Fabricius}, C. and {Jordi}, C. and {Bellazzini}, M. and {Pancino}, E. and {Harrison}, D.~L. and {Cacciari}, C. and {van Leeuwen}, F. and {Hambly}, N.~C. and {Hodgkin}, S.~T. and {Osborne}, P.~J. and {Altavilla}, G. and {Barstow}, M.~A. and {Brown}, A.~G.~A. and {Castellani}, M. and {Cowell}, S. and {De Luise}, F. and {Gilmore}, G. and {Giuffrida}, G. and {Hidalgo}, S. and {Holland}, G. and {Marinoni}, S. and {Pagani}, C. and {Piersimoni}, A.~M. and {Pulone}, L. and {Ragaini}, S. and {Rainer}, M. and {Richards}, P.~J. and {Sanna}, N. and {Walton}, N.~A. and {Weiler}, M. and {Yoldas}, A.},
        title = "{Gaia Early Data Release 3. Photometric content and validation}",
      journal = {\aap},
         year = 2021,
        month = may,
       volume = {649},
          eid = {A3},
        pages = {A3},
          doi = {10.1051/0004-6361/202039587},
archivePrefix = {arXiv},
       eprint = {2012.01916},
 primaryClass = {astro-ph.IM},
       adsurl = {https://ui.adsabs.harvard.edu/abs/2021A&A...649A...3R}
}

@ARTICLE{Golovin2023,
       author = {{Golovin}, Alex and {Reffert}, Sabine and {Just}, Andreas and {Jordan}, Stefan and {Vani}, Akash and {Jahrei{\ss}}, Hartmut},
        title = "{The Fifth Catalogue of Nearby Stars (CNS5)}",
      journal = {\aap},
         year = 2023,
        month = feb,
       volume = {670},
          eid = {A19},
        pages = {A19},
          doi = {10.1051/0004-6361/202244250},
archivePrefix = {arXiv},
       eprint = {2211.01449},
 primaryClass = {astro-ph.SR},
       adsurl = {https://ui.adsabs.harvard.edu/abs/2023A&A...670A..19G}
}

@ARTICLE{Nordstrom2004,
       author = {{Nordstr{\"o}m}, B. and {Mayor}, M. and {Andersen}, J. and {Holmberg}, J. and {Pont}, F. and {J{\o}rgensen}, B.~R. and {Olsen}, E.~H. and {Udry}, S. and {Mowlavi}, N.},
        title = "{The Geneva-Copenhagen survey of the Solar neighbourhood. Ages, metallicities, and kinematic properties of {\ensuremath{\sim}}14 000 F and G dwarfs}",
      journal = {\aap},
         year = 2004,
        month = may,
       volume = {418},
        pages = {989-1019},
          doi = {10.1051/0004-6361:20035959},
archivePrefix = {arXiv},
       eprint = {astro-ph/0405198},
 primaryClass = {astro-ph},
       adsurl = {https://ui.adsabs.harvard.edu/abs/2004A&A...418..989N}
}

@ARTICLE{Lee2019,
       author = {{Lee}, Aaron T. and {Offner}, Stella S.~R. and {Kratter}, Kaitlin M. and {Smullen}, Rachel A. and {Li}, Pak Shing},
        title = "{The Formation and Evolution of Wide-orbit Stellar Multiples In Magnetized Clouds}",
      journal = {\apj},
         year = 2019,
        month = dec,
       volume = {887},
       number = {2},
          eid = {232},
        pages = {232},
          doi = {10.3847/1538-4357/ab584b},
archivePrefix = {arXiv},
       eprint = {1911.07863},
 primaryClass = {astro-ph.GA},
       adsurl = {https://ui.adsabs.harvard.edu/abs/2019ApJ...887..232L}
}

@ARTICLE{Moe2017,
       author = {{Moe}, Maxwell and {Di Stefano}, Rosanne},
        title = "{Mind Your Ps and Qs: The Interrelation between Period (P) and Mass-ratio (Q) Distributions of Binary Stars}",
      journal = {\apjs},
         year = 2017,
        month = jun,
       volume = {230},
       number = {2},
          eid = {15},
        pages = {15},
          doi = {10.3847/1538-4365/aa6fb6},
archivePrefix = {arXiv},
       eprint = {1606.05347},
 primaryClass = {astro-ph.SR},
       adsurl = {https://ui.adsabs.harvard.edu/abs/2017ApJS..230...15M}
}

@ARTICLE{Andersen1991,
       author = {{Andersen}, J.},
        title = "{Accurate masses and radii of normal stars}",
      journal = {\aapr},
         year = 1991,
        month = jan,
       volume = {3},
       number = {2},
        pages = {91-126},
          doi = {10.1007/BF00873538},
       adsurl = {https://ui.adsabs.harvard.edu/abs/1991A&ARv...3...91A}
}

@ARTICLE{Henry1999,
       author = {{Henry}, Todd J. and {Franz}, Otto G. and {Wasserman}, Lawrence H. and {Benedict}, G. Fritz and {Shelus}, Peter J. and {Ianna}, Philip A. and {Kirkpatrick}, J. Davy and {McCarthy}, Jr., Donald W.},
        title = "{The Optical Mass-Luminosity Relation at the End of the Main Sequence (0.08-0.20 M$_{solar}$)}",
      journal = {\apj},
         year = 1999,
        month = feb,
       volume = {512},
       number = {2},
        pages = {864-873},
          doi = {10.1086/306793},
       adsurl = {https://ui.adsabs.harvard.edu/abs/1999ApJ...512..864H}
}

@ARTICLE{Harpsoe2012,
       author = {{Harps{\o}e}, K.~B.~W. and {Andersen}, M.~I. and {Kj{\ae}gaard}, P.},
        title = "{Bayesian photon counting with electron-multiplying charge coupled devices (EMCCDs)}",
      journal = {\aap},
         year = 2012,
        month = jan,
       volume = {537},
          eid = {A50},
        pages = {A50},
          doi = {10.1051/0004-6361/201117089},
archivePrefix = {arXiv},
       eprint = {1111.2066},
 primaryClass = {astro-ph.IM},
       adsurl = {https://ui.adsabs.harvard.edu/abs/2012A&A...537A..50H}
}

@ARTICLE{Simbad2000,
       author = {{Wenger}, M. and {Ochsenbein}, F. and {Egret}, D. and {Dubois}, P. and {Bonnarel}, F. and {Borde}, S. and {Genova}, F. and {Jasniewicz}, G. and {Lalo{\"e}}, S. and {Lesteven}, S. and {Monier}, R.},
        title = "{The SIMBAD astronomical database. The CDS reference database for astronomical objects}",
      journal = {\aaps},
         year = 2000,
        month = apr,
       volume = {143},
        pages = {9-22},
          doi = {10.1051/aas:2000332},
archivePrefix = {arXiv},
       eprint = {astro-ph/0002110},
 primaryClass = {astro-ph},
       adsurl = {https://ui.adsabs.harvard.edu/abs/2000A&AS..143....9W}
}

@ARTICLE{Ciardi2015,
       author = {{Ciardi}, David R. and {Beichman}, Charles A. and {Horch}, Elliott P. and {Howell}, Steve B.},
        title = "{Understanding the Effects of Stellar Multiplicity on the Derived Planet Radii from Transit Surveys: Implications for Kepler, K2, and TESS}",
      journal = {\apj},
         year = 2015,
        month = may,
       volume = {805},
       number = {1},
          eid = {16},
        pages = {16},
          doi = {10.1088/0004-637X/805/1/16},
archivePrefix = {arXiv},
       eprint = {1503.03516},
 primaryClass = {astro-ph.EP},
       adsurl = {https://ui.adsabs.harvard.edu/abs/2015ApJ...805...16C}
}

@ARTICLE{Furlan2017,
       author = {{Furlan}, E. and {Howell}, S.~B.},
        title = "{The Densities of Planets in Multiple Stellar Systems}",
      journal = {\aj},
         year = 2017,
        month = aug,
       volume = {154},
       number = {2},
          eid = {66},
        pages = {66},
          doi = {10.3847/1538-3881/aa7b70},
archivePrefix = {arXiv},
       eprint = {1707.01942},
 primaryClass = {astro-ph.EP},
       adsurl = {https://ui.adsabs.harvard.edu/abs/2017AJ....154...66F}
}

@ARTICLE{Ricker2015,
       author = {{Ricker}, George R. and {Winn}, Joshua N. and {Vanderspek}, Roland and {Latham}, David W. and {Bakos}, G{\'a}sp{\'a}r {\'A}. and {Bean}, Jacob L. and {Berta-Thompson}, Zachory K. and {Brown}, Timothy M. and {Buchhave}, Lars and {Butler}, Nathaniel R. and {Butler}, R. Paul and {Chaplin}, William J. and {Charbonneau}, David and {Christensen-Dalsgaard}, J{\o}rgen and {Clampin}, Mark and {Deming}, Drake and {Doty}, John and {De Lee}, Nathan and {Dressing}, Courtney and {Dunham}, Edward W. and {Endl}, Michael and {Fressin}, Francois and {Ge}, Jian and {Henning}, Thomas and {Holman}, Matthew J. and {Howard}, Andrew W. and {Ida}, Shigeru and {Jenkins}, Jon M. and {Jernigan}, Garrett and {Johnson}, John Asher and {Kaltenegger}, Lisa and {Kawai}, Nobuyuki and {Kjeldsen}, Hans and {Laughlin}, Gregory and {Levine}, Alan M. and {Lin}, Douglas and {Lissauer}, Jack J. and {MacQueen}, Phillip and {Marcy}, Geoffrey and {McCullough}, Peter R. and {Morton}, Timothy D. and {Narita}, Norio and {Paegert}, Martin and {Palle}, Enric and {Pepe}, Francesco and {Pepper}, Joshua and {Quirrenbach}, Andreas and {Rinehart}, Stephen A. and {Sasselov}, Dimitar and {Sato}, Bun'ei and {Seager}, Sara and {Sozzetti}, Alessandro and {Stassun}, Keivan G. and {Sullivan}, Peter and {Szentgyorgyi}, Andrew and {Torres}, Guillermo and {Udry}, Stephane and {Villasenor}, Joel},
        title = "{Transiting Exoplanet Survey Satellite (TESS)}",
      journal = {Journal of Astronomical Telescopes, Instruments, and Systems},
         year = 2015,
        month = jan,
       volume = {1},
          eid = {014003},
        pages = {014003},
          doi = {10.1117/1.JATIS.1.1.014003},
       adsurl = {https://ui.adsabs.harvard.edu/abs/2015JATIS...1a4003R}
}

@ARTICLE{Belokurov2020,
       author = {{Belokurov}, Vasily and {Penoyre}, Zephyr and {Oh}, Semyeong and {Iorio}, Giuliano and {Hodgkin}, Simon and {Evans}, N. Wyn and {Everall}, Andrew and {Koposov}, Sergey E. and {Tout}, Christopher A. and {Izzard}, Robert and {Clarke}, Cathie J. and {Brown}, Anthony G.~A.},
        title = "{Unresolved stellar companions with Gaia DR2 astrometry}",
      journal = {\mnras},
         year = 2020,
        month = aug,
       volume = {496},
       number = {2},
        pages = {1922-1940},
          doi = {10.1093/mnras/staa1522},
archivePrefix = {arXiv},
       eprint = {2003.05467},
 primaryClass = {astro-ph.SR},
       adsurl = {https://ui.adsabs.harvard.edu/abs/2020MNRAS.496.1922B}
}

@ARTICLE{Castro-G2024,
       author = {{Castro-Ginard}, Alfred and {Penoyre}, Zephyr and {Casey}, Andrew R. and {Brown}, Anthony G.~A. and {Belokurov}, Vasily and {Cantat-Gaudin}, Tristan and {Drimmel}, Ronald and {Fouesneau}, Morgan and {Khanna}, Shourya and {Kurbatov}, Evgeny P. and {Price-Whelan}, Adrian M. and {Rix}, Hans-Walter and {Smart}, Richard L.},
        title = "{Gaia DR3 detectability of unresolved binary systems}",
      journal = {\aap},
         year = 2024,
        month = aug,
       volume = {688},
          eid = {A1},
        pages = {A1},
          doi = {10.1051/0004-6361/202450172},
archivePrefix = {arXiv},
       eprint = {2404.14127},
 primaryClass = {astro-ph.GA},
       adsurl = {https://ui.adsabs.harvard.edu/abs/2024A&A...688A...1C}
}

@ARTICLE{ElBadry2024,
       author = {{El-Badry}, Kareem},
        title = "{Gaia's binary star renaissance}",
      journal = {\nar},
         year = 2024,
        month = jun,
       volume = {98},
          eid = {101694},
        pages = {101694},
          doi = {10.1016/j.newar.2024.101694},
archivePrefix = {arXiv},
       eprint = {2403.12146},
 primaryClass = {astro-ph.SR},
       adsurl = {https://ui.adsabs.harvard.edu/abs/2024NewAR..9801694E}
}

@ARTICLE{Mason2023,
       author = {{Mason}, Brian D. and {Tokovinin}, Andrei and {Mendez}, Rene A. and {Costa}, Edgardo},
        title = "{Speckle Interferometry at SOAR in 2022}",
      journal = {\aj},
         year = 2023,
        month = oct,
       volume = {166},
       number = {4},
          eid = {139},
        pages = {139},
          doi = {10.3847/1538-3881/acedaf},
archivePrefix = {arXiv},
       eprint = {2405.10071},
 primaryClass = {astro-ph.SR},
       adsurl = {https://ui.adsabs.harvard.edu/abs/2023AJ....166..139M}
}

@ARTICLE{Tokovinin2024,
       author = {{Tokovinin}, Andrei and {Mason}, Brian D. and {Mendez}, Rene A. and {Costa}, Edgardo},
        title = "{Speckle Interferometry at SOAR Telescope in 2023}",
      journal = {\aj},
         year = 2024,
        month = jul,
       volume = {168},
       number = {1},
          eid = {28},
        pages = {28},
          doi = {10.3847/1538-3881/ad4d56},
       adsurl = {https://ui.adsabs.harvard.edu/abs/2024AJ....168...28T}
}

@ARTICLE{Raghavan2010,
       author = {{Raghavan}, Deepak and {McAlister}, Harold A. and {Henry}, Todd J. and {Latham}, David W. and {Marcy}, Geoffrey W. and {Mason}, Brian D. and {Gies}, Douglas R. and {White}, Russel J. and {ten Brummelaar}, Theo A.},
        title = "{A Survey of Stellar Families: Multiplicity of Solar-type Stars}",
      journal = {\apjs},
         year = 2010,
        month = sep,
       volume = {190},
       number = {1},
        pages = {1-42},
          doi = {10.1088/0067-0049/190/1/1},
archivePrefix = {arXiv},
       eprint = {1007.0414},
 primaryClass = {astro-ph.SR},
       adsurl = {https://ui.adsabs.harvard.edu/abs/2010ApJS..190....1R}
}

@ARTICLE{Roberts2017,
       author = {{Roberts}, Jr., Lewis C. and {Tokovinin}, Andrei and {Mason}, Brian D. and {Marinan}, Anne D.},
        title = "{Continued Kinematic and Photometric Investigations of Hierarchical Solar-type Multiple Star Systems}",
      journal = {\aj},
         year = 2017,
        month = mar,
       volume = {153},
       number = {3},
          eid = {100},
        pages = {100},
          doi = {10.3847/1538-3881/153/3/100},
archivePrefix = {arXiv},
       eprint = {1701.05865},
 primaryClass = {astro-ph.SR},
       adsurl = {https://ui.adsabs.harvard.edu/abs/2017AJ....153..100R}
}

@ARTICLE{Clark2024,
       author = {{Clark}, Catherine A. and {van Belle}, Gerard T. and {Horch}, Elliott P. and {Ciardi}, David R. and {von Braun}, Kaspar and {Skiff}, Brian A. and {Winters}, Jennifer G. and {Lund}, Michael B. and {Everett}, Mark E. and {Hartman}, Zachary D. and {Llama}, Joe},
        title = "{The POKEMON Speckle Survey of Nearby M Dwarfs. III. The Stellar Multiplicity Rate of M Dwarfs within 15 pc}",
      journal = {\aj},
         year = 2024,
        month = apr,
       volume = {167},
       number = {4},
          eid = {174},
        pages = {174},
          doi = {10.3847/1538-3881/ad267d},
archivePrefix = {arXiv},
       eprint = {2401.14703},
 primaryClass = {astro-ph.SR},
       adsurl = {https://ui.adsabs.harvard.edu/abs/2024AJ....167..174C}
}

@ARTICLE{Mendez2025,
       author = {{Mendez}, Rene A. and {Tokovinin}, Andrei and {Costa}, Edgardo and {Dirk}, Maximiliano},
        title = "{Southern Binaries with the Zorro Speckle Camera @ Gemini South}",
      journal = {\aj},
         year = 2025,
        month = apr,
       volume = {169},
       number = {4},
          eid = {226},
        pages = {226},
          doi = {10.3847/1538-3881/adbbb4},
archivePrefix = {arXiv},
       eprint = {2503.08721},
 primaryClass = {astro-ph.SR},
       adsurl = {https://ui.adsabs.harvard.edu/abs/2025AJ....169..226M}
}

@ARTICLE{Delorme2012,
       author = {{Delorme}, P. and {Lagrange}, A.~M. and {Chauvin}, G. and {Bonavita}, M. and {Lacour}, S. and {Bonnefoy}, M. and {Ehrenreich}, D. and {Beust}, H.},
        title = "{High-resolution imaging of young M-type stars of the solar neighbourhood: probing for companions down to the mass of Jupiter}",
      journal = {\aap},
         year = 2012,
        month = mar,
       volume = {539},
          eid = {A72},
        pages = {A72},
          doi = {10.1051/0004-6361/201118223},
archivePrefix = {arXiv},
       eprint = {1112.3008},
 primaryClass = {astro-ph.SR},
       adsurl = {https://ui.adsabs.harvard.edu/abs/2012A&A...539A..72D}
}

@ARTICLE{Janson2012,
       author = {{Janson}, Markus and {Hormuth}, Felix and {Bergfors}, Carolina and {Brandner}, Wolfgang and {Hippler}, Stefan and {Daemgen}, Sebastian and {Kudryavtseva}, Natalia and {Schmalzl}, Eva and {Schnupp}, Carolin and {Henning}, Thomas},
        title = "{The AstraLux Large M-dwarf Multiplicity Survey}",
      journal = {\apj},
         year = 2012,
        month = jul,
       volume = {754},
       number = {1},
          eid = {44},
        pages = {44},
          doi = {10.1088/0004-637X/754/1/44},
archivePrefix = {arXiv},
       eprint = {1205.4718},
 primaryClass = {astro-ph.SR},
       adsurl = {https://ui.adsabs.harvard.edu/abs/2012ApJ...754...44J}
}

@ARTICLE{Winters2021,
       author = {{Winters}, Jennifer G. and {Charbonneau}, David and {Henry}, Todd J. and {Irwin}, Jonathan M. and {Jao}, Wei-Chun and {Riedel}, Adric R. and {Slatten}, Kenneth},
        title = "{The Volume-complete Sample of M Dwarfs with Masses 0.1 {\ensuremath{\leq}} M/M$_{☉}$ {\ensuremath{\leq}} 0.3 within 15 Parsecs}",
      journal = {\aj},
         year = 2021,
        month = feb,
       volume = {161},
       number = {2},
          eid = {63},
        pages = {63},
          doi = {10.3847/1538-3881/abcc74},
archivePrefix = {arXiv},
       eprint = {2011.09409},
 primaryClass = {astro-ph.SR},
       adsurl = {https://ui.adsabs.harvard.edu/abs/2021AJ....161...63W}
}

@ARTICLE{Vrijmoet2026,
       author = {{Vrijmoet}, Eliot Halley and {Tokovinin}, Andrei and {Henry}, Todd J. and {Winters}, Jennifer G. and {Jao}, Wei-Chun and {Horch}, Elliott},
        title = "{The Solar Neighborhood LIV: 54 Orbits of M Dwarf Multiples within 30 pc with Speckle Interferometry at SOAR}",
      journal = {\aj},
         year = 2026,
        month = mar,
       volume = {171},
       number = {3},
          eid = {186},
        pages = {186},
          doi = {10.3847/1538-3881/ae3b40},
archivePrefix = {arXiv},
       eprint = {2601.18905},
 primaryClass = {astro-ph.SR},
       adsurl = {https://ui.adsabs.harvard.edu/abs/2026AJ....171..186V}
}

@ARTICLE{Tokovinin2025,
       author = {{Tokovinin}, A.},
        title = "{Speckle Interferometry at SOAR}",
      journal = {The NOIRLab Mirror},
         year = 2025,
        month = jun,
       volume = {9},
        pages = {11},
       adsurl = {https://ui.adsabs.harvard.edu/abs/2025Mirro...9...11T}
}

@ARTICLE{Salama2022,
       author = {{Salama}, Ma{\"\i}ssa and {Ziegler}, Carl and {Baranec}, Christoph and {Liu}, Michael C. and {Law}, Nicholas M. and {Riddle}, Reed and {Henry}, Todd J. and {Winters}, Jennifer G. and {Jao}, Wei-Chun and {Ou}, James and {Hermosillo Ruiz}, Arcelia},
        title = "{An Adaptive Optics Census of Companions to Northern Stars Within 25 pc with Robo-AO}",
      journal = {\aj},
         year = 2022,
        month = may,
       volume = {163},
       number = {5},
          eid = {200},
        pages = {200},
          doi = {10.3847/1538-3881/ac53fc},
archivePrefix = {arXiv},
       eprint = {2203.11250},
 primaryClass = {astro-ph.SR},
       adsurl = {https://ui.adsabs.harvard.edu/abs/2022AJ....163..200S}
}

@ARTICLE{Ziegler2020,
       author = {{Ziegler}, Carl and {Tokovinin}, Andrei and {Brice{\~n}o}, C{\'e}sar and {Mang}, James and {Law}, Nicholas and {Mann}, Andrew W.},
        title = "{SOAR TESS Survey. I. Sculpting of TESS Planetary Systems by Stellar Companions}",
      journal = {\aj},
         year = 2020,
        month = jan,
       volume = {159},
       number = {1},
          eid = {19},
        pages = {19},
          doi = {10.3847/1538-3881/ab55e9},
archivePrefix = {arXiv},
       eprint = {1908.10871},
 primaryClass = {astro-ph.EP},
       adsurl = {https://ui.adsabs.harvard.edu/abs/2020AJ....159...19Z}
}

@ARTICLE{Vizier,
       author = {{Ochsenbein}, F. and {Bauer}, P. and {Marcout}, J.},
        title = "{The VizieR database of astronomical catalogues}",
      journal = {\aaps},
         year = 2000,
        month = apr,
       volume = {143},
        pages = {23-32},
          doi = {10.1051/aas:2000169},
archivePrefix = {arXiv},
       eprint = {astro-ph/0002122},
 primaryClass = {astro-ph},
       adsurl = {https://ui.adsabs.harvard.edu/abs/2000A&AS..143...23O}
}

@ARTICLE{Penoyre2020,
       author = {{Penoyre}, Zephyr and {Belokurov}, Vasily and {Wyn Evans}, N. and {Everall}, A. and {Koposov}, S.~E.},
        title = "{Binary deviations from single object astrometry}",
      journal = {\mnras},
         year = 2020,
        month = jun,
       volume = {495},
       number = {1},
        pages = {321-337},
          doi = {10.1093/mnras/staa1148},
archivePrefix = {arXiv},
       eprint = {2003.05456},
 primaryClass = {astro-ph.GA},
       adsurl = {https://ui.adsabs.harvard.edu/abs/2020MNRAS.495..321P}
}

@ARTICLE{Kervella2019,
       author = {{Kervella}, Pierre and {Arenou}, Fr{\'e}d{\'e}ric and {Mignard}, Fran{\c{c}}ois and {Th{\'e}venin}, Fr{\'e}d{\'e}ric},
        title = "{Stellar and substellar companions of nearby stars from Gaia DR2. Binarity from proper motion anomaly}",
      journal = {\aap},
         year = 2019,
        month = mar,
       volume = {623},
          eid = {A72},
        pages = {A72},
          doi = {10.1051/0004-6361/201834371},
archivePrefix = {arXiv},
       eprint = {1811.08902},
 primaryClass = {astro-ph.SR},
       adsurl = {https://ui.adsabs.harvard.edu/abs/2019A&A...623A..72K}
}

@ARTICLE{Vrijmoet2020,
       author = {{Vrijmoet}, Eliot Halley and {Henry}, Todd J. and {Jao}, Wei-Chun and {Dieterich}, Serge B.},
        title = "{The Solar Neighborhood. XLVI. Revealing New M Dwarf Binaries and Their Orbital Architectures}",
      journal = {\aj},
         year = 2020,
        month = nov,
       volume = {160},
       number = {5},
          eid = {215},
        pages = {215},
          doi = {10.3847/1538-3881/abb4e9},
archivePrefix = {arXiv},
       eprint = {2009.00121},
 primaryClass = {astro-ph.SR},
       adsurl = {https://ui.adsabs.harvard.edu/abs/2020AJ....160..215V}
}


\section{ACKNOWLEDGEMENTS}

This research is supported by the Direcci\'on General de Asuntos del Personal Acad\'emico (UNAM, M\'exico) under project IN114123 and IN106226. Based upon observations acquired at the Observatorio Astron\'omico Nacional in the Sierra San Pedro M\'artir (OAN-SPM), Baja California, M\'exico and on observations collected at the European Organization for Astronomical Research in the Southern Hemisphere under ESO programme(s) 114.27G (65.2 hrs) and 115.2800 (78.3 hrs) with PI Rene A. Mendez. The authors thank the daytime and night support staff at the OAN-SPM for facilitating and helping to obtain our observations; from Chile (RAMB) FONDECYT/ANID \# 124 0049 and Fondo GEMINI, Astrónomo de Soporte GEMINI-ANID grant \# 3223 AS0002.


\onecolumn
\begin{longtable}{lccccccc}
\caption{Five points of the 5$\sigma$ contrast curves for 115 data cubes corresponding to 102 stars without close companions.}\\
\label{Tab:ContCurv5p} \\

\toprule
\toprule
Gaia DR3 & Julian Year & Filter & $\Delta m$ at $0.1''$ & $\Delta m$ at $0.2''$ & $\Delta m$ at $0.3''$ & $\Delta m$ at $0.5''$ & $\Delta m$ at $ 0.8''$ \\
 &  &  & $mag$ & $mag$ & $mag$ & $mag$ & $mag$ \\
\midrule
\endfirsthead

\caption*{continued.}\\
\toprule
\toprule
Gaia DR3 & Julian Year & Filter & $\Delta m$ at $0.1''$ & $\Delta m$ at $0.2''$ & $\Delta m$ at $0.3''$ & $\Delta m$ at $0.5''$ & $\Delta m$ at $ 0.8''$ \\
\midrule
\endhead

\bottomrule
\bottomrule
\endfoot

Gaia DR3 1155276250492687616 & 2022.5482 & R & nan & 2.41 & 3.177 & 4.265 & 6.324	\\
Gaia DR3 1159745146784043264 & 2022.5454 & R & nan & 1.486 & 2.285 & 3.417 & 5.478	\\
Gaia DR3 1191518146611956096 & 2022.5455 & B & 3.016 & 5.07 & 5.946 & 7.112 & 9.451	\\
Gaia DR3 1193030490492925824 & 2022.5455 & U & 2.239 & 4.038 & 4.783 & 5.747 & 8.217	\\
Gaia DR3 1201405126966920448 & 2022.5482 & R & nan & 1.035 & 1.7 & 2.95 & 4.963	\\
Gaia DR3 1205070452055788928 & 2022.5482 & R & nan & 1.671 & 2.208 & 3.372 & 5.447	\\
Gaia DR3 1209281375431039104 & 2022.5454 & I & 0.757 & 2.717 & 3.337 & 4.228 & 5.749	\\
Gaia DR3 1214160733157163008 & 2022.5454 & I & 3.653 & 5.462 & 6.099 & 6.974 & 8.875	\\
Gaia DR3 1314438839310796800 & 2022.5427 & I & 5.025 & 6.877 & 7.575 & 8.458 & 10.095	\\
Gaia DR3 1323427759184545536 & 2022.5426 & U & 1.302 & 2.946 & 3.548 & 4.442 & 6.917	\\
Gaia DR3 1327542647092300288 & 2022.5427 & I & 1.086 & 2.741 & 3.359 & 4.231 & 5.983	\\
Gaia DR3 1333760419707327744 & 2022.5427 & B & 5.056 & 6.776 & 7.463 & 8.438 & 10.621	\\
Gaia DR3 1341113541156480768 & 2022.5427 & I & 2.617 & 4.279 & 4.993 & 5.864 & 7.438	\\
Gaia DR3 1348028919540189056 & 2022.5427 & I & 2.994 & 4.524 & 5.196 & 6.214 & 8.021	\\
Gaia DR3 1349047857580994560 & 2022.5428 & I & 3.752 & 5.403 & 6.075 & 7.137 & 8.894	\\
Gaia DR3 1350714133093019904 & 2022.5428 & I & 2.786 & 4.364 & 4.977 & 6.142 & 7.986	\\
Gaia DR3 1355264565043431040 & 2022.5428 & I & 2.754 & 4.377 & 5.076 & 6.093 & 7.9	\\
Gaia DR3 1366580223320727808 & 2022.5428 & N & nan & 1.063 & 2.047 & 3.29 & 5.048	\\
Gaia DR3 1372215976327300480 & 2022.5426 & I & 1.088 & 2.795 & 3.35 & 4.198 & 6.061	\\
Gaia DR3 1379500756257031808 & 2022.5427 & B & 3.28 & 5.246 & 5.999 & 6.992 & 9.117	\\
Gaia DR3 1379500928055726848 & 2022.5427 & I & nan & 0.48 & 1.423 & 2.574 & 4.178	\\
Gaia DR3 1380821338441693312 & 2022.5426 & I & 4.224 & 5.898 & 6.558 & 7.456 & 9.439	\\
Gaia DR3 1384195086792268800 & 2022.5426 & U & 2.289 & 3.851 & 4.478 & 5.308 & 8.033	\\
Gaia DR3 1385293808145621504 & 2022.5426 & B & 3.194 & 5.056 & 5.739 & 6.635 & 8.803	\\
Gaia DR3 1408029436569383296 & 2022.5427 & V & 3.616 & 5.607 & 6.249 & 7.3 & 9.515	\\
Gaia DR3 1412645602995521920 & 2022.5428 & N & nan & 2.599 & 3.319 & 4.574 & 6.239	\\
Gaia DR3 1428030622526979968 & 2022.5427 & I & 3.21 & 5.065 & 5.878 & 6.741 & 8.601	\\
Gaia DR3 1428427236986209024 & 2022.5426 & I & 3.669 & 5.465 & 6.034 & 6.985 & 9.062	\\
Gaia DR3 1431176943768690816 & 2022.5427 & I & 1.021 & 2.866 & 3.499 & 4.459 & 6.284	\\
Gaia DR3 1776179945695834368 & 2022.5433 & I & nan & 1.954 & 2.549 & 3.418 & 5.292	\\
Gaia DR3 1776317182780536064 & 2022.5432 & I & 3.63 & 5.259 & 5.884 & 6.651 & 8.65	\\
Gaia DR3 1872046574983497216 & 2022.5431 & B & 3.92 & 5.687 & 6.37 & 7.363 & 9.704	\\
Gaia DR3 1872046609345556480 & 2022.5431 & B & 4.837 & 6.566 & 7.234 & 8.359 & 10.6	\\
Gaia DR3 1889703529122914560 & 2022.5433 & I & 2.225 & 3.87 & 4.463 & 5.287 & 7.135	\\
Gaia DR3 1891598193816300544 & 2022.5433 & B & 5.906 & 7.488 & 8.112 & 8.919 & 11.089	\\
Gaia DR3 1909518584084205312 & 2022.5433 & I & nan & 2.042 & 2.618 & 3.488 & 5.067	\\
Gaia DR3 1917144384415057536 & 2022.5434 & I & 1.824 & 3.408 & 4.022 & 4.965 & 6.782	\\
Gaia DR3 1923896622760418176 & 2022.5434 & I & nan & 2.913 & 3.653 & 4.526 & 6.125	\\
Gaia DR3 1923896725839633024 & 2022.5434 & I & 1.333 & 3.035 & 3.662 & 4.4 & 5.857	\\
Gaia DR3 1934263333784036736 & 2022.5433 & I & 4.163 & 5.742 & 6.348 & 7.185 & 9.168	\\
Gaia DR3 1940584323054654592 & 2022.5434 & I & 0.8 & 2.682 & 3.273 & 4.055 & 5.699	\\
Gaia DR3 1941011930001137280 & 2022.5434 & I & 1.025 & 2.664 & 3.257 & 4.143 & 5.937	\\
Gaia DR3 1952802469918554368 & 2022.5431 & I & 2.819 & 4.279 & 4.947 & 6.336 & 8.095	\\
Gaia DR3 1974606167761195904 & 2022.5431 & I & 0.79 & 2.371 & 3.101 & 4.262 & 6.054	\\
Gaia DR3 2009439555118168832 & 2022.5434 & I & 3.608 & 5.363 & 6.015 & 6.971 & 9.068	\\
Gaia DR3 2009481748875806976 & 2022.5434 & B & 3.42 & 5.53 & 6.348 & 7.445 & 9.769	\\
Gaia DR3 2034628782438989440 & 2022.5429 & I & 2.577 & 4.177 & 4.812 & 6.032 & 7.954	\\
Gaia DR3 2042871958509344896 & 2022.5430 & I & 2.022 & 3.827 & 4.547 & 5.322 & 7.053	\\
Gaia DR3 2046207670631964928 & 2022.5429 & B & 4.216 & 5.885 & 6.608 & 7.689 & 10.14	\\
Gaia DR3 2047690533883897856 & 2022.5430 & I & 0.985 & 2.648 & 3.275 & 4.249 & 6.087	\\
Gaia DR3 2050589430625453824 & 2022.5429 & I & 2.02 & 3.661 & 4.283 & 5.427 & 7.265	\\
Gaia DR3 2051984436005124480 & 2022.5429 & I & 2.622 & 4.2 & 4.884 & 6.058 & 7.91	\\
Gaia DR3 2063967910147595008 & 2022.5430 & N & 0.515 & 2.845 & 3.595 & 4.67 & 6.508	\\
Gaia DR3 2064100569188230144 & 2022.5430 & I & nan & 2.683 & 3.584 & 5.05 & 6.834	\\
Gaia DR3 2066768499405895296 & 2022.5431 & I & 0.946 & 2.61 & 3.335 & 4.464 & 6.158	\\
Gaia DR3 2070115588901082368 & 2022.5431 & I & 3.665 & 5.236 & 5.855 & 7.102 & 9.065	\\
Gaia DR3 2080285727861817088 & 2022.5430 & I & 1.171 & 2.926 & 3.559 & 4.459 & 6.309	\\
Gaia DR3 2090085056523454464 & 2022.5429 & I & 3.103 & 4.831 & 5.484 & 6.507 & 8.541	\\
Gaia DR3 2098413719661573248 & 2022.5429 & N & 0.346 & 2.706 & 3.421 & 4.507 & 6.35	\\
Gaia DR3 2110165780975185792 & 2022.5428 & I & 3.218 & 4.786 & 5.415 & 6.481 & 8.329	\\
Gaia DR3 2110336583235233536 & 2022.5428 & I & 2.641 & 4.202 & 4.769 & 5.937 & 7.734	\\
Gaia DR3 2115454019586473600 & 2022.5428 & I & 3.547 & 5.28 & 5.944 & 6.888 & 8.781	\\
Gaia DR3 2118424350249708800 & 2022.5428 & I & 4.202 & 5.862 & 6.528 & 7.775 & 9.783	\\
Gaia DR3 2118429538570209920 & 2022.5428 & I & 1.409 & 2.971 & 3.601 & 4.858 & 6.607	\\
Gaia DR3 2118429538570209920 & 2022.5428 & N & 1.411 & 3.366 & 4.149 & 5.439 & 7.298	\\
Gaia DR3 2136665454312416128 & 2022.5429 & I & 1.471 & 3.101 & 3.732 & 5.042 & 6.756	\\
Gaia DR3 2145277550935525760 & 2022.5428 & I & 5.334 & 7.052 & 7.713 & 8.669 & 10.866	\\
Gaia DR3 2171727299589482112 & 2022.5431 & I & 2.742 & 4.283 & 4.924 & 6.29 & 8.16	\\
Gaia DR3 2185710338703934976 & 2022.5431 & I & 1.533 & 3.28 & 4.026 & 5.168 & 6.94	\\
Gaia DR3 2719475542667772416 & 2022.5432 & U & 2.299 & 4.052 & 4.714 & 5.553 & 8.031	\\
Gaia DR3 2817999068780218368 & 2022.5433 & I & 2.894 & 4.437 & 4.985 & 5.982 & 8.034	\\
Gaia DR3 2825274743381064320 & 2022.5433 & I & 1.061 & 3.119 & 3.716 & 4.526 & 6.362	\\
Gaia DR3 2828928008202069376 & 2022.5432 & I & 4.97 & 6.574 & 7.168 & 8 & 10.041	\\
Gaia DR3 2830099469122381056 & 2022.5432 & I & 3.19 & 4.74 & 5.328 & 6.22 & 8.249	\\
Gaia DR3 2835207319109249920 & 2022.5432 & B & 4.148 & 5.961 & 6.625 & 7.496 & 9.652	\\
Gaia DR3 2864382138598091648 & 2022.5433 & I & 0.58 & 2.292 & 2.829 & 3.84 & 5.689	\\
Gaia DR3 2866851710433326208 & 2022.5433 & I & 0.831 & 2.416 & 3.011 & 4.05 & 5.92	\\
Gaia DR3 2881313793031884544 & 2022.5434 & I & nan & 1.682 & 2.239 & 3.082 & 4.797	\\
Gaia DR3 4330690742322011520 & 2022.5482 & R & 1.482 & 4.129 & 5.043 & 6.556 & 8.727	\\
Gaia DR3 4345775217221821312 & 2022.5482 & B & 3.353 & 5.642 & 6.798 & 8.258 & 10.906	\\
Gaia DR3 4347870370990204928 & 2022.5481 & R & 0.881 & 3.708 & 4.51 & 5.851 & 7.982	\\
Gaia DR3 4358031335898505472 & 2022.5482 & B & 3.267 & 5.442 & 6.544 & 8.03 & 10.591	\\
Gaia DR3 4358031335898505472 & 2022.5509 & B & 3.723 & 6.077 & 7.177 & 8.657 & 11.181	\\
Gaia DR3 4375233191015944192 & 2022.5509 & I & 6.275 & 8.101 & 8.871 & 10.351 & 12.13	\\
Gaia DR3 4375233191015944192 & 2022.5509 & R & 5.097 & 7.493 & 8.293 & 9.601 & 11.649	\\
Gaia DR3 4375233191015944192 & 2022.5509 & V & 4.039 & 6.347 & 7.229 & 8.566 & 10.683	\\
Gaia DR3 4389844948935164544 & 2022.5509 & I & 4.88 & 6.689 & 7.437 & 8.635 & 10.613	\\
Gaia DR3 4389844948935164544 & 2022.5509 & R & 3.759 & 5.957 & 6.659 & 8.116 & 10.153	\\
Gaia DR3 4389844948935164544 & 2022.5509 & V & 2.647 & 4.801 & 5.62 & 6.9 & 8.957	\\
Gaia DR3 4410702581435191296 & 2022.5454 & I & 1.014 & 2.866 & 3.534 & 4.393 & 6.338	\\
Gaia DR3 4418203587193988480 & 2022.5454 & B & 2.042 & 4.158 & 5.167 & 6.47 & 9.272	\\
Gaia DR3 4419140645978987904 & 2022.5454 & B & 1.62 & 3.749 & 4.73 & 6.106 & 8.918	\\
Gaia DR3 4423865487962417024 & 2022.5454 & B & 2.645 & 4.745 & 5.708 & 7.045 & 9.606	\\
Gaia DR3 4430238051199001216 & 2022.5454 & B & 4.511 & 6.561 & 7.512 & 8.737 & 11.171	\\
Gaia DR3 4439625337884136064 & 2022.5482 & R & 3.414 & 5.792 & 6.496 & 7.671 & 9.732	\\
Gaia DR3 4446574040918651904 & 2022.5482 & R & nan & 0.251 & 1.519 & 2.525 & 4.469	\\
Gaia DR3 4453239765733463936 & 2022.5509 & I & 0.524 & 2.665 & 3.345 & 4.725 & 6.299	\\
Gaia DR3 4453239765733463936 & 2022.5509 & R & nan & 2.667 & 3.349 & 4.764 & 6.525	\\
Gaia DR3 4453239765733463936 & 2022.5509 & V & nan & 1.013 & 1.815 & 3.021 & 5.03	\\
Gaia DR3 4467057362719443072 & 2022.5482 & R & nan & 1.214 & 1.813 & 3.033 & 5.059	\\
Gaia DR3 4547881191164080768 & 2022.5510 & I & 2.198 & 3.811 & 4.389 & 5.73 & 7.452	\\
Gaia DR3 4547881191164080768 & 2022.5510 & R & nan & 2.522 & 3.16 & 4.494 & 6.287	\\
Gaia DR3 4547881191164080768 & 2022.5510 & V & nan & nan & 0.954 & 2.281 & 4.322	\\
Gaia DR3 4550763526539955712 & 2022.5510 & I & 4.799 & 6.462 & 7.053 & 8.465 & 10.424	\\
Gaia DR3 4550763526539955712 & 2022.5510 & R & 3.996 & 5.888 & 6.605 & 7.834 & 9.894	\\
Gaia DR3 4550763526539955712 & 2022.5510 & V & 2.778 & 4.623 & 5.398 & 6.617 & 8.68	\\
Gaia DR3 4558856584955186432 & 2022.5482 & R & 0.979 & 3.216 & 3.843 & 5.039 & 6.811	\\
Gaia DR3 4572835535272627712 & 2022.5509 & I & 4.719 & 6.377 & 6.903 & 8.36 & 10.275	\\
Gaia DR3 4572835535272627712 & 2022.5509 & R & 3.849 & 5.634 & 6.363 & 7.623 & 9.676	\\
Gaia DR3 4572835535272627712 & 2022.5509 & V & 2.852 & 4.487 & 5.225 & 6.655 & 8.62	\\
Gaia DR3 4594497769766809216 & 2022.5510 & U & 3.925 & 5.485 & 6.225 & 7.643 & 10.144	\\
Gaia DR3 4610177733290768896 & 2022.5429 & B & 3.806 & 5.626 & 6.318 & 7.495 & 9.89	\\
Gaia DR3 6265472113586221696 & 2022.5454 & I & nan & 2.196 & 2.932 & 3.9 & 5.739	\\
Gaia DR3 6313600142709958144 & 2022.5453 & R & 1.494 & 4.193 & 5.133 & 6.25 & 8.614	\\
Gaia DR3 6322070093095493504 & 2022.5454 & I & 3.189 & 5.243 & 5.907 & 6.967 & 9.253	\\

\end{longtable}
\twocolumn

\end{document}